\documentclass[11pt,twoside,a4paper]{article}

\RequirePackage[OT1]{fontenc}
\RequirePackage{amsthm,amsmath}
\RequirePackage[numbers]{natbib}
\RequirePackage[colorlinks,citecolor=blue,urlcolor=blue]{hyperref}

\usepackage{mathtools}
\usepackage{amsmath}
\usepackage{graphicx}
\usepackage{grffile}
\usepackage{helvet} 
\usepackage{relsize}
\usepackage{multirow}
\usepackage{tabularx}
\usepackage{amssymb}
\usepackage{parskip}
\usepackage{natbib} 
\usepackage{inputenc}
\usepackage{amsthm} 
\usepackage{booktabs}
\usepackage{lscape}
\usepackage{hyperref}
\usepackage{array}
\usepackage{amsmath}
\usepackage{graphicx,psfrag,epsf}
\usepackage{enumerate}
\usepackage{natbib}
\usepackage{url} 
\usepackage{subfig}
\usepackage{wrapfig}
\usepackage{afterpage}
\usepackage{algpseudocode}
\usepackage{algorithm}

\newcommand{\blind}{0}

\newcolumntype{H}{>{\setbox0=\hbox\bgroup}c<{\egroup}@{}}
\newcommand{\s}{\ensuremath{s}}
\newcommand{\pkg}[1]{{\fontseries{b}\selectfont #1}} 
\newcommand{\vect}[1]{\boldsymbol{#1}}

\graphicspath{ {Image_New/} }
\numberwithin{equation}{section}
\newtheorem{theorem}{Theorem}[section]
\newtheorem{lemma}{Lemma}[section]
\newtheorem{property}{Property}[section]
\newtheorem{exmp}{Example}[section]

\begin{document}
	
	\def\spacingset#1{\renewcommand{\baselinestretch}%
	{#1}\small\normalsize} \spacingset{1}


	\title{\bf Testing the equality of multivariate means when $p>n$ by combining the Hotelling and Simes tests}
\author{Tzviel Frostig, Yoav Benjamini\thanks{
		The research leading to these results has received funding from the European Research Council: ERC grant agreement \textit{PSARPS\{294519\}.}}\hspace{.2cm}\\
	Department of Statsistics and Operation Research, University of Tel Aviv}
\maketitle

\begin{abstract}
	We propose a method of testing the shift between mean vectors of two multivariate Gaussian random variables in a high-dimensional setting incorporating the possible dependency and allowing $p > n$. This method is a combination of two well-known tests: the Hotelling test and the Simes test. The tests are integrated by sampling several dimensions at each iteration, testing each using the Hotelling test, and combining their results using the Simes test. We prove that this procedure is valid asymptotically. This procedure can be extended to handle non-equal covariance matrices by plugging in the appropriate extension of the Hotelling test. Using a simulation study, we show that the proposed test is advantageous over state-of-the-art tests in many scenarios and robust to violation of the Gaussian assumption. 
	
\end{abstract}

\noindent%
\vfill


	\section{Introduction}
	A common problem in statistics is testing the equality of two mean vectors $\vect{\mu}_X\in \mathbb{R}^p$ and $\vect{\mu}_Y \in \mathbb{R}^p$ based on two independent random samples, $X_i \; i = 1,\ldots, n_X$ and $Y_j \; j=1, \ldots, n_Y$.
	Formally, the hypothesis being tested is 
	\begin{equation} \label{eq:1}
	H_0: \vect{\mu_X} - \vect{\mu_Y} = \vect{\delta}  \quad \quad \quad H_1: \vect{\mu_X} - \vect{\mu_Y} \neq \vect{\delta}. 
	\end{equation} 
	This problem appears in many fields of science, including but not limited to analyzing EEG data \cite{hemmelmann2004multivariate}, fMRI data \cite{xiong2002generalized}, Neurobiology \cite{yang2016chromatin} and more.  
	At the first stage of the analysis the interest lies in detecting a difference and not necessarily identifying where this difference lies.
	A common approach to test the hypothesis is using the classical Hotelling $T^2$ test \cite{hotblling1931generalization} which assumes that $X$ and $Y$ follow the same multivariate normal distribution $X,Y \sim N\left( \vect{\mu}, \Sigma \right)$.

	\subsection{Hotelling $T^2$ test} \label{Hotelling}
	The Hotelling $T^2$ test statistic is defined as follows: 
	\begin{center}
		{$  T^2(X, Y) = \frac{n - p - 1}{(n - 2)p} \cdot \frac{n_X n_Y}{n}(\bar{X} - \bar{Y})'\hat{\Sigma}^{-1}(\bar{X}-\bar{Y})$},
	\end{center} 
	where $\bar{X}$ and $\bar{Y}$ are the estimated mean vectors, $n = n_X + n_Y$ and $\hat{\Sigma} =  (n_X + n_Y - 2)^{-1}\big(\sum_{i=1}^{n_X}(X_{i} - {\bar{X}})(X_{i} - {\bar{X}})'  + \sum_{j=1}^{n_Y}(Y_{j} - {\bar{Y}})(Y_{j} - {\bar{Y}})'\big)$ is the pooled covariance matrix estimator.  Under $H_0$, the test statistic follows the $F_{p, n -1 - p}$ distribution, the p-value of the test is given by $1 - F\left( T^2(X, Y) \right)_{p, n -1 - p}$, the null hypothesis is rejected for large value of the test statistic only. 
	
	The test is a natural generalization of the one-dimensional t-test. It is invariant under any non-singular linear transformations and is the uniformly most powerful test among all tests which share this invariance property \cite{bibby1979multivariate}. 
	The test cannot be used when $p > n$, a common situation in modern scientific research, as the estimated covariance matrix is not of full rank, and therefore its inverse is undefined. 
	Moreover, even for $p < n$, as $p$ approaches $n$, the power of the test deteriorates \cite{bai1996effect}.
	Several approaches have been suggested for testing \ref{eq:1} in the $p > n$ setting. Many of them rely on a different method to estimate $\Sigma$, which will guarantee that it would be of full rank. 
	
	\subsection{Simes test} \label{simes}
	
	A very different approach is to rely on the logical fact that $\vect{\mu}_X = \vect{\mu}_Y$ $\iff$ $\mu_{X,l} = \mu_{Y,l} \; \forall l ,  l = 1,\dots,p$. This allows us to test each dimension marginally and reject the global-null hypothesis using a combination test, such as Fisher, Simes, or Stouffer. 
	
		
	Assuming that $X$ and $Y$ follow the multivariate normal distribution, each one of their dimensions also follows the normal distribution $X_{l} \sim  N(\mu_{X_{l}}, \Sigma_{l,l})$ and $Y_{l} \sim N(\mu_{{Y}_l} , \Sigma_{l,l})$. 
	Under these conditions we can test the marginal hypotheses $H_{0_l}: \mu_{X_l} - \mu_{Y_l} = \delta_l$, by applying the t-test. 
	To complete the procedure, we need to find a test of the global null hypothesis, \begingroup\makeatletter\def\f@size{8}\check@mathfonts ${H_0 = \bigcap\limits_{l=1}^{p} H_{0_l}}$\endgroup. Since the tests are dependent, a good candidate is the Simes test \cite{simes1986improved}, which does not rely on the assumption of independence. 
	
	The Simes test, is a test of the global null, making use of the p-values, $\boldsymbol{p}$, which in our context, the p-values are acquired by marginally testing each dimension. The p-values are sorted, $p_{(1)} \leq p_{(2)} \leq \cdots \leq p_{(p)}$, and the test statistic is 
	
	\begin{equation*} 
		Simes(\boldsymbol{p}) =\min\limits_{l} {p_{(l)} \frac{p}{l}} \quad l \in {1,\ldots,p}.
	\end{equation*} 

	Under the null hypothesis, if the test statistics are independent then $Simes(\boldsymbol{p}) \sim U[0,1] $ \cite{simes1986improved}. While the Simes test offers a method to test the global-null, it ignores the covariance structure as each dimension is tested marginally. 	
	We suggest combining the Hotelling and Simes tests to construct a new test for the hypothesis of the two sample mean equality in the $p > n$ dependent setting.   \\

	The rest of the paper is organized as follows: In Section 2 we propose the test, in Section 3, default test parameters values are suggested.  In Section 4 we extend the test by removing the equal covariance assumption and the normality assumption using a permutation test based on the suggested test. In Section 5 we present the results of the simulation study comparing our suggested test with existing alternatives, and finally in Section 6 we apply the suggested test to two real-data examples.

	\section{The proposed test} 
	
	In order to reject the null hypothesis presented in Eq. \ref{eq:1}, it is enough to reject any marginal hypothesis $H_{0,l}: \mu_{X^*_l} - \mu_{Y^*_l} = \delta^*_l$. However, we need not limit ourselves to a single marginal hypothesis as we can test different subset of hypotheses, 	 $ \bigcap\limits_{l \in S} H_{0,l}, \; S \subset \{1,\cdots,p\}$. Any subset of hypotheses rejected indicates that the global null hypothesis should also be rejected. 
	
	Utilizing this idea, define $M$ to be the group of all possible $m$ subset of dimensions, $|M| = \binom{p}{m}$. At each iteration we sample $S_m \in M, |S| = m$ dimensions, and conduct the Hotelling $T^2$ test of the null hypothesis $\vect{\mu}_{1,S_m} = \vect{\mu}_{2,S_m}$. Repeat the process $B$ times and combine the $B$ resulting p-values from the random Hotelling statistics using the Simes test and compute $p_{simes}$, then test at level $\alpha$.   
	Algorithmically the test procedure is 

	\begin{algorithm}
		\caption{Suggested procedure}
		\begin{algorithmic}
			\State{$\boldsymbol{p} \leftarrow \emptyset$} 
			\For{$i = 1 \; \boldsymbol{to}  \;  B$} 
			\State{$S \leftarrow sample(p, m)$}
			\State{$\boldsymbol{p} \leftarrow \boldsymbol{p} \cup \left(1 -  \left( F\left( T^2(X_{S_m}, Y_{S_m}) \right)_{n, n - 1 - m} \right) \right)$}
			\EndFor 
			\State{$p_{simes} \gets Simes(\boldsymbol{p})$} \\
			\Return $p_{simes}$
		\end{algorithmic}
	\end{algorithm}

	Each Hotelling test under the intersection hypothesis ($\vect{\mu}_Y = \vect{\mu}_X$) yields a uniformly distributed p-value at step 2. We denoted the test as $SH_{m, B}(X, Y)$, for Simes-Hotelling. 
	One of the reasons to use the Simes test, is that it remains valid under various dependency structures of the test statistics. In particular, the Simes test has been proven to be conservative if the positive regression dependency on subset (PRDS) property between the test statistics holds \cite{benjamini2001control}. This is type of dependency likely to arise when two samples of dimensions overlap.
	
	\begin{property}[$\vect{PRDS}$]
		For any increasing set $D$, and for each $i\in I_0, \; P(\boldsymbol{X}\in D, X_i = x)$ is non-decreasing in $x$.
	\end{property}		
	Unfortunately, even if $X$ is PRDS, $|X|$ is not necessarily PRDS, and therefore does not assure the validity of Simes for p-values stemming from two-sided tests. For p-values originating from two-sided Gaussian tests, if $X$ is multivariate total positivity of order 2 (MTP$_2$), or if it is so after changing the sign of some of the $X_i$-s, then so is $|X|$, thereby implying the validity of the Simes test \cite{karlin1981total}, \cite{sarkar1998some}.

	\begin{property}[$\vect{MTP_2}$]
		$X$ is MTP$_2$ if for all $\boldsymbol{x}$ and $\boldsymbol{y}$, 
		\begin{equation}
		f(\boldsymbol{x}) f(\boldsymbol{y}) \leq f(\min(\boldsymbol{x}, \boldsymbol{y}) )  f(\max(\boldsymbol{x}, \boldsymbol{y}) ) , 
		\end{equation}
		where $f$ is the joint density, and the minimum and maximum are evaluated component-wise. 
	\end{property}	
	
	Even for pairwise dependency structure which is neither MTP$_2$ nor PRDS  \cite{yekutieli2008false} gave strong theoretical support for its validity. These are known theoretical results establishing the validity of Simes test for $X \sim N(0,\Sigma)$. 
	Still, PRDS is not a necessary condition for the validity of Simes test: the case of t-tests derived from PRDS normal distribution scaled by the same independent estimator $S \sim \chi^2$ is not PRDS yet the Benjamini-Hochberg (BH) procedure and the Simes test are valid for a significance level less than 0.5. Evidence of the conservatism of the Simes test based on two-sided p-values under general dependency for Gaussian distribution have accumulated since then. This is based on numerous reported simulation studies, see \cite{williams1999controlling}, \cite{keselman2002controlling}, \cite{reiner2007fdr}, \cite{lauter2013simes},
	as well as unreported ones. Even though the global bound of $\alpha (1+1/2+...+1/p)$ on the level of Simes test for any joint distribution has been shown to be attainable \cite{benjamini2001control}, a counter example based on Gaussian distributed test statistics has not been found. The distribution for which the bound is attained is far from being Gaussian, not being ellipsoidal or even unimodal. Hence Simes test is widely considered valid for two-sided tests based on Gaussian or t-distributions under any dependency.
	
	We show that the covariance between pairs of $T^2$ Hotelling statistics is asymptotically positive and they are asymptotically distributed as multivariate Gaussian. Since positively correlated multivariate Gaussian hold the PRDS property this ensures the validity of the Simes test. For further details see the supplementary material. 

	\begin{theorem} \label{mainRes}
		If  $X\sim N(\boldsymbol{\mu}, \Sigma_p)$, then $\lim\limits_{n,p \rightarrow \infty} P(SH_{m,B} (X) \leq \alpha) \leq \alpha$ for $\xi_n = \frac{m}{n}, \; \lim\limits_{n\to\infty} \xi_n = \xi \in (0, 1)$. 
	\end{theorem}

	The theorem is stated using a single sample, $X$, but can easily be extended to accommodate two-samples. We support the validity of the procedure for finite sample sizes in the Simulation section. 
	The proposed test has certain advantages over the classic Hotelling test, as it is applicable when $p > n$, and it allows to inspect smaller subsets of the dimensions. It also has advantages over the marginal Simes procedure, it combines a number of weak signals potentially improving the power and incorporates information regarding the covariance structure. 
	In some sense the procedure is like that suggested by Thulin \cite{thulin2014high}, where the sampled Hotelling statistics are averaged, and the computation of p-value relies on permutations. The suggested procedure does not rely on permutation for the computations and therefore enjoys a lower computational burden. 
	It remains to determine the values of parameter $B$ and $m$, which will be done in the next section.

	\section{Considerations for choosing $m$ and $B$ in $SH_{m,B}$}

	We follow the result of \cite{lopes2011more}, and set $m = \frac{n}{2}$. Since the proof of the property underlying their recommendation is asymptotic, we will also consider another option in our simulations study.

\begin{figure} 
		\centering
		\includegraphics[width=12cm,height=8cm,keepaspectratio]{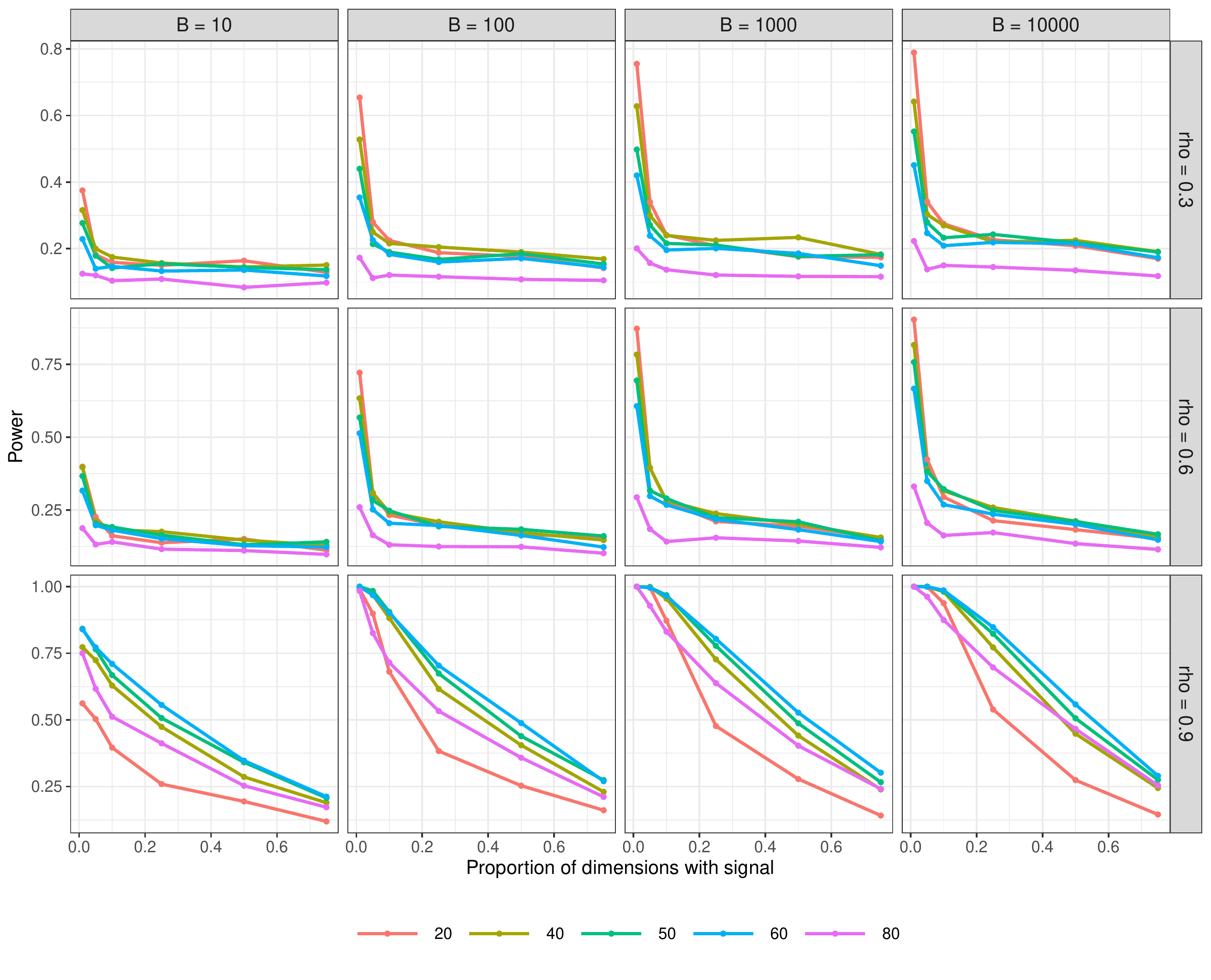}
		\caption{Comparison of various choices for $m$ and $B$. Data is simulated from multivariate Gaussian the covariance matrix $\Sigma_{i,j} = \rho^{|i - j|}, i,j = \{1, \ldots, 300\}$, $n_X = n_Y = 50$, $p = 300$ and $\boldsymbol{\mu_y} = 0$. The normalized Euclidean distance $\frac{|| \boldsymbol{\mu}_X - \boldsymbol{\mu}_Y||_2^2}{n_X + n_Y} = 5$ The x-axis is the number of non-null hypotheses $\frac{\#( \vert \vert \boldsymbol{\mu_X} \vert \vert_0 \neq 0)}{p} $, the y-axis is the power, the rows correspond to various $\rho$ values, the columns to the number of samples taken $B$, and the colors to sample sizes, $m$. The number of iterations is 1000, and the maximum standard deviation is 0.016}	
		\label{b_m_selection}
	\end{figure}

	For a given parameter $m$ there are as many as $\binom{p}{m}$ different dimension subsets of size $m$. Even for small $m$ and $p$ the number of all possible samples is too large to consider each any every possibility. Therefore, we must find a suitable value for the number of samples, $B$, to be taken. 
	Note that if we happen to sample the same set of dimensions twice the Simes test does not lose power even though the number of tests increases because of its step-up nature.
	We would like to ensure that each dimension will be chosen at least once, as to avoid situations where only null hypothesis dimensions are captured by the sampling, which could lead to a loss of power (see Fig. \ref{b_m_selection}).	For computational reason we would like to use smaller $B$. 
	Finding the number of samples needed to cover all the dimensions is known as the Coupon Collector Problem, and for choosing one dimensions at each time the expected number $K$ of samples needed to ensure it,	
	\begin{equation}
	E(K) = p \log p + 1/2 + \gamma p,
	\end{equation}
	where $\gamma$ is the Mascheroni-Euler constant. For selecting more than one dimensions at each sample see \cite{stadje1990collector}, the above equation can be viewed as an upper bound for the expected number of draws when selecting more than one dimension. Setting $B = p \log p$ seems to satisfy the condition of ensuring the selection of all dimensions with high probability. 
	
	The power increases by sampling the dimensions rather than conducting a single Simes test over all dimensions. Suppose that the proportion of dimensions with no signal $\mu_{i,1} - \mu_{i,2} = 0$ is $\beta$. If testing is conducted marginally (t-test followed by Simes) there are $\beta * p$ test statistics distributed according to the null distribution. By sampling each time $m$ dimensions and combining them using $T^2$ Hotelling, the proportion of $T^2$ statistics that are from the null distribution is
	\begin{equation}
	\frac{\binom{\beta p} {m}}{\binom{p}{m}} = \frac{\beta p\dots (\beta p - m + 1) }{p\dots (p-m+1)}< \beta^m.
	\end{equation}
	The gain from the sampling mechanism is not only by allowing the incorporation of data regarding the covariance matrix, but also by reducing the ratio of null to non-null test statistics.
	The parameter $m$ can be viewed as controlling the trade-off between including data regarding the covariance matrix and the averaging of effects. If $m = 1$ the procedure would correspond to the marginal testing of components and will include no data regarding the covariance matrix, and for $m = p$ the procedure would correspond to the standard $T^2$ Hotelling test.  
	
	The test seems to lose power for all values of $m$ when $B$ is small, for larger values of $B$ it seems to have little effect (see Figure \ref{b_m_selection}, the difference between $B = 10$ to $B= 100$ is large, yet the the difference between $B = 1000$ and $B = 10,000$ is almost negligible).  
	While $m = \frac{n}{2}= 50$, is the suggested value, it is not consistently the best option. The optimal $m$ is dependent both on $\rho$, and on $\frac{\#( \vert \vert \boldsymbol{\mu_X} \vert \vert_0 \neq 0)}{p} $. When signal is spread across more dimensions, it is better to select a larger value for $m$, but if the signal is centered only on a few dimensions it seems that a smaller value of $m$ should be considered.  Usually, this type of information is not known, and it seems that $m = \frac{n}{2}$ is a good compromise, having good power for most scenarios. Further theoretical justification for the value is given in Section "Choosing sample size $m$ in the supplementary material".

		\section{Extensions}
	\subsection{Non equal covariance matrices}
	
	The testing of Gaussian means when the variances are not equal, also known as the Behrens-Fisher problem, has been of interest from the early age of Statistics. The natural statistic in one dimension is 
	\begin{equation*}
	t^* = \frac{\bar{X} - \bar{Y}}{ \sqrt{\frac{\hat{\sigma}_X^2}{n_X} + \frac{\hat{\sigma}_Y^2}{n_Y}} }
	\end{equation*}
	The common solution is approximating $t^*$ using the t-distribution and using the Welch method to evaluate the degree of freedom.
	The corresponding statistic in high dimensions would be: 
	\begin{equation*}
	T^{2*} = (\bar{X}-\bar{Y})'\bigg(\frac{\hat{\Sigma}_X}{n_X} + \frac{\hat{\Sigma}_Y}{n_Y}\bigg)^{-1}(\bar{X}-\bar{Y}),
	\end{equation*}
	
	To find its distribution we use the approximation by \cite{krishnamoorthy2004modified} 
	
	\begin{equation*}
	\frac{vp}{v - p - 1} T^{2*} \sim F(p, v - p - 1),
	\end{equation*}
	where the degree of freedom $v$ is calculated as follows
	\begin{equation*}
	v = \frac{p + p^2}{(1/n_1)\big\{tr\big[(\tilde{\s}_X \tilde{\s}^{-1})^2\big] + \big[tr(\tilde{\s}_X \tilde{\s}^{-1})\big]^2\big\} + (1/n_2)\big\{tr\big[(\tilde{\s}_Y \tilde{\s}^{-1})^2\big] + [tr(\tilde{\s}_Y \tilde{\s}^{-1})]^2  \big\}},
	\end{equation*}
	and $\tilde{\s}_X = (n_X - 1)^{-1}\hat{\Sigma}_X$ (or Y accordingly) and $\tilde{\s} = \tilde{s}_X + \tilde{s}_Y$.
	This suggestion was shown to have the best power out of the invariant methods. It can be easily integrated into our proposed test, using $T^{2*}$ instead of the usual Hotelling statistic for the sampled dimensions. Note, that Result \ref{mainRes} still holds as the only thing changed is the number of degrees of freedom. 
	We assess the performance of the procedure and then compare the performance of three other methods in a simulation study (see Section \ref{nonEqualCov}).   
		
	\subsection{Extension to non-Gaussian distributions} \label{sec:NonGauss} 
	
	In order to overcome the reliance of the suggested test on the Gaussianity assumption, we can conduct a permutation test of the same statistic, resulting with a distribution free test. The only assumption needed is that the two samples are interchangeable. Under the global null there is no difference between the two groups, the two distributions are equal, so observations in the two groups are interchangeable. Notice, that the non-parametric extension of the procedure no longer test for equality of the mean vector but the equality of the distributions, $H_0: F_X = F_Y \quad H_1: F_X \neq F_Y$. 
	
	The permutation test is the following, for each sample $S_m^i$ obtain the Hotelling $T^2$ statistics, for each such sample shuffle the labels $L$ times, and obtain the permutation p-value, for each shuffle and sample. Now use function $G$ to combine p-values for each permutation resulting in a single vector of p-values. From there compare the shuffled p-values to the original to obtain the final p-value. Algorithmically the permutation test is the following

		\begin{algorithm}
		\caption{Extension of suggested procedure to permutation tests}
		\begin{algorithmic}

			\For{$i = 1 \; \boldsymbol{to}  \;  B$} 
			\State{$S^{i}_m \leftarrow sample(p, m)$}
			\State{$T_{i,0} \leftarrow  T^2(X_{S_m^{i}}, Y_{S_m^{i}})$} 
			\For{$j = 1 \; \boldsymbol{to}  \;  L$} 
			\State{$X^*, Y^* \leftarrow shuffle(X, Y)$}
			\State{$T_{i,j} \leftarrow  T^2 \left(X^*_{S_m^{i}}, Y^*_{S_m^{i}}\right)$}
			\EndFor 
			\EndFor
			\For{$j = 0 \; \boldsymbol{to}  \;  L$} 
			\For{$i = 1 \; \boldsymbol{to}  \;  B$}
			    \State{$P_{i,j} \leftarrow  \frac{ \Sigma_{k=0}^{L} I\left( T_{i,j} \geq T_{i,k} \right)} {L + 1 }  $}
			\EndFor
			\State{$\boldsymbol{p}_j  \leftarrow G(\boldsymbol{P}_{.,j} )$}
			\EndFor
			\State{$p_{permute} \leftarrow \frac{\Sigma_{k=0}^{L} I \big(  \vect{p}_0 \geq \vect{p}_k)\big)}{L + 1}$}\\
			\Return{$ p_{permute}$}
		\end{algorithmic}
	\end{algorithm}

	After obtaining the $B$ p-values for all the subsets, one can either combine them using any valid test (Bonferroni, the brave scientist may still use Simes), or use any function coupled with a permutation test. 
	This method is similar to the method suggested by \cite{thulin2014high} and shares the disadvantage of being computationally expensive. A small study of additional non-parametric tests and their comparison to the suggested test can be found in the supplementary material.

	\section{Simulations} 
	
	A simulation study was conducted in order to evaluate the suggested test in terms of Type I error and power in comparison to other tests. 
	The other tests considered rely on various approaches, regularization based methods \cite{srivastava2008test} (SD), \cite{chen2010two} (CQ), that are similar to Hotelling $T^2$ but the estimate of $\Sigma^{-1}$ is replaced by a matrix which can be inverted, and thereby allow for the calculation of the test statistic. Their main drawback is the loss of power when the covariance structure is not negligible \cite{lopes2011more}. The random projection type tests such as \cite{lopes2011more} (Lopes) and the non-parametric test suggested by Thulin \cite{thulin2014high}, where the data are projected into a smaller sub-space, thus allowing the calculation of Hotelling-like test statistic. The suggested method falls under this category (using Simes instead of permutation tests).
	A test based on the supremum of the standardized differences between the observed mean vectors \cite{tony2014two} (CLX). Another test is based on the smoothed t-test statistics  \cite{gregory2015two} (GCT). More information regarding the tests can be found in the supplementary material. 
	The simulation section is split into two subsections each dedicated to a different type of tests: subsection \ref{parametric}, Parametric tests, and subsection \ref{nonEqualCov} covers tests which do not require the equal covariance assumption. 
	The section regarding the performance of the non-parametric tests can be found in the supplementary material \ref{nonPara}.
	In this section we present the most interesting results, the full comprehensive tables and figures can be found in the supplementary material. 
	The simulation code can be accessed in \href{https://github.com/tfrostig/Simulations-Simes-Hotelling}{github.com/.../Simulations-Simes-Hotelling}.
	
	\subsection{Simulations parameters} 
	
	Since the goal is to evaluate the tests over a wide array of scenarios, we varied multiple parameters of the data generation. 
	\begin{enumerate}
		\item The number of observations: $n_X = n_Y = 20,50,100$. 
		\item The number of dimensions, $p$: For the equal and non-equal covariance matrix samples, $p = 600$, for the non-parametric simulation $p = 300$.  
		\item The sparsity, $ \frac{|| \mu_X - \mu_Y ||_0}{p}$, varied between $1-\beta  =0.01, 0.05, 0.15$.
	\end{enumerate}
		
	We considered 4 different types of covariance matrices. 

	\begin{enumerate} 
		\item An \textit{auto regression (AR)} matrix, that is $\Sigma = (\sigma_{ij})$, $\sigma_{ij} = \rho^{|i-j|}, \quad 1 \leq i,j  \leq p$, $\rho = 0.3, 0.5, 0.75, 0.95$.
		\item The \textit{block} covariance matrix, where we varied the in block correlation $\rho = 0.5, 0.65, 0.8$, and the size of each block $20, 50, 100$.
		\item \textit{Model 7} presented in \cite{tony2014two}, $\Sigma^{*} = (\sigma^{*}_{i,j}) $ where $\sigma^{*}_{i,i} = 1$ and  for $i \neq j$ $\sigma^{*}_{i,j} = 0.5 * |i - j|^{-5}$, $\Sigma = D^{1/2} \Sigma^{*} D^{1/2}$. $D$ is a diagonal matrix with $d_{i,i}$ sampled from a Uniform distribution $U[1,3]$. 
		\item  \textit{Equally correlated} covariance matrix , $\Sigma = (\sigma_{ij})$, $\sigma_{ij} = \rho \quad 1 \leq i \neq j  \leq p$, $\sigma_{ii} = 1$. $\rho = 0.1, 0.3, 0.5$.
	\end{enumerate}  
	
	The above list of covariance matrices represents scenario used in other matrices, as well as covariance structures that resembles those encountered in practice. 
	To simulate the signal, we kept $\vect{\mu}_Y = \vect{0}$, and vary $\vect{\mu}_X$. $\vect{\mu}_X$ is divided into 6 sub-vectors, 1 sub-vector of size ($\beta p$), whose entries are 0. The rest of the sub-vectors have a size of $\frac{(1-\beta)p}{5}$ where each sub-vector elements are $\frac{i}{5} L$. To find $L$ we determine $||\vect{\mu}_X||_2^2$, and solve  
	\begin{equation*} 
		||\vect{\mu}_X||_2^2 =\frac{(1 - \beta) p}{5}  \sum_{i=1}^{5}  \left(\frac{i * L}{5}\right)^2,
	\end{equation*} 
	for $L$. The norm $||\vect{\mu}_X||_2^2 $ is chosen to set a specific signal to noise ratio, allowing us to compare different scenarios results.  
	The non-zero entries are determined at random. We followed a full factorial design observations $\times$  1-$\beta$ $\times $ dependency structure. All tests are evaluated at level $\alpha = 0.05$.

	\subsection{Simulations - parametric tests} \label{parametric}

	The parametric tests that were chosen for comparison are SD \ref{Srivastava}, 
	CQ \ref{Chen}, Lopes \ref{Lopes}, GCT \ref{Gregory}, CLX \ref{Cai} and the suggested test (SH) with number of dimensions sampled at each iteration $m \in \{ \frac{n}{2}, \frac{n}{4} \}$. We kept the number of samples constant at $B = p * log(p)$, so we shall simply denote the test as $SH_m$.  
	The implementation of GCT and CLX used can be found in the R package \pkg{highD2pop} \cite{GCTpack}, and the implementation of SD, CQ in the R package \pkg{highmean} \cite{MeanPack}.
	Across all simulations the number of repetitions is 1,000, such that the conservative deviation of the estimated power and Type I error is $0.015$. All tests are conducted on the same simulated data. Throughout the simulation $||\vect{\mu}_X - \vect{\mu}_Y||_2^2 = 2.5 \sqrt{\frac{40}{n_1 + n_2}}$ to allow us to compare between the rest of the simulations shown.

	\subsubsection{Type I Error Comparison} 
	
	 Here we describe our main finding with some of the supporting evidence. The complete results of the Type I error are presented in the supplementary material.
	
	\begin{table}
		\centering
		\resizebox{\textwidth}{!}{\begin{tabular}{|c|c||c|c|c|c|c|c|c|c|c|}
			\hline
			$n_1 = n_2$   & $\rho$ & SD & CQ  & Lopes & Simes & GCT   & CLX   & KNN   & SH$_\frac{n}{2}$ &  SH$_\frac{n}{4}$ \\
			\hline
			\multirow{5}[2]{*}{20} & 0.3   & 0.034 & 0.051 & 0.051 & 0.066 & 0.077 & 0.19  & 0.114 & 0.054 & 0.051 \\
			& 0.45  & 0.034 & 0.056 & 0.052 & 0.048 & 0.074 & 0.178 & 0.113 & 0.045 & 0.046 \\
			& 0.6   & 0.035 & 0.057 & 0.052 & 0.049 & 0.074 & 0.19  & 0.119 & 0.052 & 0.056 \\
			& 0.75  & 0.029 & 0.047 & 0.07  & 0.05  & 0.077 & 0.188 & 0.111 & 0.049 & 0.044 \\
			& 0.95  & 0.019 & 0.05  & 0.046 & 0.03  & 0.164 & 0.15  & 0.124 & 0.054 & 0.032 \\
			\hline
			\multirow{5}[2]{*}{50} & 0.3   & 0.046 & 0.06  & 0.046 & 0.049 & 0.065 & 0.08  & 0.101 & 0.051 & 0.05 \\
			& 0.45  & 0.044 & 0.061 & 0.047 & 0.043 & 0.068 & 0.1   & 0.1   & 0.048 & 0.052 \\
			& 0.6   & 0.041 & 0.053 & 0.041 & 0.045 & 0.064 & 0.057 & 0.095 & 0.044 & 0.047 \\
			& 0.75  & 0.034 & 0.056 & 0.047 & 0.043 & 0.077 & 0.093 & 0.103 & 0.052 & 0.038 \\
			& 0.95  & 0.034 & 0.074 & 0.043 & 0.024 & 0.191 & 0.137 & 0.12  & 0.042 & 0.028 \\
			\hline
			\multirow{5}[2]{*}{100} & 0.3   & 0.043 & 0.049 & 0.046 & 0.046 & 0.056 & 0.072 & 0.094 & 0.04  & 0.04 \\
			& 0.45  & 0.04  & 0.05  & 0.036 & 0.048 & 0.056 & 0.054 & 0.098 & 0.045 & 0.028 \\
			& 0.6   & 0.04  & 0.053 & 0.048 & 0.06  & 0.068 & 0.031 & 0.087 & 0.042 & 0.042 \\
			& 0.75  & 0.043 & 0.054 & 0.045 & 0.045 & 0.075 & 0.095 & 0.092 & 0.043 & 0.031 \\
			& 0.95  & 0.029 & 0.067 & 0.052 & 0.025 & 0.178 & 0.099 & 0.113 & 0.036 & 0.029 \\
			\hline
		\end{tabular}}%
		\caption{Parametric tests, type I error rates for the \textit{AR} covariance matrix scenario for each of the methods. $\rho$ is the correlation parameter.  The number of iterations is 1000, and the conservative s.e of the simulation is 0.016.} 
		\label{tab:Cov1_Alpha}%
	\end{table}%
	
	\afterpage{

	\begin{table}
		\centering
		\resizebox{\textwidth}{0.4\textheight}{\begin{tabular}{|c|c|c||c|c|c|c|H c|c|c|c|}
				\hline
				$n_1 = n_2$   & Block Size & $\rho$   & SD & CQ  & Lopes & Simes & GCT   & CLX   & KNN   &  SH$_\frac{n}{2}$ &  SH$_\frac{n}{4}$  \\
				\hline
				\multirow{9}[2]{*}{20} & \multirow{3}[1]{*}{20} & 0.5   & 0.033 & 0.055 & 0.050 & 0.032 & 0.154 & 0.193 & 0.116 & 0.048 & 0.049 \\
				&       & 0.65  & 0.029 & 0.056 & 0.050 & 0.033 & 0.156 & 0.204 & 0.109 & 0.059 & 0.031 \\
				&       & 0.8   & 0.024 & 0.056 & 0.046 & 0.023 & 0.158 & 0.228 & 0.118 & 0.048 & 0.034 \\
				& \multirow{3}[0]{*}{50} & 0.5   & 0.024 & 0.059 & 0.045 & 0.044 & 0.301 & 0.241 & 0.128 & 0.044 & 0.042 \\
				&       & 0.65  & 0.017 & 0.057 & 0.037 & 0.029 & 0.324 & 0.251 & 0.122 & 0.035 & 0.030 \\
				&       & 0.8   & 0.010 & 0.059 & 0.047 & 0.025 & 0.338 & 0.305 & 0.121 & 0.052 & 0.024 \\
				& \multirow{3}[1]{*}{100} & 0.5   & 0.029 & 0.070 & 0.049 & 0.052 & 0.511 & 0.288 & 0.142 & 0.046 & 0.040 \\
				&       & 0.65  & 0.021 & 0.069 & 0.040 & 0.030 & 0.532 & 0.318 & 0.143 & 0.049 & 0.031 \\
				&       & 0.8   & 0.012 & 0.072 & 0.030 & 0.021 & 0.551 & 0.431 & 0.133 & 0.053 & 0.039 \\
				\hline
				\multirow{9}[2]{*}{50} & \multirow{3}[1]{*}{20} & 0.5   & 0.028 & 0.060 & 0.049 & 0.045 & 0.132 & 0.295 & 0.126 & 0.042 & 0.051 \\
				&       & 0.65  & 0.027 & 0.062 & 0.035 & 0.026 & 0.152 & 0.290 & 0.119 & 0.047 & 0.041 \\
				&       & 0.8   & 0.023 & 0.067 & 0.051 & 0.031 & 0.156 & 0.593 & 0.109 & 0.035 & 0.033 \\
				& \multirow{3}[0]{*}{50} & 0.5   & 0.038 & 0.069 & 0.044 & 0.058 & 0.314 & 0.537 & 0.123 & 0.051 & 0.046 \\
				&       & 0.65  & 0.030 & 0.068 & 0.055 & 0.029 & 0.337 & 0.546 & 0.116 & 0.046 & 0.038 \\
				&       & 0.8   & 0.019 & 0.070 & 0.041 & 0.029 & 0.354 & 1.000 & 0.112 & 0.048 & 0.035 \\
				& \multirow{3}[1]{*}{100} & 0.5   & 0.017 & 0.064 & 0.054 & 0.041 & 0.469 & 0.757 & 0.098 & 0.045 & 0.046 \\
				&       & 0.65  & 0.014 & 0.069 & 0.058 & 0.032 & 0.493 & 0.878 & 0.098 & 0.038 & 0.044 \\
				&       & 0.8   & 0.010 & 0.065 & 0.055 & 0.022 & 0.515 & 1.000 & 0.101 & 0.052 & 0.047 \\
				\hline
				\multirow{9}[2]{*}{100} & \multirow{3}[1]{*}{20} & 0.5   & 0.049 & 0.067 & 0.046 & 0.051 & 0.135 & 0.164 & 0.111 & 0.054 & 0.035 \\
				&       & 0.65  & 0.039 & 0.059 & 0.066 & 0.039 & 0.144 & 0.240 & 0.107 & 0.044 & 0.046 \\
				&       & 0.8   & 0.033 & 0.060 & 0.045 & 0.039 & 0.147 & 0.242 & 0.100 & 0.042 & 0.024 \\
				& \multirow{3}[0]{*}{50} & 0.5   & 0.032 & 0.064 & 0.049 & 0.047 & 0.309 & 0.373 & 0.103 & 0.034 & 0.049 \\
				&       & 0.65  & 0.023 & 0.069 & 0.049 & 0.034 & 0.331 & 0.828 & 0.109 & 0.042 & 0.038 \\
				&       & 0.8   & 0.017 & 0.068 & 0.053 & 0.022 & 0.348 & 0.842 & 0.106 & 0.045 & 0.031 \\
				& \multirow{3}[1]{*}{100} & 0.5   & 0.025 & 0.070 & 0.051 & 0.045 & 0.505 & 0.833 & 0.106 & 0.040 & 0.039 \\
				&       & 0.65  & 0.011 & 0.070 & 0.052 & 0.023 & 0.544 & 1.000 & 0.100 & 0.054 & 0.037 \\
				&       & 0.8   & 0.006 & 0.069 & 0.048 & 0.014 & 0.554 & 1.000 & 0.100 & 0.045 & 0.041 \\
				\hline
			\end{tabular}}
		\caption{Parametric tests, type I error rates for the \textit{block} covariance matrix. $\rho$ is in block correlation, GCT was removed as the test assumptions do not hold in the scenario.  The number of iterations is 1000, and the conservative s.e of the simulation is 0.016.} 
		\label{tab:Cov3_Alpha} 
	\end{table}%
}
	
	Across the various scenarios it can be seen that both SH variants keep the required error rate. As the correlation structure is stronger (Table \ref{tab:Cov3_Alpha} and Table \ref{tab:Cov1_Alpha}), the Simes and SD tests are becoming more conservative. CQ test has a slightly elevated error rates ($>0.05$) under all the covariance structures. It seems that a GCT, and CLX fail to keep type I error, especially when the covariance between the dimensions decays slowly (Table \ref{tab:Cov3_Alpha}).
	CLX performance was worst for the block covariance structure, due to the difficulty in estimating the covariance matrix. The covariance matrix was estimated using the adaptive threshold estimator, indeed, it can be expected that for sparse covariance matrices using the CLIME would have yielded better performance. The type I error of the GCT test holds asymptotically as $p \rightarrow \infty$ and should have therefore performed better for larger $p$. KNN also does not seem to keep the type I error, this is possibly due to the asymptotic nature of the test. It can be expected to perform better when $n \rightarrow \infty$. 
	Due to these results, GCT, KNN and CLX tests were omitted from the rest of the simulations study. 
	
	\subsubsection{Power comparison - parametric tests} \label{parasim}
	
	We compare the power curves of the different tests in different scenarios. The power of the global null tests is affected by the sparsity of the signal, the covariance structure and number of observations. Considering the covariance structure can change the sparsity of the problem: if a signal is sparse in the original dimensions considered, it could be dense when considering it across the principle components. As can be seen in Lemma \ref{lemma1}, in the supplementary material, Hotelling's test sums the signal across the principal components, increasing the power of tests that take into consideration the covariance structure in certain situations. 
	
	\begin{figure} 
		\centering
		\includegraphics[width=15cm,height=15cm,keepaspectratio]{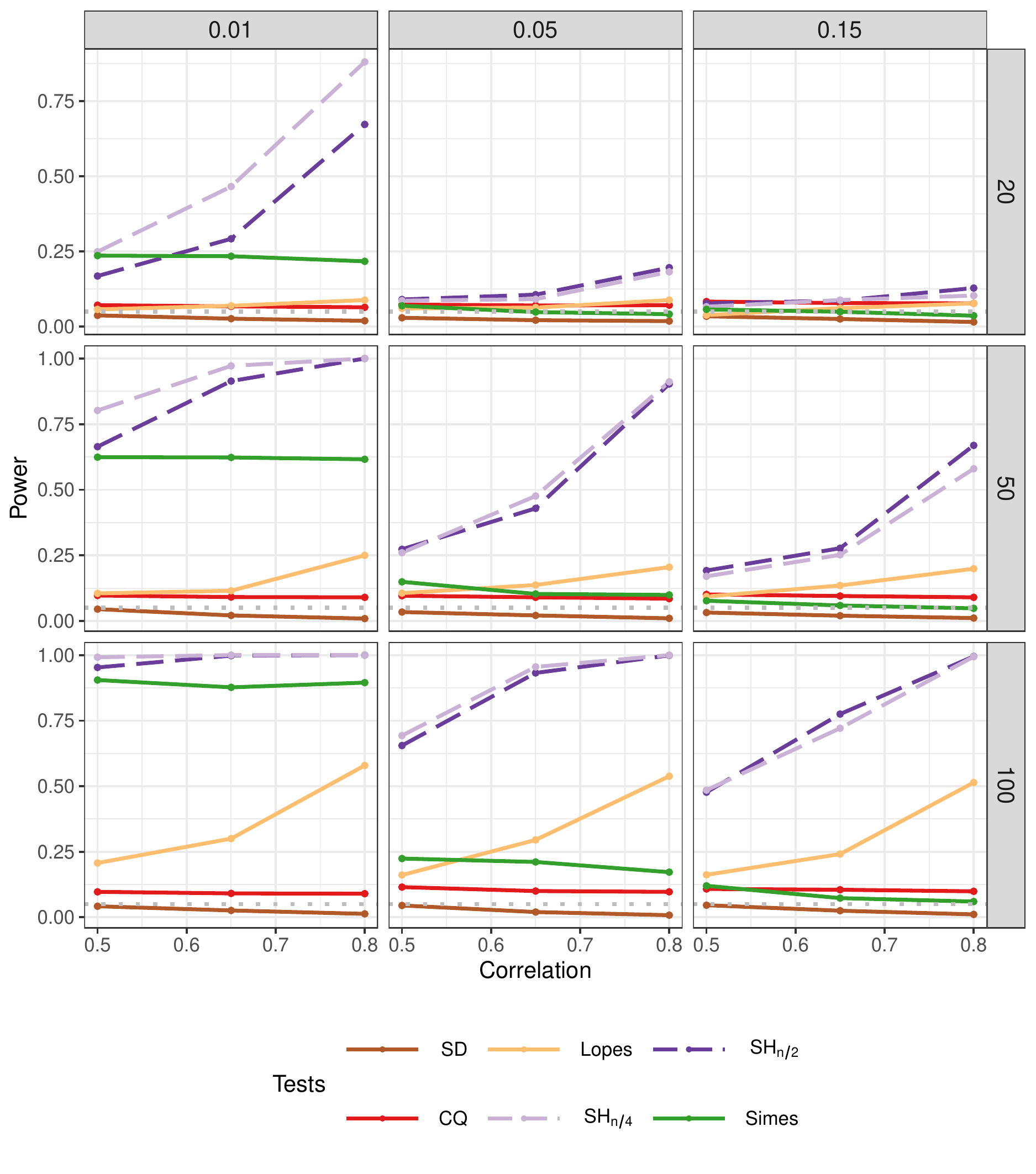}
		\caption{Parametric tests power comparison, \textit{block} covariance matrix,  block size is 100. Rows indicate number of observations ($n_1 = n_2$), the columns indicate $1-\beta$, the proportion of dimensions where $\delta \neq 0$.  The number of iterations is 1000, and the conservative s.e of the simulation is 0.016.}
		\label{Block100_Parametric}
	\end{figure}

	As seen in Figure \ref{Block100_Parametric} the SH variants and Lopes tests are the only tests that do not lose power as the correlation increases as evident by their increasing power curves across the x-axis in each of the plot. This is an example of the difference between tests which utilize the sum of signal across the principle components versus tests such as SD and CQ that mainly rely on summing the signal across the original dimensions. It is also worth noting that in the sparse case the Simes procedure performs well as expected. 
	
\begin{figure} 
		\includegraphics[width=15cm,height=16cm,keepaspectratio]{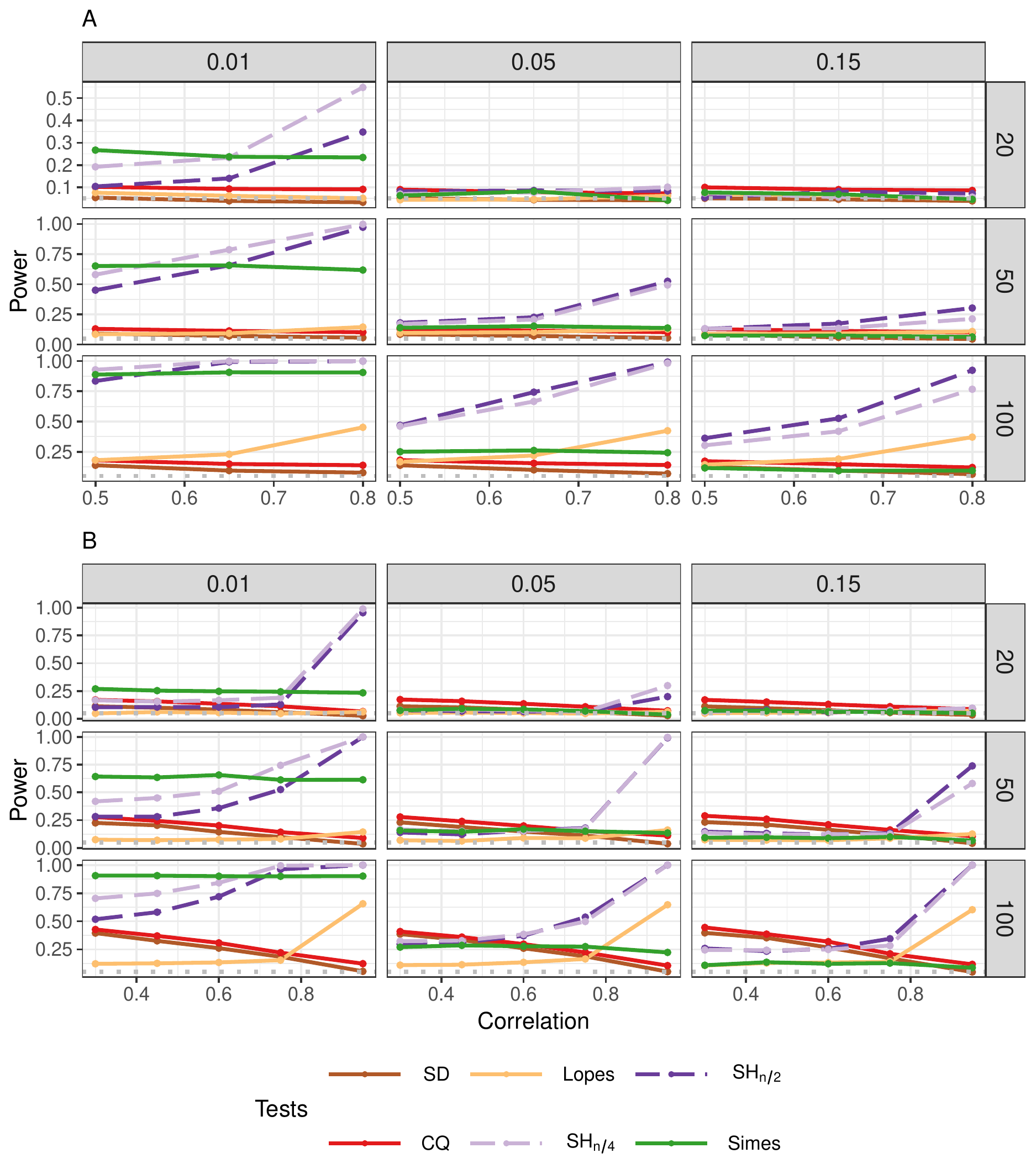}
		\caption{Parametric tests power comparison, (a) \textit{block} covariance matrix, block size is 20 and (b) \textit{AR} covariance matrix. The x-axis is the covariance parameter and the y-axis is power. The number of iterations is 1000, and the conservative s.e of the simulation is 0.016.}
		\label{combplot} 
	\end{figure}

	In Figure \ref{combplot}a, we see the power curves of the tests under a milder covariance structure, where the block size of equally correlated dimensions size is 20, rather than 100. One of the downsides of the SH procedure becomes apparent: if the number of observations is small then the number of dimensions sampled at each iteration is also small, which makes it harder for the test to detect a signal that is dense. This can be seen in the downward trend of the SH power curves across the columns of the figure.  Notice that the trend decreases as the number of observations increases. This is unlike the rest of the tests such as the CQ, SD and Lopes which power does not change as a result of changes in the number of false null hypotheses.  It is important to mention that the Simes test is better than all other tests except for SH in the sparse setting ($1-\beta = 0.01$). Still, the SH variants dominate the other tests except for the instance where there the signal is dense ($1-\beta = 0.15$) and the number of observations is small ($n_X = n_Y = 20$).

	Figure \ref{combplot}b displays the results of the least favorable scenario for the SH variants, the signal is relatively dense ($1-\beta= 0.15$) and the covariance structure is close to independence ($\rho = 0.3$ with exponential decay). Indeed, the power of the SH and Lopes tests is lower than the power of SD and CQ.  However when the number of observations or $\rho$ increases, so that either more dimensions are sampled or there is more signal across the principle components, the SH tests becomes more competitive, as can be seen in the power curves trend upwards across rows and the x-axis. 
	Overall, it seems that each test has its preferable scenarios: for the Simes test it is a strong and sparse signal across independent dimensions, for the CQ and SD it is when the signal is spread across independent dimensions, and for the Lopes test it is when the signal is dense and spread across strongly dependent dimensions. A way to encompass the comparisons across all scenarios jointly is presented next.

	\subsubsection{Min-min power comparison (A-la Princeton)}
	
	The SH test dominates in some scenarios but even in its least favorable scenarios retains a reasonable amount of power. It performs best when the dimensions are strongly correlated with a sparse signal, but even when the correlation is weak, if the number of observations, $n$, is large enough for $m = \frac{n}{2}$ to capture most of the signal, it returns decent results.
	This performance can be quantified by using empirical min-min criterion (A-la Princeton simulation study), 
	
	\begin{equation}
	S_M = \min\limits_{c_j \in \vect{C}} \left( \frac{ Power(M_i, c_j) }{\max\limits_{M_l}Power(M_l, c_j)} \right),
	\end{equation}
	
	$c_j$ being a scenario, $M \in \{$SD, CQ, SH, Lopes, Simes$\}$. The score indicates how well a test performs relative to the best among the compared tests under its least favorable condition. $S_M$ is calculated separately for each $n_1=n_2=20,50,100$, the number of observations, which is a known parameter of the problem when testing the global null. 
	
	\begin{table} 
		\centering
		\label{tab: Princeton Parametric} 
		\begin{tabular}{|p{1.5cm}||p{1cm}|p{1.0cm}|p{1.0cm}|p{1.0cm}|p{2.0cm}|p{2.0cm}|p{2.0cm}|p{2.0cm}|} 
				\hline
				$n_1 = n_2$ &  SD & CQ & Lopes &  Simes & SH$_{n/2}$ &  SH$_{n/4}$ \\
				\hline
				20  & 0.022 & 0.066 & 0.063 & 0.124 & 0.383 & 0.371 \\
				50  & 0.009 & 0.089 & 0.108 & 0.072 & 0.439 & 0.456 \\
				100 & 0.008 & 0.090 & 0.131 & 0.060 & 0.551 & 0.551 \\
				\hline
		\end{tabular}
		\caption{A-la Princeton simulation study results, finite sample min-min results.} 
	\end{table}
	
  	In the equal-correlated covariance structure (4) the correlation between the dimensions is high and the SH test variants have very high power. We therefore removed the equal correlation case from the following simulation.  
  	The SH variants dominate the other tests. This is because most of the tests specialize in a certain kind of signal, while the SH test power is not always the highest, it does not plummet in any scenario.

	\subsection{Simulation - non-equal covariance matrices} \label{nonEqualCov}
	
	In this section we study the SH variant for the non-equal covariance matrices and compare its power to that of the other available tests. The simulation is conducted as in \ref{parametric}, but with $X_1 \sim N(\vect{\mu}_1, 2 * \Sigma)$ and $X_2 \sim N(\vect{\mu}_2, \Sigma)$. We first assess the type I error for the tests that controlled the type I error in the equal covariance simulation, namely, SD, CQ, Simes (with Welch t-test), SH$_\frac{n}{2}$ and SH$_\frac{n}{4}$, and then continue to compare their power.

	\subsubsection{Type I error comparison}

	\begin{table} [!htbp]
		\centering
		\begin{tabular}{|c|c||c|c|H c|c|c|}
			\hline
			\multicolumn{1}{|l|}{$n_1 = n_2$} & \multicolumn{1}{l||}{$\rho$} & \multicolumn{1}{l|}{SD} & \multicolumn{1}{l|}{CQ} &
			 \multicolumn{1}{l}{} & \multicolumn{1}{l|}{Simes} & \multicolumn{1}{l|}{SH$_\frac{n}{2}$} & \multicolumn{1}{l|}{SH$_\frac{n}{4}$} \\
			\hline
			\multirow{5}[2]{*}{20} & 0.3   & 0.001 & 0.061 & 0.064 & 0.048 & 0.035 & 0.015 \\
			& 0.45  & 0.006 & 0.061 & 0.075 & 0.047 & 0.021 & 0.03 \\
			& 0.6   & 0.005 & 0.056 & 0.067 & 0.043 & 0.028 & 0.025 \\
			& 0.75  & 0.015 & 0.049 & 0.083 & 0.049 & 0.037 & 0.03 \\
			& 0.95  & 0.018 & 0.062 & 0.178 & 0.027 & 0.023 & 0.021 \\
			\hline
			\multirow{5}[2]{*}{50} & 0.3   & 0.014 & 0.06  & 0.064 & 0.049 & 0.046 & 0.044 \\
			& 0.45  & 0.023 & 0.061 & 0.063 & 0.046 & 0.045 & 0.046 \\
			& 0.6   & 0.027 & 0.06  & 0.073 & 0.041 & 0.046 & 0.04 \\
			& 0.75  & 0.029 & 0.062 & 0.078 & 0.052 & 0.048 & 0.024 \\
			& 0.95  & 0.03  & 0.071 & 0.202 & 0.029 & 0.039 & 0.023 \\
			\hline
			\multirow{5}[2]{*}{100} & 0.3   & 0.032 & 0.05  & 0.056 & 0.045 & 0.061 & 0.045 \\
			& 0.45  & 0.028 & 0.05  & 0.05  & 0.052 & 0.051 & 0.041 \\
			& 0.6   & 0.027 & 0.051 & 0.066 & 0.058 & 0.058 & 0.044 \\
			& 0.75  & 0.037 & 0.058 & 0.061 & 0.048 & 0.047 & 0.029 \\
			& 0.95  & 0.032 & 0.073 & 0.199 & 0.03  & 0.047 & 0.032 \\
			\hline
		\end{tabular}%
		
		\caption{Non-equal covariance matrix tests, type I error rates for the \textit{AR} covariance matrix scenario. $\rho$ is in correlation parameter.  The number of iterations is 1000, and the conservative s.e of the simulation is 0.016.} 
		\label{tab:Cov1 Alpha Non-equal covmat} 
	\end{table}%


		\begin{table}
			\centering
			\resizebox{0.8\textwidth}{0.4 \textheight}{
				\begin{tabular}{|c|c|r||r|r|r|r|r|}
				\hline
				\multicolumn{1}{|l|}{$n_1 = n_2$} & \multicolumn{1}{l|}{Block Size} & \multicolumn{1}{l||}{$\rho$} & \multicolumn{1}{l|}{SD} & \multicolumn{1}{l|}{CQ} & \multicolumn{1}{l|}{Simes} & \multicolumn{1}{l|}{SH$_\frac{n}{2}$} & \multicolumn{1}{l|}{SH$_\frac{n}{4}$} \\
				\hline
				\multirow{9}[2]{*}{20} & \multirow{3}[1]{*}{20} & 0.5   & 0.025 & 0.057 & 0.048 & 0.029 & 0.031 \\
				&       & 0.65  & 0.021 & 0.054 & 0.029 & 0.019 & 0.029 \\
				&       & 0.8   & 0.019 & 0.052 & 0.025 & 0.025 & 0.012 \\
				& \multirow{3}[0]{*}{50} & 0.5   & 0.02  & 0.058 & 0.043 & 0.015 & 0.015 \\
				&       & 0.65  & 0.017 & 0.061 & 0.032 & 0.019 & 0.015 \\
				&       & 0.8   & 0.01  & 0.058 & 0.021 & 0.022 & 0.015 \\
				& \multirow{3}[1]{*}{100} & 0.5   & 0.027 & 0.07  & 0.034 & 0.025 & 0.023 \\
				&       & 0.65  & 0.022 & 0.073 & 0.033 & 0.025 & 0.023 \\
				&       & 0.8   & 0.018 & 0.07  & 0.018 & 0.03  & 0.024 \\
				\hline
				\multirow{9}[2]{*}{50} & \multirow{3}[1]{*}{20} & 0.5   & 0.027 & 0.058 & 0.047 & 0.044 & 0.029 \\
				&       & 0.65  & 0.026 & 0.063 & 0.039 & 0.055 & 0.03 \\
				&       & 0.8   & 0.024 & 0.066 & 0.022 & 0.035 & 0.023 \\
				& \multirow{3}[0]{*}{50} & 0.5   & 0.036 & 0.077 & 0.044 & 0.046 & 0.038 \\
				&       & 0.65  & 0.025 & 0.076 & 0.034 & 0.052 & 0.036 \\
				&       & 0.8   & 0.018 & 0.074 & 0.024 & 0.052 & 0.022 \\
				& \multirow{3}[1]{*}{100} & 0.5   & 0.021 & 0.066 & 0.036 & 0.042 & 0.037 \\
				&       & 0.65  & 0.013 & 0.066 & 0.028 & 0.048 & 0.037 \\
				&       & 0.8   & 0.007 & 0.066 & 0.022 & 0.05  & 0.024 \\
				\hline
				\multirow{9}[2]{*}{100} & \multirow{3}[1]{*}{20} & 0.5   & 0.04  & 0.067 & 0.048 & 0.057 & 0.037 \\
				&       & 0.65  & 0.038 & 0.07  & 0.036 & 0.068 & 0.036 \\
				&       & 0.8   & 0.03  & 0.07  & 0.031 & 0.063 & 0.033 \\
				& \multirow{3}[0]{*}{50} & 0.5   & 0.034 & 0.072 & 0.054 & 0.063 & 0.05 \\
				&       & 0.65  & 0.028 & 0.073 & 0.036 & 0.066 & 0.039 \\
				&       & 0.8   & 0.021 & 0.073 & 0.024 & 0.057 & 0.033 \\
				& \multirow{3}[1]{*}{100} & 0.5   & 0.027 & 0.07  & 0.035 & 0.066 & 0.038 \\
				&       & 0.65  & 0.013 & 0.069 & 0.031 & 0.074 & 0.044 \\
				&       & 0.8   & 0.007 & 0.069 & 0.015 & 0.061 & 0.041 \\
				\hline
			\end{tabular}}%
			\caption{Non-equal covariance matrix tests type I error rates for the \textit{block} covariance matrix scenario. $\rho$ is the in block covariance.  The number of iterations is 1000, and the conservative s.e of the simulation is 0.016.}
		\label{tab:Cov3Alpha} 
		\end{table}%

	The type I error is kept by all tests in all of the scenarios. The type I error rates indicate that the tests are slightly more conservative compared to \ref{parasim} for small $\rho$ and $n_1 = n_2$. We are left to assess the tests power of in the different scenarios.

	\subsubsection{Power Comparison}
	
	The signal was generated by ensuring that $||\vect{\mu}_1 - \vect{\mu}_2||_2^2 = 2.5 \sqrt{\frac{40}{n_1 + n_2}}$. This results in a smaller signal to noise ratio than in the parametric simulations tests (as the covariance matrix $\Sigma_2$ was inflated). 

		\begin{figure}  
		
			\centering
			\includegraphics[width=15cm,height=15cm,keepaspectratio]{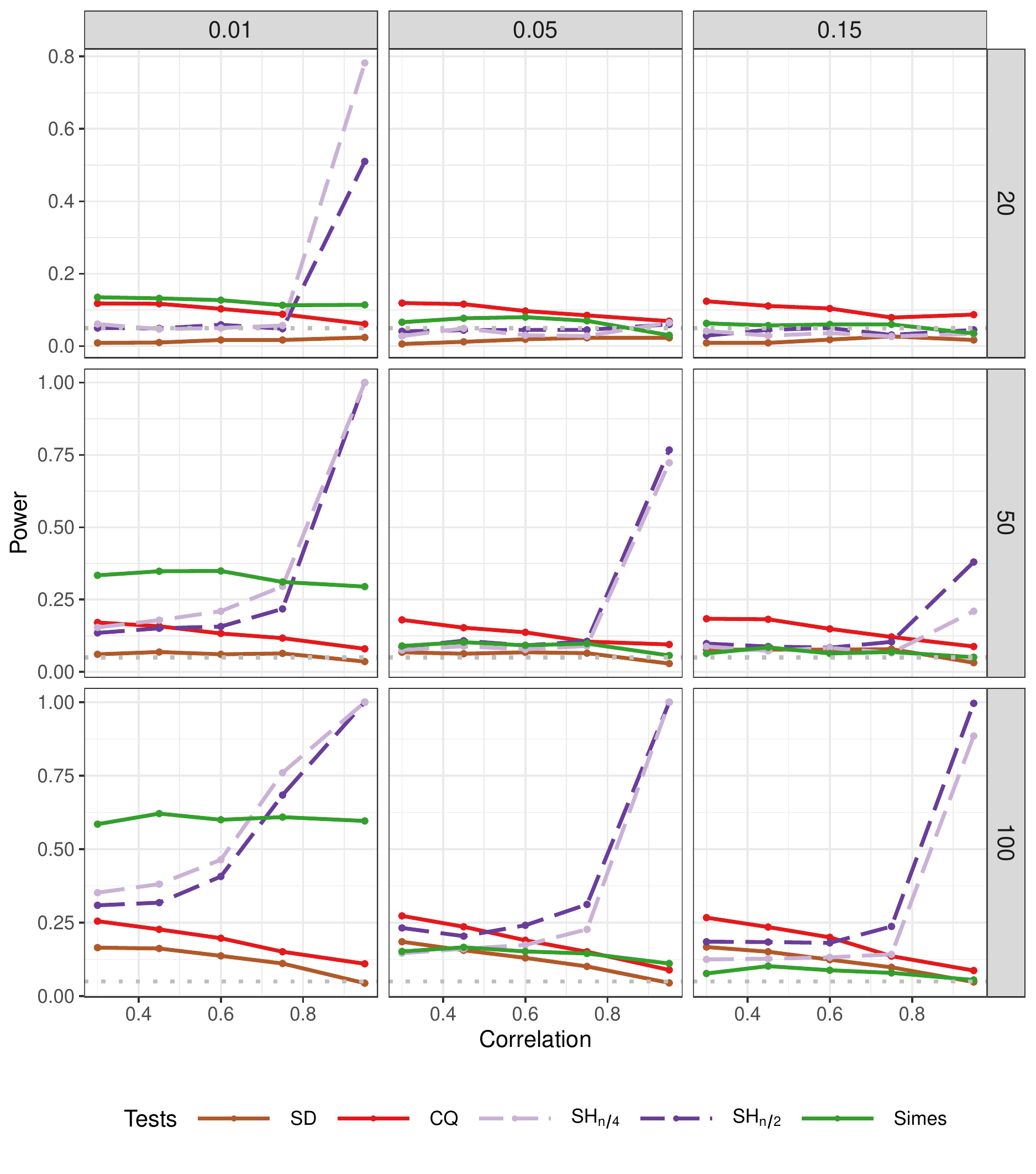}
			\caption{Non-equal covariance tests power comparison, \textit{AR} covariance matrix ($\Sigma_1 = 2 * \Sigma_2$). The number of iterations is 1000, and the conservative s.e of the simulation is 0.016.}
			\label{AR_NonEqual}
		\end{figure}

	As can be expected, the power of the procedure is lower because only $n_1$ or $n_2$ observations are available to estimate each covariance matrix rather than the $n_1 + n_2$  in the equal covariance case, with the implication on the degrees of freedom of the test statistic distribution. 
	The difference between the compared tests is similar to the differences in the parametric setting, as evident by the power curves in Figure \ref{AR_NonEqual} intersecting roughly in the same points as in Figure \ref{combplot}b.

	\subsubsection{Min - min Power Comparison (A-la Princeton)}
	Evaluating the min-min performance of the tests, after removing again, the equally correlated covariance scenario (4).

	\begin{table} [!htbp]
		\centering
		\label{tab: Princeton Non-equal covariance} 
		\begin{tabular}{|c||c|c|c|c|c|} 
			\hline
			$n_1 = n_2$ &  SD & CQ &  Simes &  SH$_{n/2}$ &  SH$_{n/4}$  \\ 
			\hline
			20  & 0.031 & 0.078 & 0.146 & 0.234 & 0.235 \\
			50  & 0.009 & 0.080 & 0.074 & 0.404 & 0.401 \\
			100 & 0.007 & 0.083 & 0.035 & 0.512 & 0.468 \\
			
			\hline
		\end{tabular}
		\caption{A-la Princeton simulation study results for the non-equal covariance tests, finite sample min-min results.} 
	\end{table}
	
	The results are similar to those seen in the equal covariance case. The SH loses power in certain scenarios, but even in the least favorable scenario its power does not deteriorate much unlike the other tests.

	\section{Real data examples} 
	
	The proposed combination of Simes and Hotelling tests (SH) was also applied to real data, found in previous papers. We show that it performs as well as other tests, in terms of getting low p-values while enjoying full guarantee of type I error.

	\subsection{Colon cancer data} 
	
	Alon et al. \cite{alon1999broad} compared the expression level of 2000 genes between 22 normal colon tissues and 40 tumor colon tissues. Same example was used in \cite{srivastava2008test} the data is available at (\url{http://genomics-pubs.princeton.edu/oncology/affydata/index.html}) and was taken from \cite{alon1999broad}.

	\begin{table}[!h]
		\centering
		\begin{tabular}{|l|l|l|l|l|l|l|} 
			\hline
			Tests   & SD   & CQ    & Lopes  & Simes & SH$_{n/2}$ & SH$_{n/4}$  \\ \hline
			P-value & $0.27$ & $1.03 * 10^{-7}$ & $9.89 * 10^{-3}$ & $7.43*10^{-5}$ & $8.47 *10^{-4}$ & $1.07 * 10^{-4}$ \\ \hline
		\end{tabular}
		\caption{The resulting p-values for the tests of the gene expression levels (the random tests SH and Lopes test were repeated ten times and averaged to lower variance)}
		\label{realdata1}
	\end{table}
	
	All tests except for SD reject the null hypothesis. CQ seems to be the most sensitive but recall that it fails to control the type I error. 
	
	\subsection{Mitochondrial calcium concentration}
	
	The next data set was used in \cite{gregory2015two} and was taken from \cite{ruiz2003cariporide}. The researcher tracked the Calcium count level every ten second during the hour, after administrating a dose of Cariporide. They repeated the experiment twice, once for intact cells, and once for permeabilized membranes. 
	The pre-processing here was conducted in the same way as in \cite{gregory2015two}. We used the non-equal covariance matrix variant of the test.

	\begin{table}[!h]
		\centering
		\begin{tabular}{|l|l|l|l|l|l|} 
			\hline
			Tests   & GCT   & CQ    & Simes & SH$_{n/2}$ & SH$_{n/4}$  \\ \hline
			P-value & $1.7*10^{-6}$ & $3.33*10^{-16}$ & $1.06 * 10^{-4}$ & $3.06 * 10^{-10}$ & $10^{-16}$ \\ \hline
		\end{tabular}
		\caption{The resulting p-values for the tests of the Calcium count levels for intact cells  (SH was repeated ten times and averaged to lower variance)}
		\label{realdata2}
	\end{table}
	
	Again, all tests reject the null hypothesis.

	\section*{Conclusions} 
	
	The suggested test seems to be powerful when there is strong correlation between dimensions. This is because the Mahalanobis distance between $\mu_{X}$ and $\mu_{Y}$ can be bigger than the Euclidean distance in such a scenario. 	

	\begin{exmp}
		Suppose $X \sim N(\boldmath{\mu}, \Sigma)$ where $\Sigma = \begin{pmatrix}
		\sigma_2^2 & - \rho  \sigma_1 \sigma_2 \\ 
		-\rho  \sigma_1 \sigma_2 & \sigma_1^2 
		\end{pmatrix}$, $\Sigma$ is known. Then $E(n\bar{X}\Sigma^{-1}\bar{X}') = 2 + \frac{n(\sigma_2^2\mu_1^2 + \sigma_1^2\mu_2^2 - 2\rho\sigma_1\sigma_2\mu_1 \mu_2)}{\sigma_1^2\sigma_2^2(1 - \rho^2)}$.
		Taking $\sigma_1 = \sigma_2 = 1, \vect{\mu} = (2,0), \rho = 0.5, n=1$, the expected Mahalanobis distance is $7.333$, while the expected Euclidean distance $E(n\bar{X}I\bar{X}') = 6$. 	
	\end{exmp}
	
	We can see that dimensions which do not contain any signal influence the Mahalanobis distance between the mean vectors. When sampling dimensions we obtain various estimates to the Mahalanobis distance between parts of the mean vector.  This is beneficial since the Simes test is powerful for a single strong effect but retains its power for many moderate effects. 
		
	To conclude, we have suggested a two-sample test for detecting a shift between the mean vectors of two multivariate normal distributions, by combining the Simes test and the Hotelling $T^2$ test.
	While the Hotelling $T^2$ test cannot be used when $p > n$, and the Simes test does not take dependency into account, combining them resolves both issues. 
	The procedure is different from other tests in two important ways: (i) It allows for the incorporation of the covariance matrix, as seen in the simulation in the scenarios where the dimensions are correlated;
	(ii) The procedure is relatively computationally efficient, as it does not require conducting permutations to obtain a p-value. 
	The test retains power across widely different scenarios, as can be seen in the min-min simulations. While many tests have a preferred scenario, the suggested test works well in most of them, often achieving the highest or second highest power. 
	The simulation results can be summarized to a rule of thumb of which test to use for various signal and dependencies strength (assuming $p>n$), as can be seen in the table below. 
		
	\begin{table*}[!h]
		\centering
		\begin{tabular}{|l|l|l|} 
			\hline
			Signal   & Dependency  & Test to use    \\ \hline
			Very sparse & Weak &  Simes \\ \hline
			Sparse & Strong & SH \\ \hline  
			Dense & Weak & CQ  \\ \hline 
			Dense & Strong & SH \\ \hline
	\end{tabular}
	\caption{Rule of thumb summary. When to use each test according to expected signal sparsity and covariance between dimensions.}
	\end{table*}

	If computational power is not an issue, then it is recommended to use PSH.
	As was mentioned earlier, the simulation study is not exhaustive, and missing several tests such as Higher Criticism \cite{donoho2004higher} which is suitable to deal with a much larger amount of hypotheses but relies on the assumption of independence, as well as non-parametric tests such as suggested in \cite{heller2016multivariate}, which is suited for testing a general difference between distributions and not a specific difference between the mean vectors. It could be interesting to check the effect of combining sampled dimensions for different tests, as this methodology can be appropriate for other problems. 

	\bibliography{refrence}

\begin{thebibliography}{}

\bibitem[\protect\citeauthoryear{Alon, Barkai, Notterman, Gish, Ybarra, Mack,
  and Levine}{Alon et~al.}{1999}]{alon1999broad}
Alon, U., N.~Barkai, D.~Notterman, K.~Gish, S.~Ybarra, D.~Mack, and A.~Levine
  ({1999}, {JUN 8}).
\newblock {Broad patterns of gene expression revealed by clustering analysis of
  tumor and normal colon tissues probed by oligonucleotide arrays}.
\newblock {\em {Proceedings of the National Academy of Sciences of the United
  States of America}\/}~{\em {96}\/}({12}), {6745--6750}.

\bibitem[\protect\citeauthoryear{Bai and Saranadasa}{Bai and
  Saranadasa}{1996}]{bai1996effect}
Bai, Z. and H.~Saranadasa ({1996}, {APR}).
\newblock {Effect of high dimension: By an example of a two sample problem}.
\newblock {\em {Statistica Sinica}\/}~{\em {6}\/}({2}), {311--329}.

\bibitem[\protect\citeauthoryear{Benjamini and Yekutieli}{Benjamini and
  Yekutieli}{2001}]{benjamini2001control}
Benjamini, Y. and D.~Yekutieli ({2001}, {AUG}).
\newblock {The control of the false discovery rate in multiple testing under
  dependency}.
\newblock {\em {Annals of Statistics}\/}~{\em {29}\/}({4}), {1165--1188}.

\bibitem[\protect\citeauthoryear{Bibby, Kent, and Mardia}{Bibby
  et~al.}{1979}]{bibby1979multivariate}
Bibby, J., J.~Kent, and K.~Mardia (1979).
\newblock Multivariate analysis.

\bibitem[\protect\citeauthoryear{Cai and Liu}{Cai and
  Liu}{2011}]{cai2011adaptive}
Cai, T. and W.~Liu ({2011}, {JUN}).
\newblock {Adaptive thresholding for sparse covariance matrix estimation}.
\newblock {\em {Journal of the American Statistical Association}\/}~{\em
  {106}\/}({494}), {672--684}.

\bibitem[\protect\citeauthoryear{Cai, Liu, and Luo}{Cai
  et~al.}{2011}]{cai2011constrained}
Cai, T., W.~Liu, and X.~Luo ({2011}, {JUN}).
\newblock {A Constrained l(1) Minimization Approach to Sparse Precision Matrix
  Estimation}.
\newblock {\em {Journal of the American Statistical Association}\/}~{\em
  {106}\/}({494}), {594--607}.

\bibitem[\protect\citeauthoryear{Cai, Liu, and Xia}{Cai
  et~al.}{2014}]{tony2014two}
Cai, T.~T., W.~Liu, and Y.~Xia ({2014}, {MAR}).
\newblock {Two-sample test of high dimensional means under dependence}.
\newblock {\em {Journal of the Royal Statistical Society Series B-Statistical
  Methodology}\/}~{\em {76}\/}({2}), {349--372}.

\bibitem[\protect\citeauthoryear{Chen and Qin}{Chen and
  Qin}{2010}]{chen2010two}
Chen, S.~X. and Y.-L. Qin ({2010}, {APR}).
\newblock {A two-sample test for high-dimensional data with applications to
  gene-set testing}.
\newblock {\em {Annals of Statistics}\/}~{\em {38}\/}({2}), {808--835}.

\bibitem[\protect\citeauthoryear{Dempster}{Dempster}{1958}]{dempster1958high}
Dempster, A.~P. (1958).
\newblock A high dimensional two sample significance test.
\newblock {\em The Annals of Mathematical Statistics\/}, 995--1010.

\bibitem[\protect\citeauthoryear{Donoho and Jin}{Donoho and
  Jin}{2004}]{donoho2004higher}
Donoho, D. and J.~Jin ({2004}, {JUN}).
\newblock {Higher criticism for detecting sparse heterogeneous mixtures}.
\newblock {\em {Annals of Statistics}\/}~{\em {32}\/}({3}), {962--994}.

\bibitem[\protect\citeauthoryear{Gregory}{Gregory}{2014}]{GCTpack}
Gregory, K. (2014).
\newblock {\em highD2pop: Two-Sample Tests for Equality of Means in High
  Dimension}.
\newblock R package version 1.0.

\bibitem[\protect\citeauthoryear{Gregory, Carroll, Baladandayuthapani, and
  Lahiri}{Gregory et~al.}{2015}]{gregory2015two}
Gregory, K.~B., R.~J. Carroll, V.~Baladandayuthapani, and S.~N. Lahiri ({2015},
  {JUN}).
\newblock {A Two-Sample Test for Equality of Means in High Dimension}.
\newblock {\em {Journal of the American Statistical Association}\/}~{\em
  {110}\/}({510}), {837--849}.

\bibitem[\protect\citeauthoryear{Heller and Heller}{Heller and
  Heller}{2016}]{heller2016multivariate}
Heller, R. and Y.~Heller ({2016}).
\newblock {Multivariate tests of association based on univariate tests}.
\newblock In {Lee, DD and Sugiyama, M and Luxburg, UV and Guyon, I and Garnett,
  R} (Ed.), {\em {Advances in Neural Information Processing Systems 29 (NIPS
  2016)}}, Volume~{29} of {\em {Advances in Neural Information Processing
  Systems}}.
\newblock {30th Conference on Neural Information Processing Systems (NIPS),
  Barcelona, SPAIN, 2016}.

\bibitem[\protect\citeauthoryear{Hemmelmann, Horn, Reiterer, Schack, Susse, and
  Weiss}{Hemmelmann et~al.}{2004}]{hemmelmann2004multivariate}
Hemmelmann, C., M.~Horn, S.~Reiterer, B.~Schack, T.~Susse, and S.~Weiss
  ({2004}, {OCT 15}).
\newblock {Multivariate tests for the evaluation of high-dimensional EEG data}.
\newblock {\em {Journal of Neuroscience Methods}\/}~{\em {139}\/}({1}),
  {111--120}.

\bibitem[\protect\citeauthoryear{Henze}{Henze}{1988}]{henze1988multivariate}
Henze, N. (1988).
\newblock A multivariate two-sample test based on the number of nearest
  neighbor type coincidences.
\newblock {\em The Annals of Statistics\/}, 772--783.

\bibitem[\protect\citeauthoryear{Hotelling}{Hotelling}{1931}]{hotblling1931generalization}
Hotelling, H. (1931).
\newblock The generalization of student's ratios.
\newblock {\em Annals of Mathematical Statistics\/}~{\em 2}, 360--378.

\bibitem[\protect\citeauthoryear{Karlin and Rinott}{Karlin and
  Rinott}{1981}]{karlin1981total}
Karlin, S. and Y.~Rinott ({1981}).
\newblock {Total positivity properties of absolute value multinormal variables
  with applications to confidence interval estimates and related probabilistic
  inequalities}.
\newblock {\em {Annals of Statistics}\/}~{\em {9}\/}({5}), {1035--1049}.

\bibitem[\protect\citeauthoryear{Karr}{Karr}{1993}]{karrProb}
Karr, A.~F. (1993).
\newblock {\em Probability}.
\newblock Springer.

\bibitem[\protect\citeauthoryear{Keselman, Cribbie, and Holland}{Keselman
  et~al.}{2002}]{keselman2002controlling}
Keselman, H., R.~Cribbie, and B.~Holland ({2002}, {MAY}).
\newblock {Controlling the rate of Type I error over a large set of statistical
  tests}.
\newblock {\em {British Journal of Mathematical \& Statistical
  Psychology}\/}~{\em {55}\/}({1}), {27--39}.

\bibitem[\protect\citeauthoryear{Krishnamoorthy and Yu}{Krishnamoorthy and
  Yu}{2004}]{krishnamoorthy2004modified}
Krishnamoorthy, K. and J.~Yu ({2004}, {JAN 15}).
\newblock {Modified Nel and Van der Merwe test for the multivariate
  Behrens-Fisher problem}.
\newblock {\em {Statistics \& Probability Letters}\/}~{\em {66}\/}({2}),
  {161--169}.

\bibitem[\protect\citeauthoryear{L{\"a}uter}{L{\"a}uter}{2013}]{lauter2013simes}
L{\"a}uter, J. (2013).
\newblock Simes’ theorem is generally valid for dependent normally
  distributed variables.
\newblock In {\em Invited talk at the international conference on simultaneous
  inference}.

\bibitem[\protect\citeauthoryear{Lin and Pan}{Lin and Pan}{2016}]{MeanPack}
Lin, L. and W.~Pan (2016).
\newblock {\em highmean: Two-Sample Tests for High-Dimensional Mean Vectors}.
\newblock R package version 3.0.

\bibitem[\protect\citeauthoryear{Lopes, Jacob, and Wainwright}{Lopes
  et~al.}{2011}]{lopes2011more}
Lopes, M., L.~Jacob, and M.~J. Wainwright (2011).
\newblock A more powerful two-sample test in high dimensions using random
  projection.
\newblock In {\em Advances in Neural Information Processing Systems}, pp.\
  1206--1214.

\bibitem[\protect\citeauthoryear{Reiner-Benaim}{Reiner-Benaim}{2007}]{reiner2007fdr}
Reiner-Benaim, A. ({2007}, {FEB}).
\newblock {FDR control by the BH procedure for two-sided correlated tests with
  implications to gene expression data analysis}.
\newblock {\em {Biometrical Journal}\/}~{\em {49}\/}({1}), {107--126}.
\newblock {4th International Conference on Multiple Comparison Procedures
  (MCP2005), Shanghai, PEOPLES R CHINA, AUG 17-19, 2005}.

\bibitem[\protect\citeauthoryear{Ruiz-Meana, Garcia-Dorado, Pina, Inserte,
  Agullo, and Soler-Soler}{Ruiz-Meana et~al.}{2003}]{ruiz2003cariporide}
Ruiz-Meana, M., D.~Garcia-Dorado, P.~Pina, J.~Inserte, L.~Agullo, and
  J.~Soler-Soler ({2003}, {SEP}).
\newblock {Cariporide preserves mitochondrial proton gradient and delays ATP
  depletion in cardiomyocytes during ischemic conditions}.
\newblock {\em {American Journal of Physiology-heart and Circulatory
  Physiology}\/}~{\em {285}\/}({3}), {H999--H1006}.

\bibitem[\protect\citeauthoryear{Sarkar}{Sarkar}{1998}]{sarkar1998some}
Sarkar, S. ({1998}, {APR}).
\newblock {Some probability inequalities for ordered MTP2 random variables: A
  proof of the Simes conjecture}.
\newblock {\em {Annals of Statistics}\/}~{\em {26}\/}({2}), {494--504}.

\bibitem[\protect\citeauthoryear{Shen and Lin}{Shen and
  Lin}{2015}]{shen2015adaptive}
Shen, Y. and Z.~Lin ({2015}, {SEP}).
\newblock {An adaptive test for the mean vector in large-p-small-n problems}.
\newblock {\em {Computational Statistics \& Data Analysis}\/}~{\em {89}},
  {25--38}.

\bibitem[\protect\citeauthoryear{Simes}{Simes}{1986}]{simes1986improved}
Simes, R.~J. ({1986}, {DEC}).
\newblock {An improved Bonferroni procedure for multiple tests of
  significance}.
\newblock {\em {Biometrika}\/}~{\em {73}\/}({3}), {751--754}.

\bibitem[\protect\citeauthoryear{Srivastava and Du}{Srivastava and
  Du}{2008}]{srivastava2008test}
Srivastava, M.~S. and M.~Du ({2008}, {MAR}).
\newblock {A test for the mean vector with fewer observations than the
  dimension}.
\newblock {\em {Journal of Multivariate Analysis}\/}~{\em {99}\/}({3}),
  {386--402}.

\bibitem[\protect\citeauthoryear{Stadje}{Stadje}{1990}]{stadje1990collector}
Stadje, W. (1990).
\newblock The collector's problem with group drawings.
\newblock {\em Advances in Applied Probability\/}, 866--882.

\bibitem[\protect\citeauthoryear{Thulin}{Thulin}{2014}]{thulin2014high}
Thulin, M. ({2014}, {JUN}).
\newblock {A high-dimensional two-sample test for the mean using random
  subspaces}.
\newblock {\em {Computational Statistics \& Data Analysis}\/}~{\em {74}},
  {26--38}.

\bibitem[\protect\citeauthoryear{Williams, Jones, and Tukey}{Williams
  et~al.}{1999}]{williams1999controlling}
Williams, V., L.~Jones, and J.~Tukey ({1999}, {SPR}).
\newblock {Controlling error in multiple comparisons, with examples from
  state-to-state differences in educational achievement}.
\newblock {\em {Journal of Educational and Behavioral Statistics}\/}~{\em
  {24}\/}({1}), {42--69}.

\bibitem[\protect\citeauthoryear{Xiong, Zhao, and Boerwinkle}{Xiong
  et~al.}{2002}]{xiong2002generalized}
Xiong, M., J.~Zhao, and E.~Boerwinkle ({2002}, {MAY}).
\newblock {Generalized T-2 test for genome association studies}.
\newblock {\em {American Journal of Human Genetics}\/}~{\em {70}\/}({5}),
  {1257--1268}.

\bibitem[\protect\citeauthoryear{Yang, Yamada, Hill, Hemberg, Reddy, Cho,
  Guthrie, Oldenborg, Heiney, Ohmae, Medina, Holy, and Bonni}{Yang
  et~al.}{2016}]{yang2016chromatin}
Yang, Y., T.~Yamada, K.~K. Hill, M.~Hemberg, N.~C. Reddy, H.~Y. Cho, A.~N.
  Guthrie, A.~Oldenborg, S.~A. Heiney, S.~Ohmae, J.~F. Medina, T.~E. Holy, and
  A.~Bonni ({2016}, {JUL 15}).
\newblock {Chromatin remodeling inactivates activity genes and regulates neural
  coding}.
\newblock {\em {SCIENCE}\/}~{\em {353}\/}({6296}), {300--305}.

\bibitem[\protect\citeauthoryear{Yekutieli}{Yekutieli}{2008}]{yekutieli2008false}
Yekutieli, D. ({2008}, {FEB 1}).
\newblock {False discovery rate control for non-positively regression dependent
  test statistics}.
\newblock {\em {Journal of Statistical Planning and Inference}\/}~{\em
  {138}\/}({2}), {405--415}.

\bibitem[\protect\citeauthoryear{Zhang and Pan}{Zhang and
  Pan}{2016}]{zhang2016high}
Zhang, J. and M.~Pan ({2016}, {MAY}).
\newblock {A high-dimension two-sample test for the mean using cluster
  subspaces}.
\newblock {\em {Computational Statistics \& Data Analysis}\/}~{\em {97}},
  {87--97}.

\end{thebibliography}
	\clearpage

	\setcounter{equation}{0}
	\setcounter{figure}{0}
	\setcounter{page}{1}
    \setcounter{table}{0}
   \setcounter{section}{0}
	\renewcommand{\thetable}{S\arabic{table}}%
	\renewcommand{\thefigure}{S\arabic{figure}}%
	\makeatletter
	\renewcommand{\theequation}{S\arabic{equation}}
	\renewcommand{\thefigure}{S\arabic{figure}}
	\renewcommand{\bibnumfmt}[1]{[S#1]}
	\renewcommand{\citenumfont}[1]{S#1}
	
	\begin{center}
		\textbf{\large Supplementary Materials}
	\end{center}
	
	\section{Theorem and Lemmas} 
	
	\begin{theorem}\label{dthero}
		$Cov(T^2_{S_m}, T^2_{S_m^*}) \geq 0 \: \:  S_m, S_m^* \in M$ for
		$\xi_n = \frac{m}{n}, \; \lim_{n\to\infty} \xi_n = \xi \in (0, 1)$.
	\end{theorem} 
	The proof start with the observation that the Hotelling $T^2$ statistic  of the $S_m$ subset can be rewritten as (as in  \cite{bibby1979multivariate})
	
	\begin{equation}
	T^2_{S_m} =  \frac{n \bar{X}_{S_m}' {\Sigma}^{-1}_{S_m} \bar{X}_{S_m}}{\frac{\bar{X}_{S_m}' {\Sigma}^{-1}_{S_m} \bar{X}_{S_m}}{\bar{X}_{S_m}' \hat{\Sigma}^{-1}_{S_m} \bar{X}_{S_m}}} = A_{S_m} / B_{S_m}, 
	\end{equation}	
	where  
	\begin{equation}
	A_{S_m} = n \bar{X}_{S_m}' {\Sigma}^{-1}_{S_m} \bar{X}_{S_m} \sim \chi^2_{p},
	\end{equation}
	and 
	\begin{equation} 
	B_{S_m} = \frac{\bar{X}_{S_m}' {\Sigma}^{-1}_{S_m} \bar{X}_{S_m}}{\bar{X}_{S_m}' \hat{\Sigma}^{-1}_{S_m} \bar{X}_{S_m}} \sim \chi_{n-p +1}.
	\end{equation} 
	
	We also require to prove the following Lemmas. 
	\begin{lemma} \label{ezlema}
		For $X \sim N(0, \Sigma)$, then 
		\begin{equation}
		E(X_i^2  X_j^2) =2 \Sigma_{i,j}^2.
		\end{equation}
	\end{lemma}
	
	Proof: 
	
	w.l.o.g assume $Var(X_i) =Var(X_j) = 1$. Let $Z \sim N(0,1)$ where $Z\perp X_i, X_j$, then 
	
	\begin{equation}
	X_j =  \rho X_i + \sqrt{1 -\rho^2} Z.  
	\end{equation}
	
	Allowing us to calculate the covariance 
	
	\begin{equation}
	\begin{split} 
	& 	E\left(X_i^2 (\rho^2 X_i^2 + 2X_i Z + Z^2) \right)= \\ & \rho^2 E(X_i^2  X_i^2) + 2 E(X_i  X_i  Z) + E(X_i^2 Z^2) = 2 \rho_{i,j}^2.
	\end{split} 
	\end{equation}

	\begin{lemma} \label{lemma1}
		If $X \sim N(0, \Sigma) $ then $\forall  S_m, S^*_m \in M $ 
		\begin{center}
			$\quad Cov(A_{S_m},A_{S_m^*}) > 0$
		\end{center} 
		Proof:
		\begin{equation*}
		\begin{split}
		& A_{S_m} = n \bar{X}_{S_m}' {\Sigma}^{-1}_{S_m} \bar{X}_{S_m} = \\ &  n\bar{X}_{S_m}'(D\Lambda^{\frac{1}{2}} D'D \Lambda^{\frac{1}{2}} D')^{-1}_{S_m} \bar{X}_{S_m} =\\ &  n\bar{X}_{S_m} 'D_{S_m}(\Lambda^{\frac{1}{2}}_{S_m} \Lambda^{\frac{1}{2}}_{S_m})^{-1} D'_{S_m}\bar{X}_{S_m} = \\ & U'_{S_m} \Lambda^{-\frac{1}{2}}_{S_m} \Lambda^{-\frac{1}{2}}_{S_m} U_{S_m},
		\end{split}  
		\end{equation*}
		$U_{S_m} = \sqrt{n}\bar{X}_{S_m}'D_{S_m}$, $D_{S_m}$ is a $m\times m$ matrix whose columns are the eigenvectors of $\Sigma_{S_m}$. $\Lambda_{S_m} $ is a diagonal matrix of eigenvalues. Under the null hypothesis $U_{S_m}$ is distributed as $N(0,\lambda_i)$, $\lambda_i$ being the eigenvalue with respect to the corresponding eigenvector in the $i$'th column of $D$. 
		Writing the quadratic form in as a sum, 
		
		\begin{equation*}
		n\bar{X}_{S_m}'\Sigma_{S_m}^{-1}\bar{X}_{S_m} = \sum_{i=1}^{m} \frac{U^2_{S_m,i}}{\lambda_i}.
		\end{equation*} 
		The covariance between $A_{S_m}$ and $A_{S^*_m}$ is 
		\begin{equation*}
		\begin{split}
		& Cov(A_{S_m}, A_{S_m^*}) = E(A_{S_m} A_{S_m^*}) - E(A_{S_m})E(A_{S_m^*}) =\\ &  E\bigg(\sum_{i=1}^{m}\frac{U^2_{S_m,i}} {\lambda_{S_m,i}} \sum_{j=1}^{m} \frac{U^2_{S_m^*,j}}{\lambda_{S_m^*,j}} \bigg) - m^2 =  \\&  \sum_{i=1}^{m} \sum_{j=1}^{m} E\bigg(\frac{U^2_{S_m,i}}{\lambda_{S_m,i}}  \frac{U^2_{S_m^*,j}}{\lambda_{S_m^*,j}} \bigg) - m^2.
		\end{split}
		\end{equation*}
		
		Using the result from Lemma \ref{ezlema} and defining $c_{i,j} = cov\bigg(\frac{U_{S_m, i}}{\sqrt{\lambda_{S_m, i}}} ,\frac{U_{S_m^*,j}}{\sqrt{\lambda_{S_m^*,j}}} \bigg) $, we obtain

		\begin{equation*}
		\begin{split}
		& \sum_{i=1}^{m} \sum_{j=1}^{m} E\left( \left(\frac{U^2_{S_m,i}}{\lambda_{S_m,i}} \right) \left(\frac{U^2_{S_m^*,j}}{\lambda_{S_m^*,j}}\right) \right) - m^2 =\\& \sum_{i=1}^{m} \sum_{j=1}^{m} (1 + 2c^2_{ij}) - m^2 = \\& 2 \sum_{i=1}^{m}  \sum_{j=1}^{m} c^2_{ij} + m^2 - m^2 = 2\sum_{i=1}^{m} \sum_{j=1}^{m} c^2_{ij}. 
		\end{split}  
		\end{equation*}
		
		Concluding $ Cov(A_{S_m},A_{S^*_m}) > 0$.
	\end{lemma} 
	
	\begin{lemma}  \label{lemma2}
		If $X \sim N(0, \Sigma) $ then 
		\begin{center}
			$A_{S_m} \perp B_{S_m^*}$.
		\end{center} 
		
		Proof: 
		
		The joint distribution of $X_{S_m}, X_{S_m^*}$ can be written as 
		\begin{equation}
		X_{S_m},X_{S^*_m} \sim N\left(0, 
		\begin{pmatrix}
		\Sigma_{S_m} & \Sigma_{S_m S_m^*} \\ 
		\Sigma_{S_m^* S_m} & \Sigma_{S_m^*} \\ 
		\end{pmatrix} \right). 
		\end{equation}  
		
		$B_{S_m} \perp \Sigma^{-0.5} \bar{X}_{S_m}$, \cite{bibby1979multivariate}. Therefore, $B_{S_m}$ is dependent solely on $\hat{\Sigma}^{-1}_{S_m}$, we can define $B_{S_m} = g \left(\hat{\Sigma}_{S_m} ^{-1} \right)$, and$A_{S^*_m} = f(\bar{X}_{S^*_m}) $. 
		
		Since $\hat{\Sigma}^{-1}_{S_m} \perp \bar{X}_{S_m^*}$ Cochran's Theorem, \cite{bibby1979multivariate} and $f$ and $g$ are Borel measurable functions, according to Theorem 3.1 \cite{karrProb}, $B_{S_m} \perp A_{S^*_m}$, concluding the proof of Lemma \ref{lemma2}.
		
	\end{lemma}
	
	We can now turn to prove Theorem \ref{dthero}, define $A_{S_m} = A_{S_m} / p$, $A_{S_m} \sim \chi^2_p / p$ and $B_{S_m}^* = B_{S_m} / (n - p + 1)$, $B_{S_m}^* \sim \chi^2_{n-p+1}/ (n-p+1)$. 
	We will use Taylor approximation to estimate a $E\left(f\left( \vect{\theta} \right) \right) = E\left(\frac{A_{S_m}}{B_{S_m}^*}\frac{A_{S^*_m}}{B_{S^*_m}} \right)$ at the expansion point $E\left( \vect{\theta} \right)$, where $\vect{\theta} = \left\{ A_{S_m},A_{S_m^*},B_{S_m},B_{S^*_m} \right\}$. 
	
	\begin{equation} \label{taylor}
	\begin{split}
	E\left( f(\vect{\theta} ) \right) &=  E\left(\frac{A_{S_m}^*}{B_{S_m}^*}\frac{A_{S_m^*}^*}{B_{S_m^*}^*} \right)\\ &  \approx 
	E \left(  \frac{E(A_{S_m}^*) E(A_{S_m^*}^*)}{E(B_{S_m}^*) E(B_{S_m^*}^*)} \right) + E \left( \nabla f(E(\boldsymbol{\theta}))  \left[  \boldsymbol{\theta} - E\left(\boldsymbol{\theta} \right) \right] \right)  + \\ & 0.5  
	E \left( \left[  \boldsymbol{\theta} - E\left(\boldsymbol{\theta} \right) \right]' H \left( f\left(E \left(\boldsymbol{\beta}\right) \right) \right)  \left[  \boldsymbol{\theta} - E\left(\boldsymbol{\theta} \right)  \right] \right) \\ & =  1 + 0.5  
	E \left( \left[  \boldsymbol{\theta} - E\left(\boldsymbol{\theta} \right) \right]' H \left( f\left(E \left(\boldsymbol{\theta}\right) \right) \right)  \left[  \boldsymbol{\theta} - E\left(\boldsymbol{\theta} \right)  \right] \right) .
	\end{split}
	\end{equation}

	We are left to find the Hessian, $H$, we start by finding the gradient 
	\begin{equation}
	\nabla f \left( \vect{\theta} \right) = \left(\frac{A_{S_m^*}^*}{B_{S_m}^* B_{S_m^*}^*}, \frac{-A_{S_m}^* A_{S_m^*}^*}{B_{S_m}^{*2} B_{S_m^*}^*}, 		\frac{A_{S_m}^*}{B_{S_m}^* B_{S_m^*}^*}, \frac{-A_{S_m}^* A_{S_m^*}^*}{B_{S_m}^* 	B_{S_m^*}^{*2}}\right).
	\end{equation} 
	The Hessian is
	\begin{equation} 
	H \left( f \left( \vect{\theta} \right) \right) = \begin{bmatrix}
	0             & -\frac{A_{S_m^*}^*}{B_{S_m}^{*2} B_{S_m^*}^* }  & \frac{A_{S_m}^*}{B_{S_m}^* B_{S_m^*}^*}    & -\frac{A_{S_m^*}^*}{B_{S_m^*}^{*2} B_{}^*} \\ 
	-\frac{A_{S_m^*}^*}{B_{S_m}^{*2} B_{S_m^*}^*} & \frac{2A_{S_m}^* A_{S_m^*}^*}{B_{S_m}^{*3} B_{S_m^*}^*}  & -\frac{A_{S_m}^*}{B_{S_m}^{*2} B_{S_m^*}^*} & \frac{A_{S_m}^* A_{S_m^*}^*}{B_{S_m}^{*2} B_{S_m^*}^{*2}} \\
	\frac{A_{S_m}^*}{B_{S_m}^* B_{S_m^*}^*}     & -\frac{A_{S_m}^*}{B_{S_m^*}^{*2} B_{S_m^*}^*}    &  0               & -\frac{A_{S_m}^*}{B_{S_m^*}^{*2} B_{S_m}^*} \\ 
	-\frac{A_{S_m^*}^*}{B_{S_m}^* B_{S_m^*}^{*2}} & \frac{A_{S_m}^* A_{S_m^*}^*}{B_{S_m}^{*2} B_{S_m^*}^{*2}} & -\frac{A_{S_m}^*}{B_{S_m}^* B_{S_m^*}^{*2} }  & \frac{2A_{S_m}^* A_{S_m^*}^*}{B_{S_m}^* B_{S_m^*}^{*3}} 
	\end{bmatrix}. 
	\end{equation} 
	
	returning to Eq. \ref{taylor}, 
	
	\begin{multline}
	E\left(\frac{A_{S_m}^*}{B_{S_m}^*} \frac{A_{S_m^*}^*}{B_{S_m^*}^*} \right) \approx 1 + 0 + Cov(A_{S_m}^*, A_{S_m^*}^*) + Cov(B_{S_m}^*, B_{S_m^*}^*) + Cov(A_{S_m}^*, B_{S_m^*}^*) + \\ Cov(A_{S_m^*}^*, B_{S_m}^*) + Cov(A_{S_m}^*, B_{S_m}^*) + Cov(A_{S_m^*}^*, B_{S_m^*}^*) + Var(B_{S_m}^*) + Var(B_{S_m^*}^*). 
	\end{multline}
	According to Lemma \ref{lemma2} $ \forall k,l$, $Cov(A_{S_m^*}^*, B_{S_m}^*) = 0$, and according to Lemma \ref{lemma1} $\forall k,l$, $Cov(A_{S_m}, A_{S_m^*}^*) > 0$, allowing us to bound 
	
	\begin{equation}
	E\left(\frac{A_{S_m}^*}{B_{S_m}^*} \frac{A_{S_m^*}^*}{B_{S_m^*}^*} \right) > 1 + Cov(B_{S_m}^*, B_{S_m^*}^*) + Var(B_{S_m}^*) + Var(B_{S_m^*}^*).
	\end{equation}
	
	Since $ \; B_{S_m}\sim \chi_{n-m+1}$, and $Cov(B_{S_m}^*, B_{S_m^*}^*)^2 \leq Var(B_{S_m}^*) Var(B_{S_m^*}^*)$, then $Cov(B_{S_m}^*, B_{S_m^*}^*) \geq -\frac{2}{\left(n-m+1\right)}$, we obtain that 
	\begin{equation}
	E\left(\frac{A_{S_m}^*}{B_{S_m}^*} \frac{A_{S_m^*}^*}{B_{S_m^*}^*} \right)  \geq 1 + \frac{2}{n-m+1} > 0 ,
	\end{equation}
	thus $Cov\left(T_{S_m}, T_{S_m^*}  \right) > 0$.

	\begin{lemma} \label{cltlemma} 
		
		Under the conditions of Theorem \ref{dthero}, 
		
		\begin{equation}
		\sqrt{\frac{(1 - \xi)^3n}{2 \xi}  }  \left(\frac{\vect{T}^2_m}{n} - \vect{\xi}_n \right) \rightarrow N(0, \Sigma^*), \; as \; n\to \infty,
		\end{equation}

		where $\boldsymbol{T}_m = \{T_{S_m}^{(1)} ,\ldots , T^{(B)}_{S_m}\}$, and $\vect{\xi}_n = \{ \frac{\xi_n}{1-\xi_n}, \ldots, \frac{\xi_n}{1-\xi_n} \} $. 
		
		Proof: 
		
		Apply the multivariate CLT to obtain the result. 
		
	\end{lemma}
	
	\begin{theorem}
		If  $X\sim N(\boldsymbol{\mu}, \Sigma_p)$, then $\lim\limits_{n,p \rightarrow \infty} P(SH_{m,B} (X) \leq \alpha) \leq \alpha$ for $\xi_n = \frac{m}{n}, \; \lim\limits_{n\to\infty} \xi_n = \xi \in (0, 1)$. 
		
		According Lemma \ref{cltlemma}, $\lim\limits_{n,p \rightarrow \infty} \boldsymbol{T}_m \sim N(0, \Sigma^*)$. According to Theorem \ref{dthero}  $\forall i,j \; \Sigma_{i,j} > 0$. Applying Theorem 1.2 from \cite{benjamini2001control}, concludes the proof.

	\end{theorem}
	
	\section{Choosing sample size $m$}
	
	The asymptotic power function of Hotelling's test, assuming that $\xi_n = \frac{m}{n}, \; \lim\limits_{n\to\infty} \xi_n = \xi \in (0, 1)$, is 
	
	\begin{center}
		$\pi \left(T^2(X, Y)\right) =  \Phi\bigg( Z_\alpha + \sqrt{\frac{n(1 - \xi)}{2\xi}} r(1 - r)||\Delta_{m}||^2  \bigg),$
	\end{center} 
	
	where $\Delta_m =  \Sigma_m^{-\frac{1}{2}} \vect{\delta}_m $. 
	Using Cauchy interlace theorem we can obtain the following lower bound for
	
	\begin{equation} \label{eq:2}
	\frac{\Delta_{m}^2}{||\vect{\delta}_m ||_2^2} \geq \frac{1}{\lambda_{min}(\Sigma_m)} \geq \frac{1}{\lambda_{min}(\Sigma_p)} \geq 
	\frac{tr(\Sigma_{p})}{\lambda_{min}(\Sigma_p)tr(\Sigma_{p})} \geq 
	\frac{m}{tr(\Sigma_{p})}
	\end{equation}

	Expression \ref{eq:2} indicate that the for the worst case scenario, the power function of the Hotelling increases linearly in the number of dimensions, implying that the maximum of the power function is achieved at $m = \frac{n}{2}$. 
	
	\section{Parametric simulations} 
	
	Here can be found the omitted parts regarding the the parametric tests, mainly a description of the various parametric tests and simulation results deemed not interesting enough to present in the main section. 
	
	\subsection{Comparing parametric tests}

	From \cite{dempster1958high} to \cite{tony2014two} , several tests were suggested in order to appropriately address the setting where $p>n$. 
	Some procedures are relatively similar to the $T^2$ Hotelling test, replacing $\Sigma^{-1}$ with some other estimator usually at the cost of ignoring some of the covariance structure. Others are designed to ensure that the estimated covariance matrix will be invertible. We give here a detailed description of some of the tests, emphasizing the more recent ones.

	\subsubsection{Bai \& Saranadasa~ (1996)} \label{Bai}
	
	\cite{bai1996effect} suggested replacing $\Sigma^{-1}$ with $I_p$ resulting with
	\begin{equation*}
	T^2_B = (\bar{X}_1 - \bar{X}_2)'(\bar{X}_1 - \bar{X}_2) -\tau tr(\hat{\Sigma}),
	\end{equation*}
	
	where $\tau = \frac{n_1 n_2}{n_1 + n_2}$.
	It was shown to be superior to Dempster test \cite{dempster1958high}, and even to that of Hotelling $T^2$ (when $p < n$) as long as the covariance structure is fairly regular, that is  $\Sigma$  does not cause too large of a difference between does $\delta' \Sigma^{-1} \delta$ $ \sqrt{n}$  and $||\delta||^2 / \sqrt{tr(\Sigma^2)}$.

	\subsubsection{Srivastava \& Du~ (2008) (SD)} \label{Srivastava}
	
	\cite{srivastava2008test} used $D^{-1}_\sigma$ a diagonal matrix, to replace the inverse covariance matrix, where $ D_\sigma = \{\sigma_{11},...,\sigma_{pp}\} $ yielding the following test statistic:
	
	\begin{equation*}
	T^2_S = \frac{\tau (\bar{X}_1 - \bar{X}_2)'D^{-1}_\sigma(\bar{X}_1 - \bar{X}_2) - \frac{Np}{N -2}}{\big[2(tr (R^2) - \frac{p^2}{N})c_{p,N}\big]^{0.5}}
	\end{equation*}
	
	where $R =  D_\sigma^{-0.5}\Sigma D_\sigma^{-0.5}$ is the correlation matrix, $N = n_1 + n_2 - 2$ and $c_{p,N} = 1 + \frac{tr (R^2)}{p^\frac{3}{2}}$. Using simulations it was shown that the test is superior to both \cite{bai1996effect} and \cite{dempster1958high}.

	\subsubsection{Chen \& Qin~ (2010) (CQ)} \label{Chen}
	\cite{chen2010two} suggested the following test statistic
	
	\begin{equation*}
	T^2_C = \frac{\sum_{i\neq j}^{n_1} X_{i,1}'X_{j,1}}{n_1 (n_1 -1 )} + \frac{\sum_{i\neq j}^{n_2} X_{i,2}'X_{j,2}}{n_1 (n_1 -1 )} - 2 \frac{\sum_{i = 1}^{n_1} \sum_{j = 1}^{n_2} X'_{i,1}X_{j,2}}{n_1 n_2}.
	\end{equation*}
	
	$E(T^2_C) = ||\vect{\mu}_1 - \vect{\mu}_2||^2$, that is, the test is the sum of squares test between of the distance between the two mean vectors. By measuring it in Euclidean distance instead of Mahalanobis distance, it ignores the covariance structure. 
	
	The above tests suggest replacing $\hat{\Sigma}^{-1}$ with some matrix $A$. Doing so allows the tests to be conducted when $p>n$, but it also ignores the covariance structure. This test can also be expanded to non-equal covariance matrices. Therefore, it can be expected that this type of test will perform better when the covariance structure is close to independence \cite{lopes2011more}.

	\subsubsection{Lopes et al~ (2011)} \label{Lopes}
	
	\cite{lopes2011more} used random projection in order to ensure that $\hat{\Sigma}$ could be inverted, the test statistic is:
	\begin{equation*}
	T^2_L = \tau(\bar{X}_1 - \bar{X}_2)'P_k(P_k'\hat{\Sigma}P_k)^{-1}P_k'(\bar{X}_1 - \bar{X}_2),
	\end{equation*}
	
	$P_k$ being the random projection matrix from $R^p$ to $R^k$, generated by simulating $i.i.d$ $N(0,1)$. The random projection procedure has a guarantee regarding the distance between observations in the projected space, thereby taking into account the covariance structure. The test has greater power than both Srivastava and Chen \& Qin when the covariance structure is not regular (in the sense of 5.1).

	\subsubsection{Gregory et al~ (2015) (GCT)} \label{Gregory}
	
	Gregory et al. suggested using a scaled version of the two sample un-pooled variance t-statistic average over the dimensions. The statistic is used in scenarios where the $p$ dimensions indices adhere to a certain order related to the dependence between them. Let
	
	\begin{equation*}
	G_n = p^{1/2}(T^2_G - \hat{\nu}) / \hat{\zeta}  	,
	\end{equation*}
	
	where $T^2_G$ being the sum of the squared un-pooled variance of the t-statistics $T^2_n = \frac{1}{p} \sum\limits_{j=1}^p \frac{(\bar{X}_{1,j} - \bar{X}_{2,j})^2}{\hat{\Sigma}_{1_{jj}}/n_1 + \hat{\Sigma}_{2_{jj}}/n_2 }$, with  $ \hat{\nu}$, $\hat{\zeta}$ are the location and scale parameters. Although the test statistic is based on marginal testing, the construction of the scaling takes into account the dependence between dimensions. \\
	
	All of the above tests are based on the sum of squares approach, combining the signal across all dimensions. Therefore they are best suited to find a difference when each dimension has a weak signal. However, if the difference is relatively large but lies in a small number of dimensions, it is expected that tests based on this approach will not perform well. 
	
	\subsubsection{Cai et al~ (2014) (CLX)} \label{Cai}
	\cite{tony2014two} suggested using $\Omega = \Sigma^{-1}$, to transform the data, so that the test is conducted on the transformed observations $ \big\{ \hat{\Omega} X_{i,1} \; 1\leq i \leq n_1  \}$ and $ \big\{ \hat{\Omega} X_{i,2} \; 1\leq i \leq n_2\}$. After applying the transformation the dimensions are independent. The test statistic is defined as the maximum between the square difference of the two transformed samples:
	
	\begin{equation*}
	M_n  = \tau \underset{1\leq j \leq p}{\max} \frac{\bar{Z}_j^2}{\Omega_{jj}} ,
	\end{equation*}
	
	$Z$ is defined as $\bar{Z} = \Omega(\bar{X}_1 - \bar{X}_2)$. There is still the problem of the undefined covariance matrix. When $\Omega$ is known to be sparse it was suggested to estimate it directly using the constrained $l_1$-minimization for inverse matrix estimator (CLIME) \cite{cai2011constrained}. When $\Omega$ cannot be assumed to be sparse, it was suggested to use the inverse of the adaptive thresholding estimator \cite{cai2011adaptive}. The test statistic is distributed as a type-1 extreme value distribution. 
	The test is unique in the sense that it does not combine signals in order to detect them, but instead is tailored to 'hunt' stronger and sparser signals. 
	In the simulation study we used the adaptive thresholding estimator, since in some situations no such extra assumption can be made.

	\subsubsection{Type I error simulations} 
	
	Here can be found the following type I errors tables for covariance structures missing from the parametric simulation section, \textit{Model 7} and \textit{equally correlated}.  
	
	
	\begin{table}[!htbp]
		\tabcolsep=0.07cm
		\centering
		\label{tab: model 7 Full}
		\begin{tabular}{|c|c|c|c|c|c|c|c|c|c|}
			\hline
			$n_1 = n_2$     & SD & CQ  & Lopes & Simes & KNN   & GCT   & CLX   & SH$_\frac{n}{2}$& SH$_\frac{n}{4}$ \\
			\hline
			20    & 0.037 & 0.055 & 0.042 & 0.044 & 0.113 & 0.076 & 0.175 & 0.044 & 0.035 \\
			50    & 0.033 & 0.056 & 0.046 & 0.058 & 0.106 & 0.075 & 0.187 & 0.053 & 0.045 \\
			100   & 0.029 & 0.057 & 0.058 & 0.051 & 0.111 & 0.07  & 0.176 & 0.049 & 0.06 \\
			\hline
		\end{tabular}%
		\caption{Parametric tests type I error rates for the \textit{model 7} covariance matrix scenario for each of the methods.}
	\end{table}%

	\begin{table}[!htbp]
		\tabcolsep=0.07cm
		\centering
		\label{tab: equicovariance Full}
			\begin{tabular}{|c|c||c|c|c|c|c|c|c|c|c|}
				\hline
				$n_1 = n_2$ & $\rho$ & SD & CQ  & Lopes & Simes & KNN   & GCT   & CLX   & SH$_\frac{n}{2}$& SH$_\frac{n}{4}$  \\
				\hline
				20    & 0.1   & 0.053 & 0.072 & 0.043 & 0.039 & 0.393 & 0.163 & 0.11  & 0.055 & 0.05 \\
				50    & 0.1   & 0.052 & 0.074 & 0.033 & 0.065 & 0.367 & 0.092 & 0.093 & 0.054 & 0.04 \\
				100   & 0.1   & 0.051 & 0.07  & 0.052 & 0.051 & 0.398 & 0.059 & 0.091 & 0.05  & 0.038 \\
				20    & 0.3   & 0.037 & 0.078 & 0.046 & 0.043 & 0.829 & 0.252 & 0.141 & 0.05  & 0.042 \\
				50    & 0.3   & 0.034 & 0.078 & 0.063 & 0.04  & 0.802 & 0.661 & 0.099 & 0.057 & 0.046 \\
				100   & 0.3   & 0.022 & 0.073 & 0.053 & 0.036 & 0.824 & 0.996 & 0.11  & 0.047 & 0.059 \\
				20    & 0.5   & 0.014 & 0.079 & 0.056 & 0.032 & 0.913 & 0.401 & 0.136 & 0.043 & 0.043 \\
				50    & 0.5   & 0.011 & 0.08  & 0.049 & 0.033 & 0.887 & 0.986 & 0.108 & 0.039 & 0.044 \\
				100   & 0.5   & 0.003 & 0.072 & 0.041 & 0.025 & 0.91  & 1     & 0.11  & 0.041 & 0.041 \\
				\hline
			\end{tabular}%
		\caption{Parametric tests type I error rates for the \textit{equally correlated} covariance matrix scenario. $\rho$ is the correlation parameter.} 
	\end{table}%

	\subsubsection{Power simulations}
	
\begin{center}
		\begin{figure} 
			\includegraphics[width=15cm,height=16cm,keepaspectratio]{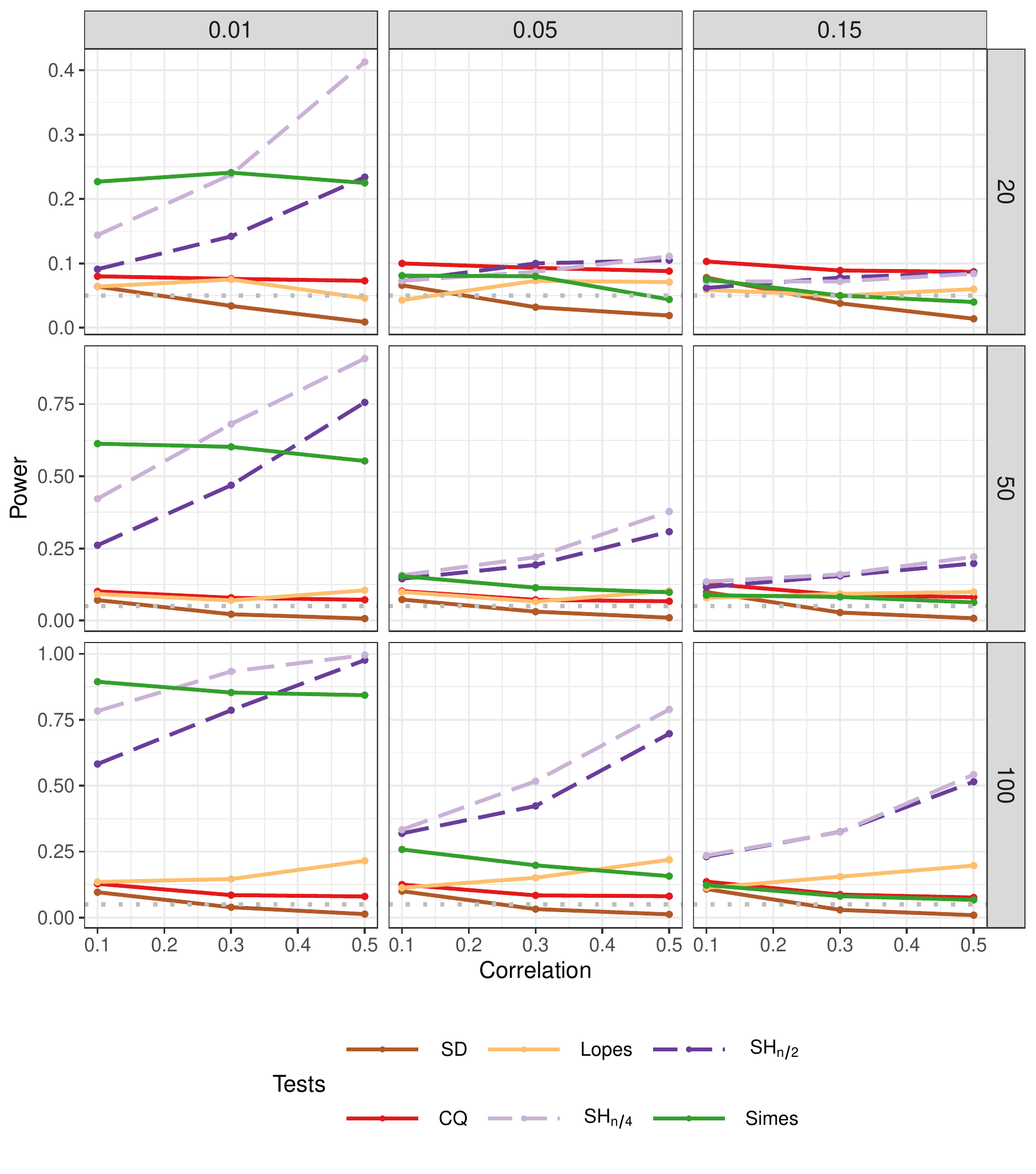}
			\caption{Parametric tests power comparison, \textit{equally correlated} covariance matrix.}
			\label{Equicorr_para}
		\end{figure} 
	\end{center}

	Shown here the rest of the simulations for the power of the functions in different scenarios. Rows indicate number of observations ($n_1 = n_2$), columns the proportion of dimensions where $\mu_i \neq 0$ unless specified otherwise. Also it is worth noticing that the plots are scales free across the rows, so notice is needed when comparing between curves in different rows. $||\vect{\mu}_1 - \vect{\mu}_2||_2^2 = 2.5 \sqrt{\frac{40}{n}}$ across all simulations. 
	Across all simulations it can noticed that SH performs uniformly better then the Lopes test. In Figure \ref{Scenario2_para}, a major different appears between test that include the covariance matrix, specifically Lopes and SH to the rest of the tests which are unable to detect any difference. 
	The rest of the simulations confirm our previous statements, the SH performs the best when the signal is sparse, and the covariance structure is such that there are dimensions that are strongly correlated or many dimensions weakly correlated (Figure \ref{Equicorr_para}). The least favorable scenario for it is complete independence. 
	For each set of parameters the power is estimated from 1000 repetitions. The number of dimensions is 600.

	\begin{center}
		\begin{figure} 
			\includegraphics[width=15cm,height=16cm,keepaspectratio]{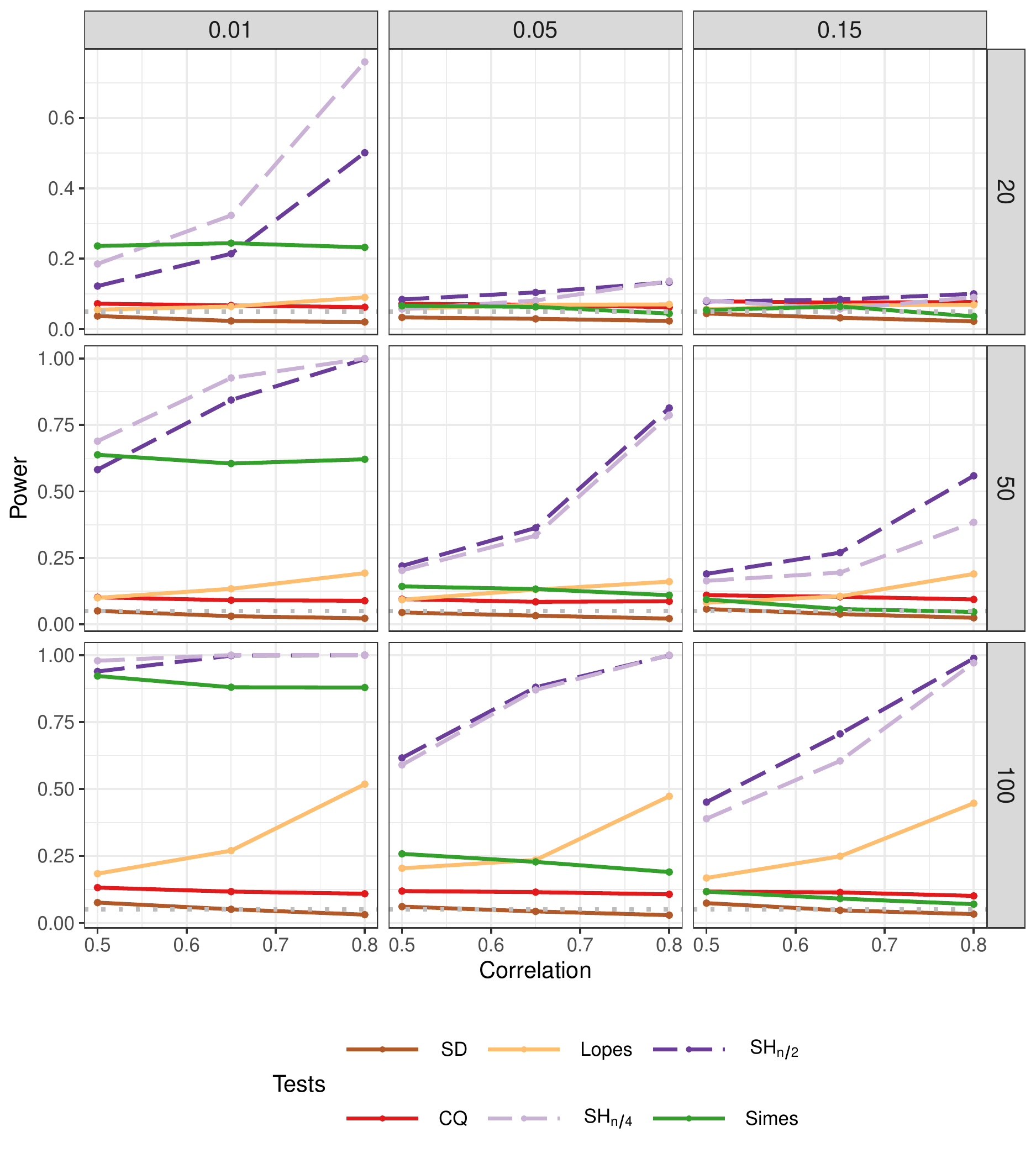}
			\caption{Parametric tests power comparison, \textit{block} covariance matrix, block size is 50.}
			\label{Block50_para}
		\end{figure} 
	\end{center}

	\begin{center}
		\begin{figure} 
			\includegraphics[width=15cm,height=16cm,keepaspectratio]{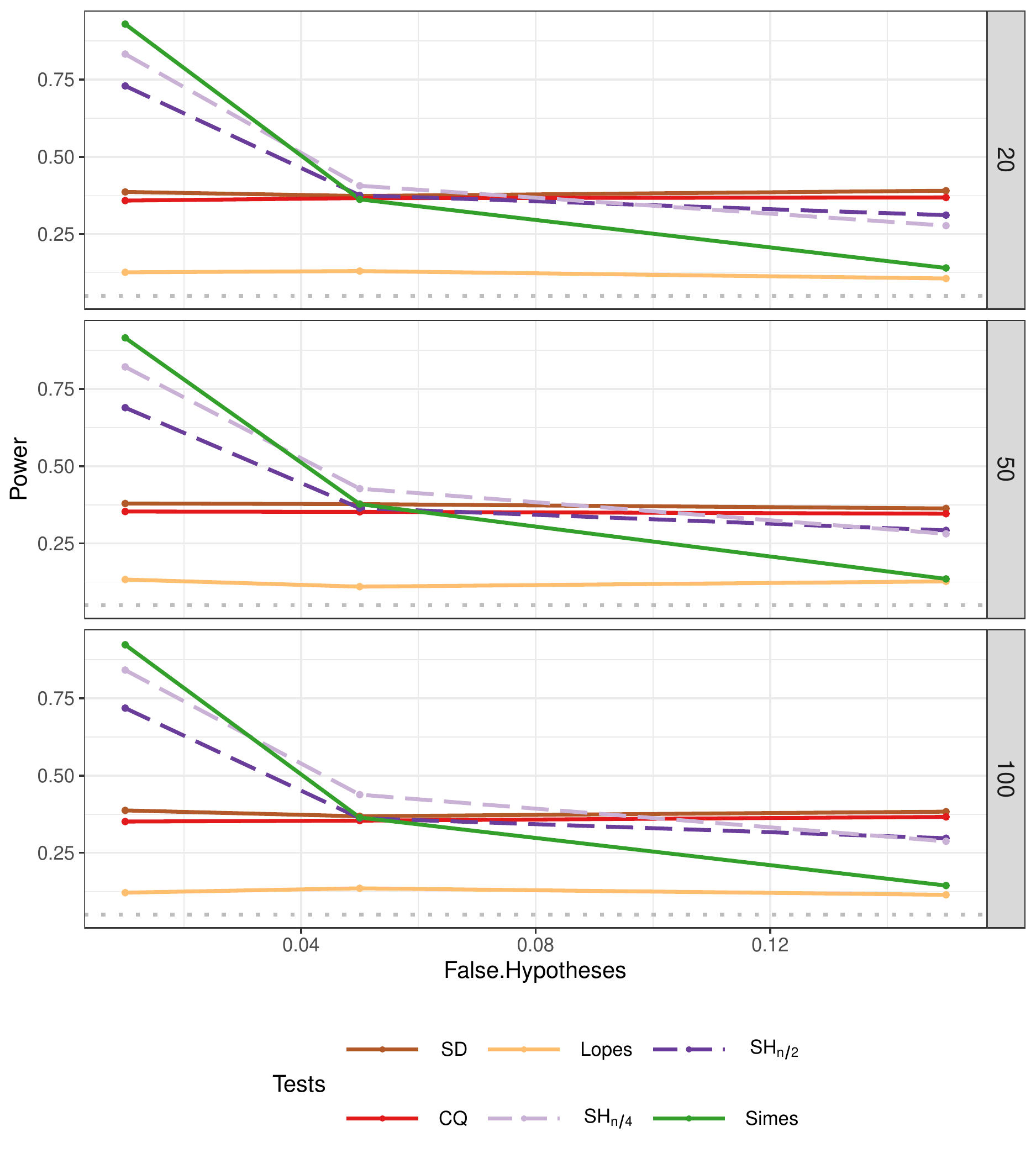}
			\caption{Parametric tests power comparison, \textit{model 7} covariance matrix.}
		\end{figure} \label{Model7_para}
	\end{center}

	\section{Non-equal covariance matrices}
	
	The following sections hold the rest of the simulations results for the non-equal covariance matrices tests. 
	
	\subsection{Type I error}
	
	Type I error rates for tests dealing with non-equal covariance matrices, the number of repetition is 1000, such the conservative standard deviation is $0.015$. It seems that all the tests except for GCT keep the required level (the tests are conducted at $\alpha = 0.05$).
	
	\begin{table}[!htbp]
		\centering
		
	 \begin{tabular}{|c||c|c|c|c|c|c|}
		\hline
		$n_1 = n_2$ & SD & CQ & Simes & SH$_\frac{n}{2}$& SH$_\frac{n}{4}$ \\
		\hline
		20    & 0.028 & 0.045 & 0.057 & 0.057 & 0.039 \\
		50    & 0.031 & 0.058 & 0.055 & 0.069 & 0.034 \\
		100   & 0.024 & 0.048 & 0.062 & 0.063 & 0.053 \\
		\hline
	\end{tabular}%
		\caption{Non-equal covariance tests type I error rate for the \textit{model 7} covariance matrix scenario ($\Sigma_1 = 2 * \Sigma_2$).}
		\label{Model7NonEqual}%
	\end{table}%

	\begin{table}[!htbp]
		\centering
	\begin{tabular}{|c|r||r|r|r|r|r|}
		\hline
		$n_1 = n_2$ & $\rho$ & SD & CQ &  Simes & SH$_\frac{n}{2}$& SH$_\frac{n}{4}$\\
		\hline
		\multirow{3}[2]{*}{20} & 0.1   & 0.045 & 0.076 & 0.057 & 0.036 & 0.023 \\
		& 0.3   & 0.036 & 0.074 & 0.051 & 0.03  & 0.02 \\
		& 0.5   & 0.019 & 0.077 & 0.037 & 0.024 & 0.021 \\
		\hline
		\multirow{3}[2]{*}{50} & 0.1   & 0.053 & 0.073 & 0.043 & 0.047 & 0.041 \\
		& 0.3   & 0.029 & 0.083 & 0.051 & 0.046 & 0.032 \\
		& 0.5   & 0.009 & 0.083 & 0.033 & 0.051 & 0.038 \\
		\hline
		\multirow{3}[2]{*}{100} & 0.1   & 0.051 & 0.072 & 0.06  & 0.072 & 0.047 \\
		& 0.3   & 0.021 & 0.071 & 0.033 & 0.057 & 0.055 \\
		& 0.5   & 0.004 & 0.068 & 0.025 & 0.057 & 0.037 \\
		\hline
	\end{tabular}%
	
		\caption{Non-equal covariance matrix tests type I error rates for the \textit{equally correlated} covariance matrix scenario ($\Sigma_1 = 2 * \Sigma_2$). $\rho$ is the covariance parameter.}
		\label{CovNonEqual}%
	\end{table}%
	
	We can see in similar vain of earlier results, it is better to sample a smaller number of dimensions since SH$_\frac{n}{4}$ performance better than SH$_\frac{n}{2}$.  
	
	\subsection{Power simulations}
	
	The results are similar to what is seen in the non-parametric case. For the sparse scenarios Simes has the highest power, and as the covariance between dimension increases (the block size or the correlation parameter increases) the SH tests begin to overtake it. CQ, SD are almost powerless in those scenarios. As the scenarios become more dense, SD and CQ became the most powerful tests, yet SH is not without power, and as was with the Simes test, as the covariance increases the SH begins to overtake them. This is especially notable in for all block covariance scenarios. 
	
	\begin{figure} [!htbp]
			\includegraphics[width=15cm,height=16cm,keepaspectratio]{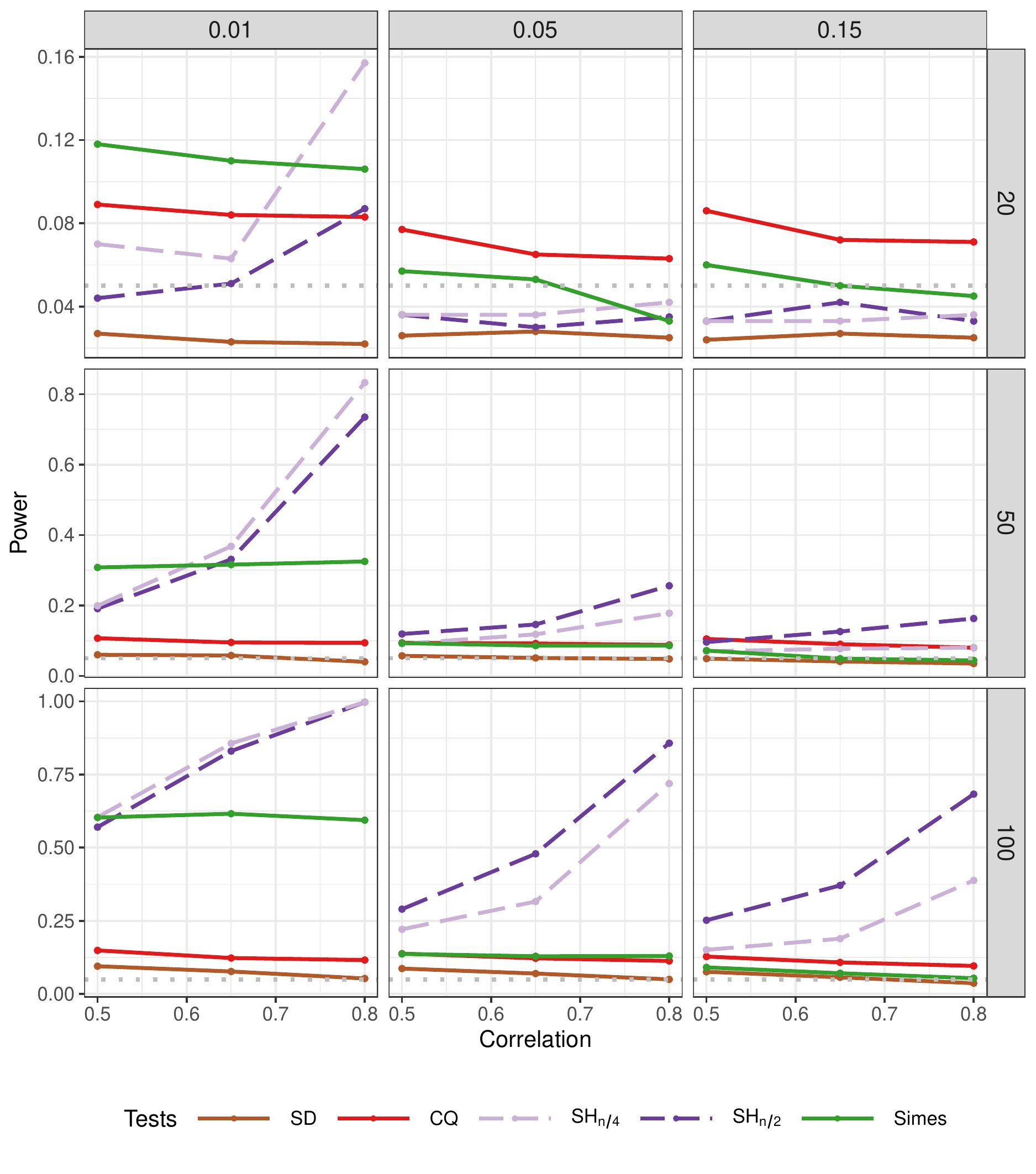}
			\caption{Non-equal covariance tests power comparison, \textit{block} covariance matrix, block size is 20 ($\Sigma_1 = 2 * \Sigma_2$). }
	\end{figure} \label{Block20_NonEqual}

	\begin{figure} [!htbp]
			\includegraphics[width=15cm,height=16cm,keepaspectratio]{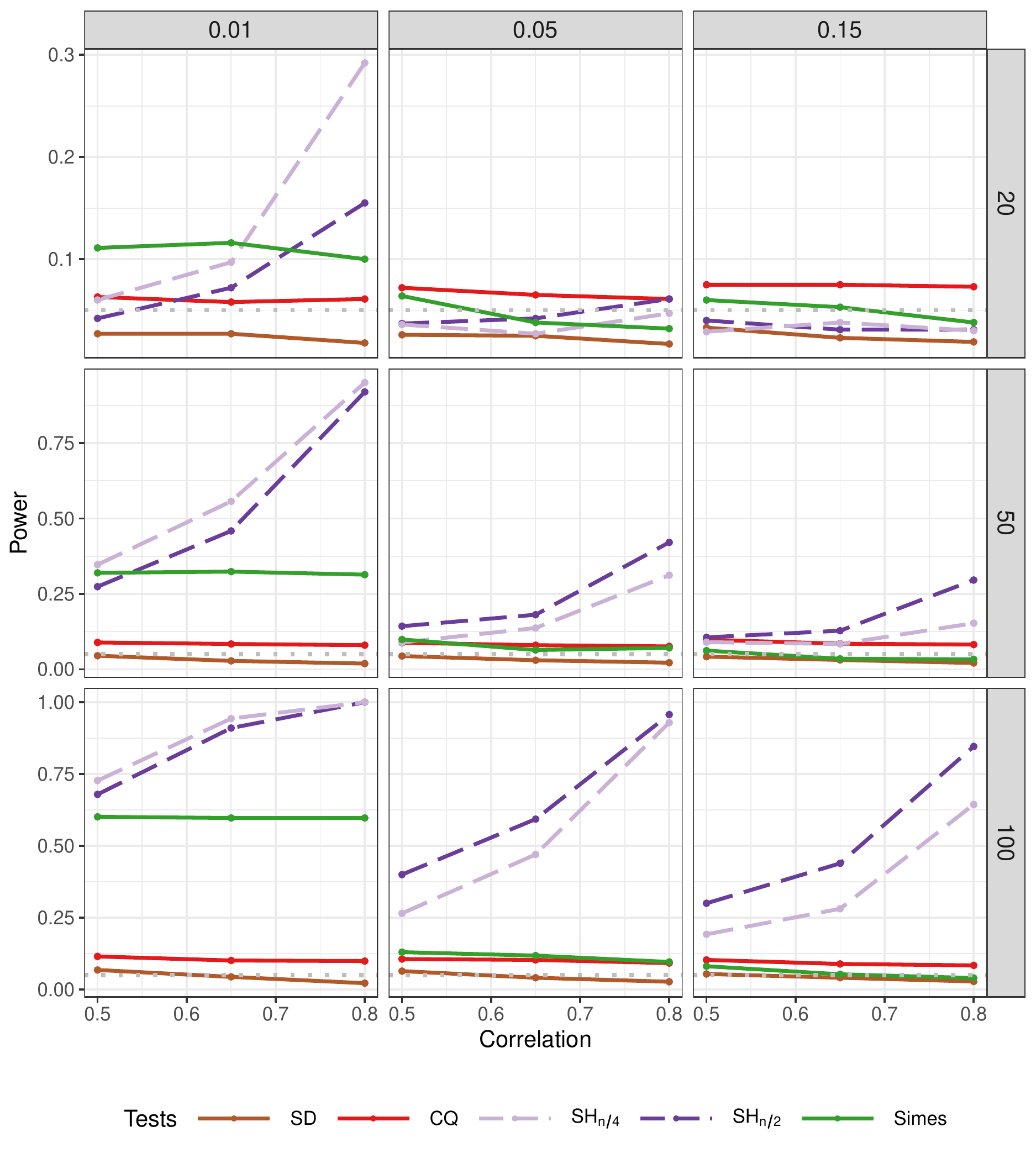}
			\caption{Non-equal covariance tests power comparison, \textit{block} covariance matrix, block size is 50 ($\Sigma_1 = 2 * \Sigma_2$). }
	\end{figure} \label{Block50_NonEqual}

	\begin{figure} [!htbp]
			\includegraphics[width=15cm,height=16cm,keepaspectratio]{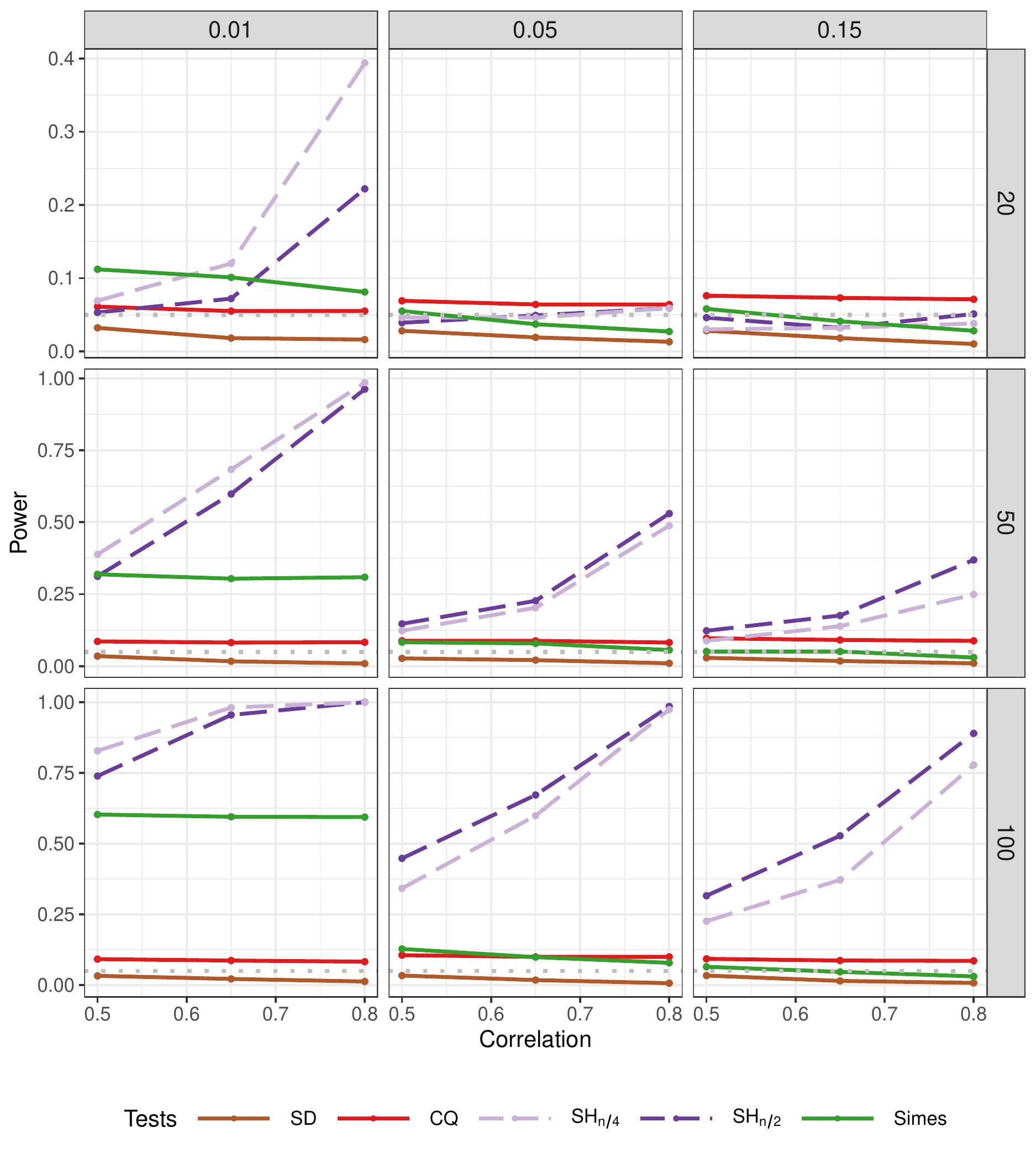}
			\caption{Non-equal covariance tests power comparison, \textit{block} covariance matrix, block size is 100 ($\Sigma_1 = 2 * \Sigma_2$).}
	\end{figure} \label{Block100_NonEqual}

	\begin{figure} [!htbp]
			\includegraphics[width=15cm,height=16cm,keepaspectratio]{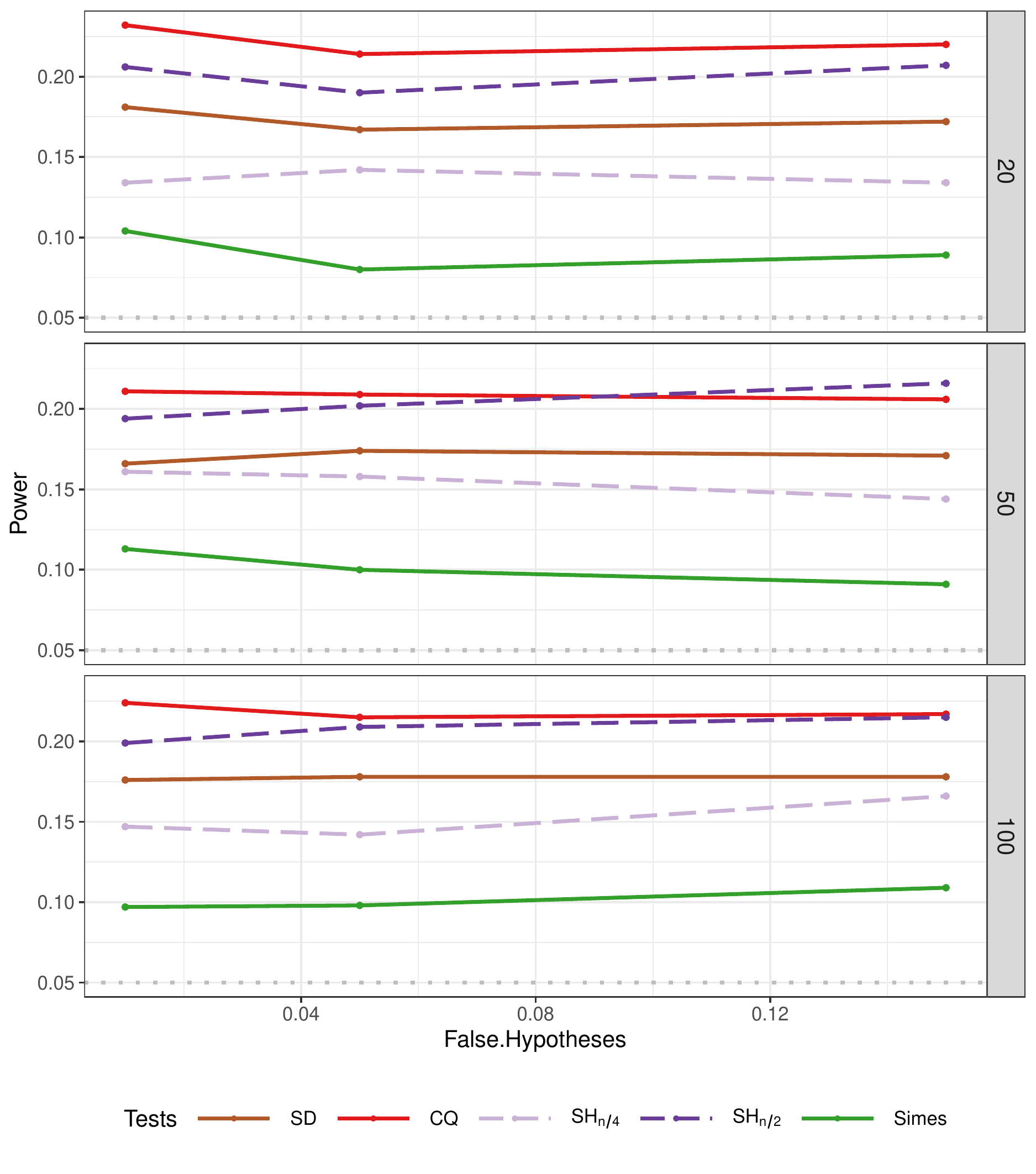}
			\caption{Non-equal covariance tests power comparison, \textit{model 7} covariance matrix. }
	\end{figure} \label{CaiMode7_NonEqual} 
	
	\section{Non-parametric tests}
	
	In the following section we will investigate the performance of the non-parametric version of the suggested test \ref{sec:NonGauss}. We begin by examining a few non-parametric alternative tests.  
	
	\subsection{Comparing non-parametric tests} 
	
	\subsubsection{KNN (1988)} 
	
	\cite{henze1988multivariate} studied the distribution of the $k$ nearest neighbors, the test statistics looks at the sum of the $k$ nearest neighbors for all $n$ observations. It is possible to use different types of distance in order to determine the nearest neighbors, we used euclidean distance. The test statistic is the following:
	
	\begin{equation}
	KN_{n,k} = \sum\limits_{i=1}^n \sum\limits_{r=1}^k I_i(r),
	\end{equation}
	
	$I_i(r) = 1$ if the $r$ nearest neighbor of observation $i$ is from the same group as observation $i$ and zero otherwise. It was proven that the test statistics is asymptoticly (in $n$) normally distributed and thus does not require permutations.

	\subsubsection{Thulin~ (2014)} \label{Thulin}
	\cite{thulin2014high} suggested a test similar to the proposed test, sampling subsets of dimensions and computing the $T^2$ Hotelling test statistic for each subset and averaging them to obtain the test statistic. Since the distribution of the test statistic is not known, it is evaluated using a permutation test. The procedure is as following, sample $m$ random Hotelling statistics and average them to obtain $T^{2^{(i)}}$, repeating the process $B$ iterations (while randomizing the observations label vector) to approximate the null distribution.

	\subsubsection{Shen \& Lin (2015)} \label{Shen}
	
	Shen \& Lin method is to try and find a subset of variables which would result in the highest statistic value \cite{shen2015adaptive}. 
	The algorithm to find the maximum is the following: 
	\begin{enumerate}
		\item Start with an empty active set of $A=\emptyset$. 
		\item Find the dimension $j^*$ for which the t-statistic is maximal.  Denote $T^2_1 = max_{j\in l} \frac{(\bar{X}_{1,j} - \bar{X}_{2,j})^2}{\hat{\sigma}_{jj}}$. Add $j^*$ to $A$. 
		\item For $k=2,3,...,r$. Find one remaining variable with the maximum asymptotic power, when included with the already-selected subset A. Let $j^* = arg max_{j \notin A}  \sqrt{\frac{n-1}{2}}  ||\Delta_{A \cup j}||^2 $, and $T^2_k = max_{j \notin A}  \sqrt{\frac{n-k}{2k}} ||\Delta_{A \cup j}||^2$. 
		Further, add $j^*$ to $A$.
		\item Calculate $T^* = max_{1\leq k \leq r} T^2_k$. 
	\end{enumerate}
	
	$r$ is suggested to be to be $\frac{n}{2}$. At each permutation shuffle the samples label and repeat the process (step 1-4) to calculate the maximal Hotelling statistic for that permutation. Comparing the original Hotelling statistics to the permutation yields the the test p-value. 
	
	\subsubsection{Zhang \& Pan (2016)} \label{Zhang}
	
	Zhang \& Pan suggested to use hierarchical clustering \cite{zhang2016high}, and for each of the cluster to compute the Hotelling $T^2$ statistic, summing the resulting test statistics. 
	
	The algorithm for obtaining the test statistic is the following:  
	\begin{enumerate}
		\item Set parameters $d_c,k_c$. 
		\item Hierarchical top-down clustering is performed (Pearson correlation coefficient and average linkage are used). Clusters are calculated based on cutoff distance $d_c$; A cluster is split into two clusters as long as it has more than $k_c$ dimensions.
		\item Calculate the Hotelling $T^2$ statistic for each cluster. 
		\item Sum the Hotelling $T^2$ statistic.
	\end{enumerate}
	
	In their paper, $k_c$ was suggested to be $\frac{2}{3} * (n_1 + n_2)$ and $d_c$ is based on the standard error of fisher transformation for the correlation coefficient. 
	After obtaining the initial hierarchical it is repeated across permutation so there is no need to repeat the process of obtaining the clusters at each permutation.
	The p-value is computed using the permutations distribution.

	The tests we chose to describe above represent the state of the art procedures (as we understand it) for the problem of testing the equality of the means of two Gaussian vectors, enriched by some non parametric tests. While, we compare several non-parametric tests this is by no mean an exhaustive comparison of all known non-parametric methods.

	\subsection{Simulation - non-parametric tests} \label{nonPara} 
	
	A simulation study is conducted to assess the power of the different non-parametric tests, as well as testing how sensitive the parametric tests to the Gaussianity assumption are. This simulation study is somewhat smaller (since the permutation tests are computationally expensive). In order to do so, we used the covariance structures, \textit{AR} and \textit{block covariance} out of the four covariance structures  considered for the parametric simulation. The number of observations is $n_1 = n_2 = 20,50$ and the sparsity parameter $1-\beta = 0.01,0.05,0.15$.  
	In this simulation we will evaluate the parametric tests, SH, SD, CQ and Lopes. Non-parametric versions of the Simes,SH and KNN. As well as non-parametric tests (permutation based), Zhang \& Pan \ref{Zhang}, Shen \& Lin \ref{Shen} and Thulin \ref{Thulin}. 
	First we check the performance in the Gaussian case and then we added to each dimensions, marginally, Laplace or Exponential noise. It is important to note that after adding the noise the variance of each dimensions $\sigma_{ii}$ increases to 2, weakening the correlation between every two dimensions dimensions. 
	Due to the increased variance we adjust  $||\vect{\mu}_1 - \vect{\mu}_2||_2^2 = 5 \sqrt{\frac{40}{n_1 +  n_2}}$.
	Each scenario was repeated 1000 times and the number of permutation for each test was 250. The conservative standard deviation remains $0.015$.
	
	\begin{table}
		\centering
		\begin{tabular}{|c|c|}
			\hline 
			Abbreviation & Tests  \\
			\hline 
			$SH$   & Simes Hotelling \\ 
			$CLX$  & Cai et al. \\
			$GCT$  & Gregory et al. \\
			$PSH$  & Permutation Simes Hotelling \\
			$SL$   & Shen \& Lin \\ 
			$ZP$   & Zhang \& Pan \\ 
			$SD$   & Srivastava \& Du \\ 
			$CQ$   & Chen \& Qin \\
			$KNN$  & K-Nearest Neighbors\\ 
			\hline
		\end{tabular} 
		\caption{Abbreviations used in parametric and non parametric tests}  
	\end{table}

	\subsubsection{Type I error comparison}
	
	We only present the type I error table for the first \textit{AR} covariance structure, the type I error rates in the \textit{block covariance} case share similar behavior and can be found in the supplementary material. 
	
	\spacingset{1} 
	\begin{table}[htbp]
		\centering
		\resizebox{\textwidth}{!}{\begin{tabular}{|c|c|c||c|c|c|c|c|c|c|c|c|c|c|}
				\hline
				Type  & $n_1 = n_2$  & \multicolumn{1}{l||}{$\rho$} & \multicolumn{1}{l|}{Simes} & \multicolumn{1}{l|}{SD} & \multicolumn{1}{l|}{CQ} & \multicolumn{1}{l|}{Lopes} & \multicolumn{1}{l|}{SH$_{\frac{n}{2}}$} & \multicolumn{1}{l|}{PSH$_{\frac{n}{4}}$} & \multicolumn{1}{l|}{PSH$_{\frac{n}{2}}$} & \multicolumn{1}{l|}{Thulin} & \multicolumn{1}{l|}{SL} & \multicolumn{1}{l|}{ZP} & \multicolumn{1}{l|}{KNN} \\
				\hline
				\multirow{10}[1]{*}{Exponential} & \multirow{5}[0]{*}{20} & 0.3   & 0.042 & 0.037 & 0.059 & 0.05  & 0.054 & 0.035 & 0.042 & 0.044 & 0.051 & 0.712 & 0.036 \\
				&       & 0.45  & 0.041 & 0.037 & 0.048 & 0.044 & 0.038 & 0.034 & 0.044 & 0.05  & 0.035 & 0.666 & 0.045 \\
				&       & 0.6   & 0.048 & 0.041 & 0.051 & 0.049 & 0.048 & 0.049 & 0.047 & 0.052 & 0.046 & 0.64  & 0.036 \\
				&       & 0.75  & 0.039 & 0.035 & 0.053 & 0.052 & 0.045 & 0.04  & 0.04  & 0.043 & 0.052 & 0.483 & 0.046 \\
				&       & 0.95  & 0.036 & 0.038 & 0.062 & 0.045 & 0.029 & 0.04  & 0.034 & 0.036 & 0.055 & 0.102 & 0.041 \\
				& \multirow{5}[1]{*}{50} & 0.3   & 0.041 & 0.041 & 0.046 & 0.049 & 0.046 & 0.041 & 0.041 & 0.042 & 0.05  & 0.234 & 0.051 \\
				&       & 0.45  & 0.029 & 0.04  & 0.052 & 0.056 & 0.048 & 0.049 & 0.043 & 0.047 & 0.038 & 0.213 & 0.049 \\
				&       & 0.6   & 0.033 & 0.039 & 0.057 & 0.048 & 0.049 & 0.043 & 0.04  & 0.036 & 0.045 & 0.138 & 0.059 \\
				&       & 0.75  & 0.036 & 0.049 & 0.071 & 0.051 & 0.047 & 0.051 & 0.045 & 0.048 & 0.049 & 0.077 & 0.049 \\
				&       & 0.95  & 0.04  & 0.058 & 0.07  & 0.052 & 0.036 & 0.048 & 0.046 & 0.049 & 0.038 & 0.048 & 0.046 \\
				\hline
				\multirow{10}[2]{*}{Laplace} & \multirow{5}[1]{*}{20} & 0.3   & 0.04  & 0.036 & 0.052 & 0.04  & 0.037 & 0.041 & 0.038 & 0.042 & 0.039 & 0.702 & 0.03 \\
				&       & 0.45  & 0.037 & 0.046 & 0.058 & 0.04  & 0.039 & 0.045 & 0.047 & 0.047 & 0.046 & 0.698 & 0.035 \\
				&       & 0.6   & 0.047 & 0.043 & 0.053 & 0.047 & 0.041 & 0.046 & 0.039 & 0.048 & 0.039 & 0.624 & 0.048 \\
				&       & 0.75  & 0.054 & 0.034 & 0.045 & 0.048 & 0.044 & 0.048 & 0.043 & 0.039 & 0.044 & 0.464 & 0.038 \\
				&       & 0.95  & 0.042 & 0.037 & 0.065 & 0.05  & 0.039 & 0.046 & 0.041 & 0.045 & 0.038 & 0.091 & 0.042 \\
				& \multirow{5}[1]{*}{50} & 0.3   & 0.028 & 0.042 & 0.05  & 0.052 & 0.043 & 0.042 & 0.037 & 0.038 & 0.039 & 0.227 & 0.049 \\
				&       & 0.45  & 0.04  & 0.054 & 0.064 & 0.055 & 0.049 & 0.054 & 0.054 & 0.058 & 0.062 & 0.206 & 0.043 \\
				&       & 0.6   & 0.041 & 0.051 & 0.059 & 0.052 & 0.046 & 0.047 & 0.049 & 0.052 & 0.043 & 0.141 & 0.045 \\
				&       & 0.75  & 0.03  & 0.049 & 0.056 & 0.053 & 0.046 & 0.044 & 0.055 & 0.048 & 0.051 & 0.085 & 0.046 \\
				&       & 0.95  & 0.044 & 0.06  & 0.073 & 0.041 & 0.066 & 0.059 & 0.052 & 0.055 & 0.054 & 0.064 & 0.046 \\
				\hline
				\multirow{10}[2]{*}{Gaussian} & \multirow{5}[1]{*}{20} & 0.3   & 0.047 & 0.044 & 0.059 & 0.046 & 0.061 & 0.047 & 0.043 & 0.049 & 0.038 & 0.633 & 0.031 \\
				&       & 0.45  & 0.042 & 0.042 & 0.06  & 0.053 & 0.05  & 0.038 & 0.037 & 0.04  & 0.054 & 0.43  & 0.036 \\
				&       & 0.6   & 0.048 & 0.043 & 0.052 & 0.044 & 0.062 & 0.04  & 0.044 & 0.045 & 0.051 & 0.214 & 0.037 \\
				&       & 0.75  & 0.042 & 0.036 & 0.056 & 0.032 & 0.05  & 0.04  & 0.047 & 0.045 & 0.041 & 0.099 & 0.034 \\
				&       & 0.95  & 0.024 & 0.028 & 0.074 & 0.047 & 0.036 & 0.041 & 0.045 & 0.049 & 0.036 & 0.039 & 0.048 \\
				& \multirow{5}[1]{*}{50} & 0.3   & 0.038 & 0.039 & 0.046 & 0.034 & 0.044 & 0.054 & 0.043 & 0.044 & 0.043 & 0.177 & 0.045 \\
				&       & 0.45  & 0.034 & 0.043 & 0.05  & 0.044 & 0.044 & 0.04  & 0.039 & 0.035 & 0.047 & 0.097 & 0.048 \\
				&       & 0.6   & 0.031 & 0.045 & 0.057 & 0.05  & 0.041 & 0.042 & 0.044 & 0.041 & 0.041 & 0.088 & 0.061 \\
				&       & 0.75  & 0.045 & 0.046 & 0.063 & 0.049 & 0.051 & 0.052 & 0.041 & 0.049 & 0.049 & 0.057 & 0.043 \\
				&       & 0.95  & 0.023 & 0.032 & 0.075 & 0.049 & 0.039 & 0.057 & 0.049 & 0.051 & 0.051 & 0.042 & 0.047 \\
				\hline
		\end{tabular}}%
		\caption{Type I error rates in the \textit{AR} covariance matrix scenario for each of the tests (parametric and non-parametric tests). $\rho$ is the correlation parameter.}
		\label{alphaNonparaAR}%
	\end{table}%
	
	\spacingset{1.45}
	
	All tests except for ZP, keep the nominal level across all scenarios. This is surprising considering that SH,SD,CQ and Lopes make use of the assumption of Gaussianty of the data. It seems that all of the parametric tests are robust to a violation of the Gaussian assumption. 
	ZP computes the cluster once, on the original data and does not repeat the computation at each permutation (it relies on the fact the covariance structure should not change between permutation).
	It seems that when the covariance between dimensions is close to independence, the clusters are over fitted on the original data leading to increased type I error. 
	As seen in Table \ref{alphaNonparaAR} as the correlation increase the test achieve better error rates, until converging to the required $\alpha = 0.05$ level. Due to these results the test is omitted from the rest of the simulations.

	\subsubsection{Power comparison}
	\begin{figure} 
		\includegraphics[width=15cm,height=16cm,keepaspectratio]{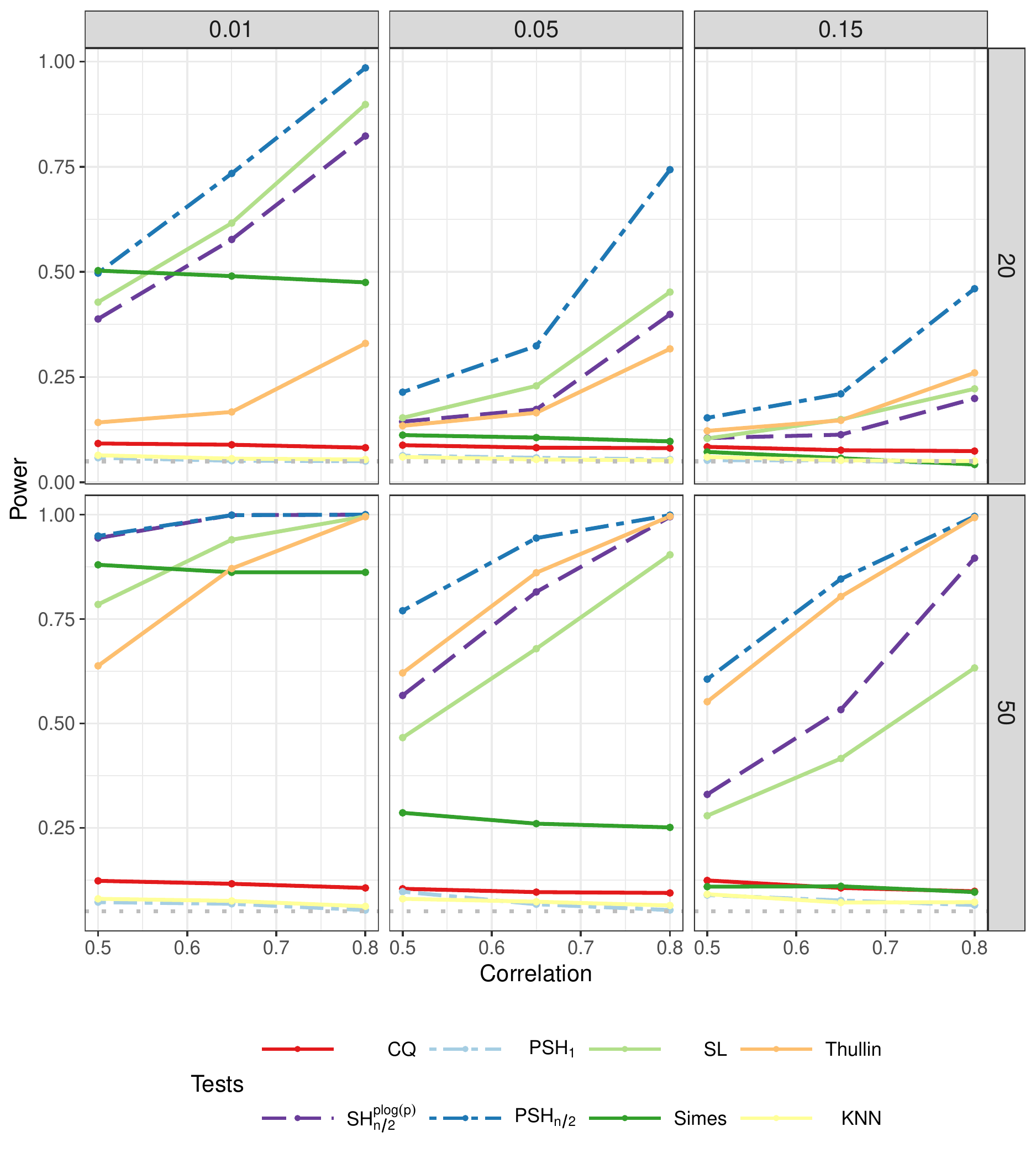}
		\caption{Non-parametric tests power comparison, \textit{block} covariance matrix, block size is 20, Laplace noise was added to each dimension independently.}
		\label{nonParaBlock25}
	\end{figure}

	Overall, throughout the non-parametric simulation we can see that permutation tests have superior power when comparing them to the parametric tests, there are also vast differences between the different p-value combination functions (Thulin vs PSH). The $PSH_{n/2}$ appears to be powerful across almost all configurations, being beaten by Simes when the signal is sparse ($1-\beta = 0.01$) and by CQ only in the \textit{block} covariance matrix, block size 10, $n_1=n_2=20$ and $1-\beta =0.15$. Almost everywhere else PSH is the most powerful test. 
	In Figure \ref{nonParaBlock25} we can see that tests which are based on Mahalnobis distances (SL, Thulin, Lopes, SH and PSH) power increases as the correlation parameter becomes larger. Thulin is more sensitive to the number of observations, as evident with the increase of the power curves between the two rows. Thulin seems to be robust to changes in sparsity (almost no change in power curves between columns). 
	SL, SH and PSH are effected negatively by the sparsity parameter, with a larger decrease in the power curve of SL than in that of PSH.  
	The Simes test is powerful in the scenarios where the signal is sparse ($1- \beta$ is 0.01), which is also the scenario where PSH obtains its maximum power. The CQ test does not perform so well even while considering the fact that the covariance structure is weaker than in the previous considered simulations. It seems that the effect of the larger marginal variance harms its power more than it does to the SH test. 
	
	\subsubsection{Min - min power comparison (A-la Princeton)}
	
	\begin{table}[!htbp]
		\centering
		\begin{tabular}{|r||r|r|r|H r|r|r|r|r|r|r|r|r|r|r|r|}
				\hline
				$n_1 = n_2$ & Simes & SD & CQ & Lopes & SH$_{\frac{n}{2}}$ & PSH$_{\frac{n}{4}}$ & PSH$_{\frac{n}{2}}$ & Thulin & SL & KNN \\
				\hline
				20    & 0.091 & 0.020 & 0.083 & 0.084 & 0.356 & 0.422 & 0.563 & 0.191 & 0.310 & 0.054 \\ 
				50    & 0.088 & 0.025 & 0.094 & 0.115 & 0.441 & 0.703 & 0.715 & 0.396 & 0.343 & 0.062 \\
				\hline
		\end{tabular}
		\caption{A-la Princeton simulation study results for the non-parametric and selected parametric tests, finite sample min-min results.}
	\end{table}%
	
	It seems that that PSH performs the best out of all the tests, with SH being a close second followed by Thulin and SL. That is it appears that the dimensions sampling methods perform well, it is interesting that SH while not being the best, is competitive as it takes substantially less time compute it (allowing to use it for larger data-sets). 
	The KNN test does not work well in any scenario, but this could be a result of the distance metric chosen (Euclidean), a different distance metric might have been better suited for the task leading to an increase power for the test. 
	In general, the permutation tests have superior power to that of the parametric tests (Thulin, PSH and SL compared to Lopes and CQ). Among the non-parametric tests those who are based on sampling of dimensions perform better than those who do not (Thulin and PSH compared to SL and KNN). 
	
	
	\subsection{Remaining simulation results} 
	
	Here can be found the the rest of the simulation results for the non-parametric tests. 
	
	\subsubsection{Type I error} 
	
	\spacingset{1}
	\begin{table}[htbp]
		\centering
		\resizebox{\textwidth}{0.45\textheight}{ \begin{tabular}{|c|c|c|c||r|r|r|r|r|r|r|r|r|r|r|}
				\hline
				\multicolumn{1}{|c|}{Type} & \multicolumn{1}{c|}{$n_1 = n_2$} & \multicolumn{1}{c|}{Block Size} & \multicolumn{1}{c||}{$\rho$}   &  \multicolumn{1}{l|}{Simes} & \multicolumn{1}{l|}{SD} & \multicolumn{1}{l|}{CQ} & \multicolumn{1}{l|}{Lopes} & \multicolumn{1}{l|}{SH$_{\frac{n}{2}}$} & \multicolumn{1}{l|}{PSH$_{\frac{n}{4}}$} & \multicolumn{1}{l|}{PSH$_{\frac{n}{2}}$} & \multicolumn{1}{l|}{Thulin} & \multicolumn{1}{l|}{SL} & \multicolumn{1}{l|}{ZP} & \multicolumn{1}{l|}{KNN} \\
				\hline
				\multirow{18}[2]{*}{Exponential} & \multirow{9}[1]{*}{20} & \multirow{3}[1]{*}{10} & 0.5   & 0.037 & 0.048 & 0.057 & 0.037 & 0.046 & 0.037 & 0.044 & 0.046 & 0.037 & 0.52  & 0.046 \\
				&       &       & 0.65  & 0.039 & 0.062 & 0.055 & 0.047 & 0.054 & 0.049 & 0.053 & 0.044 & 0.045 & 0.361 & 0.043 \\
				&       &       & 0.8   & 0.034 & 0.054 & 0.041 & 0.041 & 0.036 & 0.032 & 0.035 & 0.037 & 0.039 & 0.179 & 0.041 \\
				&       & \multirow{3}[0]{*}{25} & 0.5   & 0.037 & 0.058 & 0.06  & 0.048 & 0.051 & 0.052 & 0.047 & 0.049 & 0.05  & 0.276 & 0.034 \\
				&       &       & 0.65  & 0.035 & 0.064 & 0.045 & 0.043 & 0.038 & 0.031 & 0.032 & 0.038 & 0.047 & 0.105 & 0.042 \\
				&       &       & 0.8   & 0.036 & 0.057 & 0.048 & 0.039 & 0.043 & 0.04  & 0.045 & 0.049 & 0.05  & 0.061 & 0.032 \\
				&       & \multirow{3}[0]{*}{50} & 0.5   & 0.036 & 0.045 & 0.033 & 0.04  & 0.045 & 0.032 & 0.029 & 0.038 & 0.04  & 0.138 & 0.039 \\
				&       &       & 0.65  & 0.033 & 0.052 & 0.043 & 0.035 & 0.049 & 0.043 & 0.042 & 0.046 & 0.044 & 0.098 & 0.043 \\
				&       &       & 0.8   & 0.03  & 0.056 & 0.045 & 0.034 & 0.024 & 0.031 & 0.039 & 0.033 & 0.046 & 0.083 & 0.034 \\
				& \multirow{9}[1]{*}{50} & \multirow{3}[0]{*}{10} & 0.5   & 0.05  & 0.056 & 0.055 & 0.042 & 0.035 & 0.041 & 0.038 & 0.039 & 0.048 & 0.089 & 0.046 \\
				&       &       & 0.65  & 0.044 & 0.059 & 0.05  & 0.049 & 0.032 & 0.045 & 0.044 & 0.045 & 0.043 & 0.067 & 0.037 \\
				&       &       & 0.8   & 0.049 & 0.068 & 0.059 & 0.039 & 0.039 & 0.055 & 0.046 & 0.046 & 0.054 & 0.052 & 0.053 \\
				&       & \multirow{3}[0]{*}{25} & 0.5   & 0.055 & 0.067 & 0.057 & 0.049 & 0.029 & 0.039 & 0.044 & 0.049 & 0.053 & 0.049 & 0.045 \\
				&       &       & 0.65  & 0.057 & 0.074 & 0.053 & 0.051 & 0.043 & 0.057 & 0.048 & 0.051 & 0.034 & 0.051 & 0.045 \\
				&       &       & 0.8   & 0.049 & 0.071 & 0.053 & 0.042 & 0.039 & 0.049 & 0.05  & 0.05  & 0.045 & 0.045 & 0.05 \\
				&       & \multirow{3}[1]{*}{50} & 0.5   & 0.055 & 0.069 & 0.043 & 0.046 & 0.035 & 0.043 & 0.04  & 0.042 & 0.038 & 0.052 & 0.041 \\
				&       &       & 0.65  & 0.054 & 0.083 & 0.05  & 0.037 & 0.029 & 0.056 & 0.052 & 0.061 & 0.05  & 0.056 & 0.031 \\
				&       &       & 0.8   & 0.052 & 0.081 & 0.054 & 0.042 & 0.041 & 0.057 & 0.046 & 0.048 & 0.052 & 0.053 & 0.046 \\
				\hline
				\multirow{18}[1]{*}{Laplace} & \multirow{9}[1]{*}{20} & \multirow{3}[1]{*}{10} & 0.5   & 0.037 & 0.052 & 0.042 & 0.037 & 0.021 & 0.039 & 0.037 & 0.048 & 0.049 & 0.516 & 0.036 \\
				&       &       & 0.65  & 0.039 & 0.058 & 0.043 & 0.04  & 0.043 & 0.046 & 0.037 & 0.042 & 0.043 & 0.37  & 0.053 \\
				&       &       & 0.8   & 0.04  & 0.061 & 0.056 & 0.042 & 0.041 & 0.045 & 0.048 & 0.052 & 0.049 & 0.199 & 0.048 \\
				&       & \multirow{3}[0]{*}{25} & 0.5   & 0.039 & 0.056 & 0.036 & 0.046 & 0.041 & 0.04  & 0.04  & 0.042 & 0.044 & 0.289 & 0.042 \\
				&       &       & 0.65  & 0.048 & 0.066 & 0.052 & 0.041 & 0.037 & 0.049 & 0.045 & 0.05  & 0.043 & 0.112 & 0.037 \\
				&       &       & 0.8   & 0.04  & 0.064 & 0.051 & 0.035 & 0.041 & 0.042 & 0.041 & 0.04  & 0.045 & 0.072 & 0.042 \\
				&       & \multirow{3}[0]{*}{50} & 0.5   & 0.038 & 0.051 & 0.048 & 0.023 & 0.032 & 0.034 & 0.035 & 0.039 & 0.053 & 0.155 & 0.039 \\
				&       &       & 0.65  & 0.033 & 0.052 & 0.046 & 0.046 & 0.048 & 0.046 & 0.043 & 0.037 & 0.039 & 0.086 & 0.048 \\
				&       &       & 0.8   & 0.032 & 0.065 & 0.042 & 0.037 & 0.046 & 0.042 & 0.04  & 0.04  & 0.059 & 0.115 & 0.041 \\
				& \multirow{9}[0]{*}{50} & \multirow{3}[0]{*}{10} & 0.5   & 0.05  & 0.057 & 0.039 & 0.051 & 0.04  & 0.05  & 0.059 & 0.054 & 0.054 & 0.094 & 0.048 \\
				&       &       & 0.65  & 0.054 & 0.067 & 0.054 & 0.051 & 0.045 & 0.051 & 0.043 & 0.044 & 0.06  & 0.064 & 0.062 \\
				&       &       & 0.8   & 0.047 & 0.061 & 0.06  & 0.047 & 0.037 & 0.052 & 0.05  & 0.049 & 0.059 & 0.047 & 0.053 \\
				&       & \multirow{3}[0]{*}{25} & 0.5   & 0.055 & 0.065 & 0.051 & 0.048 & 0.028 & 0.047 & 0.05  & 0.05  & 0.052 & 0.051 & 0.048 \\
				&       &       & 0.65  & 0.047 & 0.065 & 0.055 & 0.058 & 0.028 & 0.05  & 0.049 & 0.051 & 0.047 & 0.053 & 0.042 \\
				&       &       & 0.8   & 0.05  & 0.067 & 0.052 & 0.048 & 0.038 & 0.048 & 0.048 & 0.051 & 0.059 & 0.054 & 0.05 \\
				&       & \multirow{3}[0]{*}{50} & 0.5   & 0.06  & 0.078 & 0.057 & 0.046 & 0.043 & 0.046 & 0.052 & 0.054 & 0.051 & 0.045 & 0.052 \\
				&       &       & 0.65  & 0.052 & 0.072 & 0.042 & 0.04  & 0.039 & 0.055 & 0.06  & 0.054 & 0.057 & 0.049 & 0.038 \\
				&       &       & 0.8   & 0.049 & 0.079 & 0.053 & 0.049 & 0.031 & 0.051 & 0.056 & 0.057 & 0.038 & 0.053 & 0.041 \\
				\hline
				\multirow{18}[1]{*}{Gaussian} & \multirow{9}[0]{*}{20} & \multirow{3}[0]{*}{10} & 0.5   & 0.04  & 0.054 & 0.049 & 0.049 & 0.041 & 0.052 & 0.045 & 0.046 & 0.055 & 0.079 & 0.037 \\
				&       &       & 0.65  & 0.035 & 0.059 & 0.041 & 0.038 & 0.034 & 0.04  & 0.038 & 0.042 & 0.041 & 0.054 & 0.046 \\
				&       &       & 0.8   & 0.031 & 0.062 & 0.039 & 0.044 & 0.023 & 0.043 & 0.037 & 0.041 & 0.042 & 0.04  & 0.046 \\
				&       & \multirow{3}[0]{*}{25} & 0.5   & 0.035 & 0.06  & 0.042 & 0.036 & 0.035 & 0.034 & 0.037 & 0.043 & 0.048 & 0.04  & 0.035 \\
				&       &       & 0.65  & 0.034 & 0.064 & 0.054 & 0.016 & 0.036 & 0.04  & 0.048 & 0.048 & 0.053 & 0.036 & 0.032 \\
				&       &       & 0.8   & 0.029 & 0.063 & 0.044 & 0.034 & 0.029 & 0.042 & 0.043 & 0.048 & 0.046 & 0.044 & 0.036 \\
				&       & \multirow{3}[0]{*}{50} & 0.5   & 0.032 & 0.064 & 0.051 & 0.04  & 0.032 & 0.043 & 0.044 & 0.048 & 0.05  & 0.087 & 0.038 \\
				&       &       & 0.65  & 0.023 & 0.066 & 0.042 & 0.05  & 0.031 & 0.046 & 0.045 & 0.046 & 0.053 & 0.096 & 0.045 \\
				&       &       & 0.8   & 0.017 & 0.066 & 0.045 & 0.036 & 0.017 & 0.039 & 0.042 & 0.042 & 0.045 & 0.093 & 0.043 \\
				& \multirow{9}[1]{*}{50} & \multirow{3}[0]{*}{10} & 0.5   & 0.041 & 0.055 & 0.053 & 0.038 & 0.034 & 0.044 & 0.04  & 0.046 & 0.05  & 0.048 & 0.043 \\
				&       &       & 0.65  & 0.037 & 0.057 & 0.044 & 0.043 & 0.042 & 0.048 & 0.055 & 0.049 & 0.046 & 0.048 & 0.043 \\
				&       &       & 0.8   & 0.032 & 0.058 & 0.044 & 0.03  & 0.027 & 0.051 & 0.05  & 0.048 & 0.051 & 0.043 & 0.05 \\
				&       & \multirow{3}[0]{*}{25} & 0.5   & 0.051 & 0.069 & 0.057 & 0.043 & 0.024 & 0.051 & 0.042 & 0.047 & 0.042 & 0.035 & 0.043 \\
				&       &       & 0.65  & 0.043 & 0.076 & 0.057 & 0.039 & 0.026 & 0.05  & 0.046 & 0.051 & 0.049 & 0.041 & 0.041 \\
				&       &       & 0.8   & 0.028 & 0.078 & 0.039 & 0.045 & 0.024 & 0.05  & 0.051 & 0.051 & 0.046 & 0.048 & 0.038 \\
				&       & \multirow{3}[1]{*}{50} & 0.5   & 0.042 & 0.081 & 0.041 & 0.039 & 0.03  & 0.042 & 0.041 & 0.044 & 0.042 & 0.037 & 0.045 \\
				&       &       & 0.65  & 0.026 & 0.083 & 0.044 & 0.04  & 0.025 & 0.056 & 0.047 & 0.052 & 0.048 & 0.039 & 0.047 \\
				&       &       & 0.8   & 0.014 & 0.082 & 0.058 & 0.046 & 0.021 & 0.052 & 0.044 & 0.049 & 0.058 & 0.041 & 0.042 \\
				\hline
			\end{tabular}%
		}%
		\label{tab:BlockNonParametric}%
		\caption{Type I error rates in the \textit{block} covariance matrix scenario for each of the tests (parametric and non-parametric tests). $\rho$ is the correlation parameter.} 
	\end{table}%
	\spacingset{1.45}

	\subsubsection{Power simulations}
	

		\begin{figure} 
			\includegraphics[width=15cm,height=16cm,keepaspectratio]{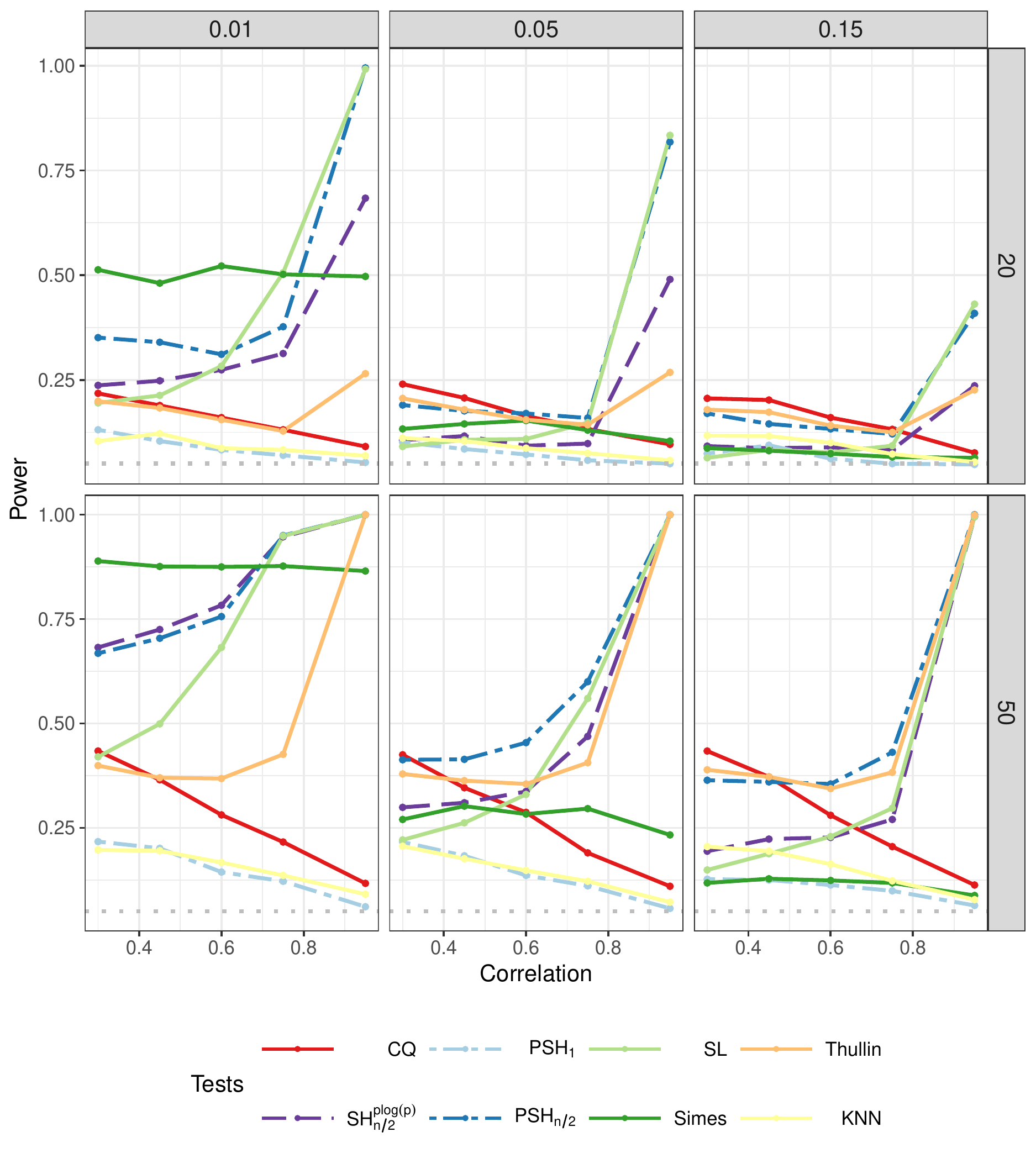}
			\caption{Non-parametric tests power comparison, \textit{AR} covariance matrix, Gaussian noise was added to each dimension independently.}
			\label{AR_1_NonPara_None}
		\end{figure}

		\begin{figure} 
			\includegraphics[width=15cm,height=16cm,keepaspectratio]{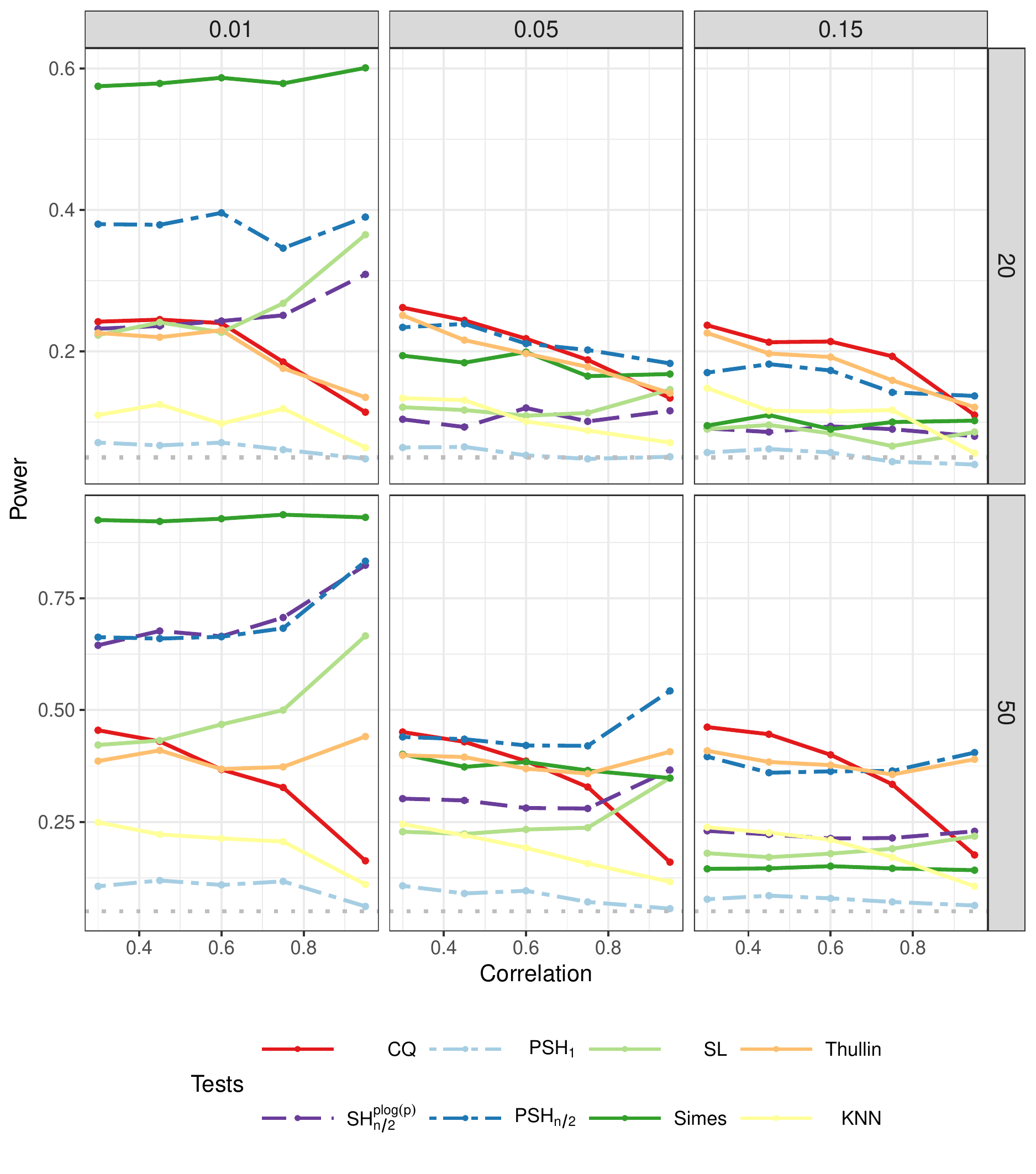}
			\caption{Non-parametric tests power comparison, \textit{AR} covariance matrix, Exponential noise was added to each dimension independently.}
		\end{figure} \label{AR_1_NonPara_Exp}

		\begin{figure} 
			\includegraphics[width=\linewidth,height=16cm,keepaspectratio]{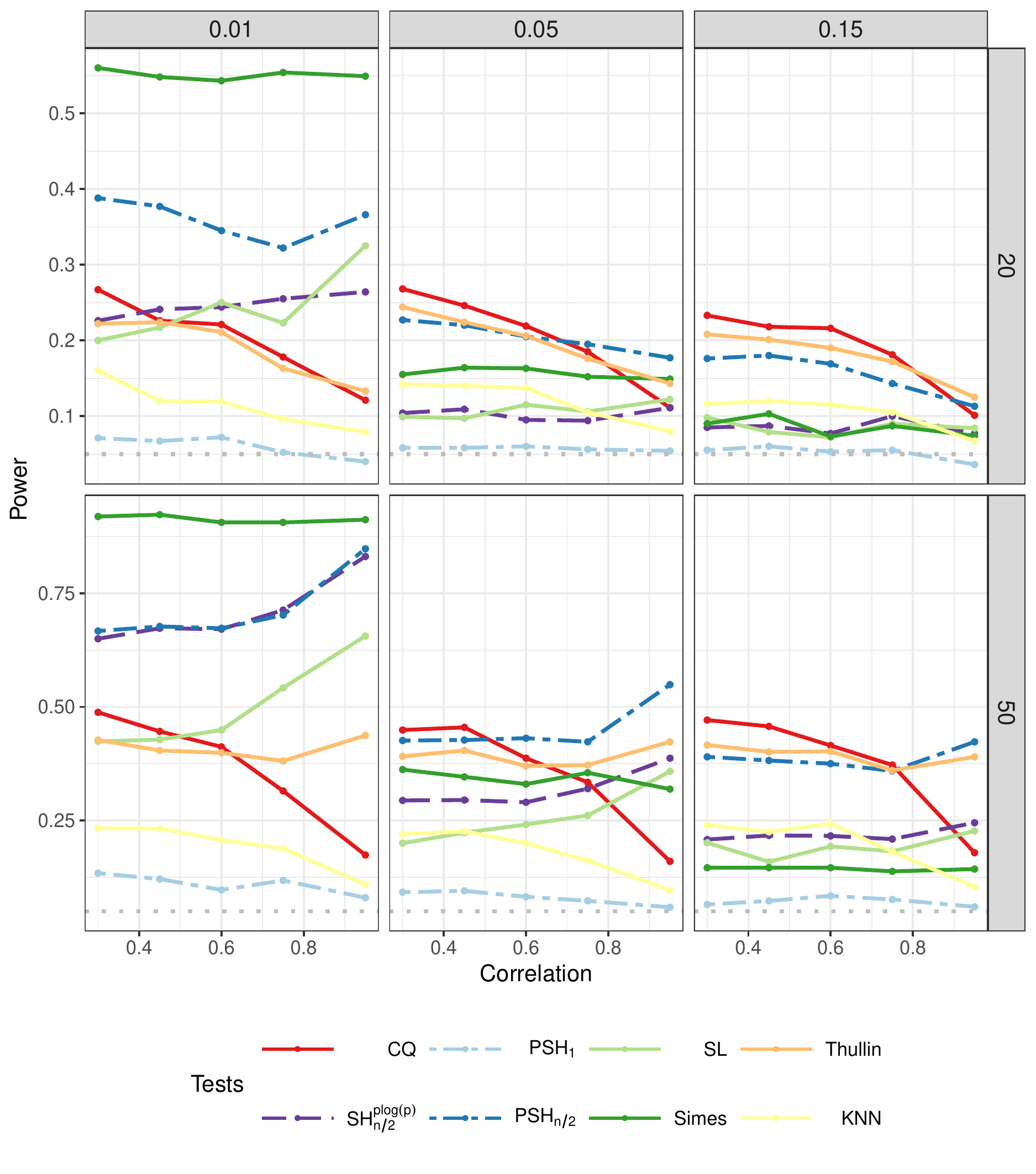}
			\caption{Non-parametric tests power comparison, \textit{AR} covariance matrix, Laplace noise was added to each dimension independently.}
		\end{figure} \label{AR_1_NonPara_Laplace}

		\begin{figure} 
			\includegraphics[width=\linewidth,height=16cm,keepaspectratio]{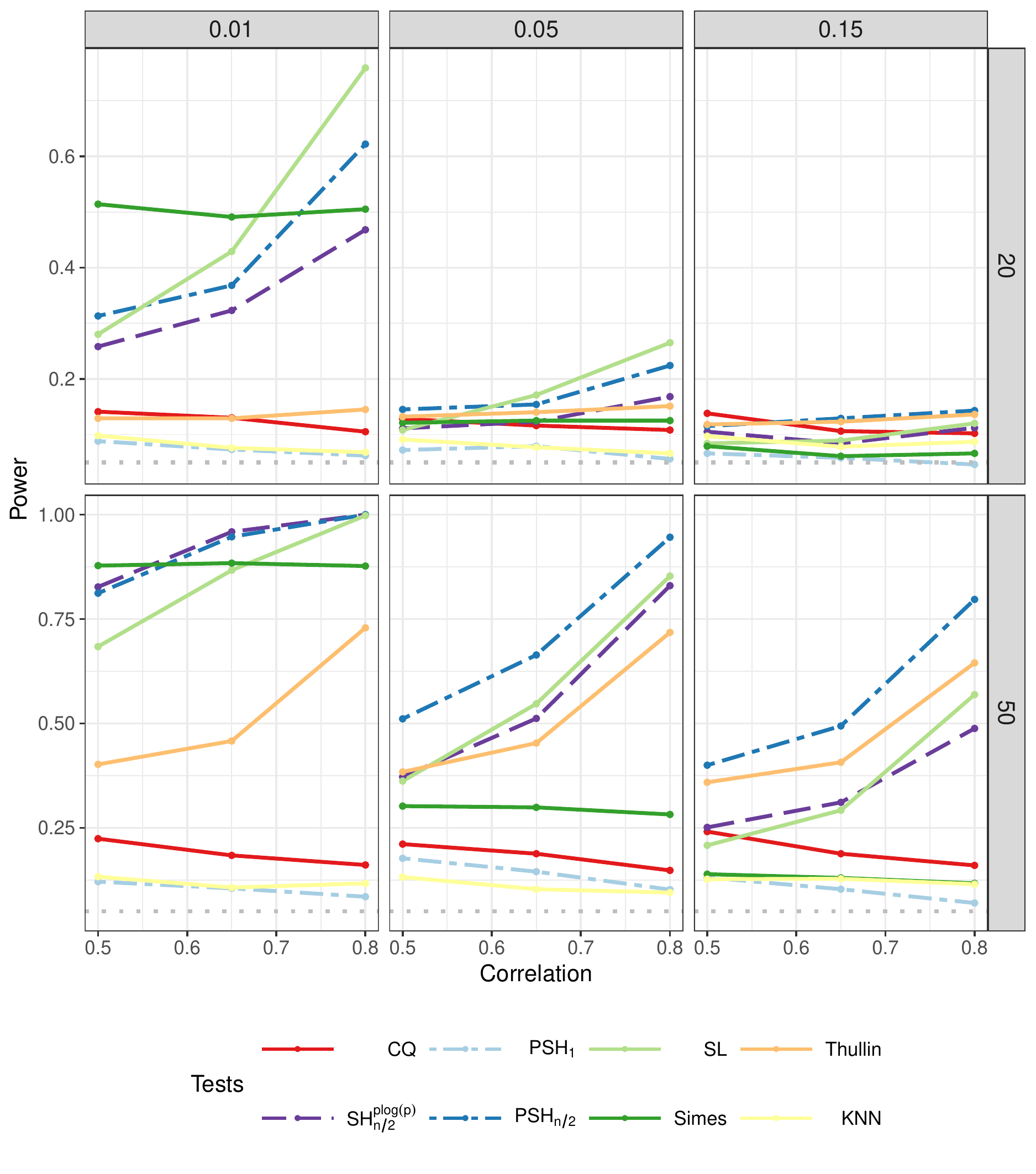}
			\caption{Non-parametric tests power comparison, \textit{block} covariance matrix, block size is 10, Gaussian noise was added to each dimension independently.}
		\end{figure} \label{Block10_NonPara_None}

		\begin{figure} 
			\includegraphics[width=\linewidth,height=16cm,keepaspectratio]{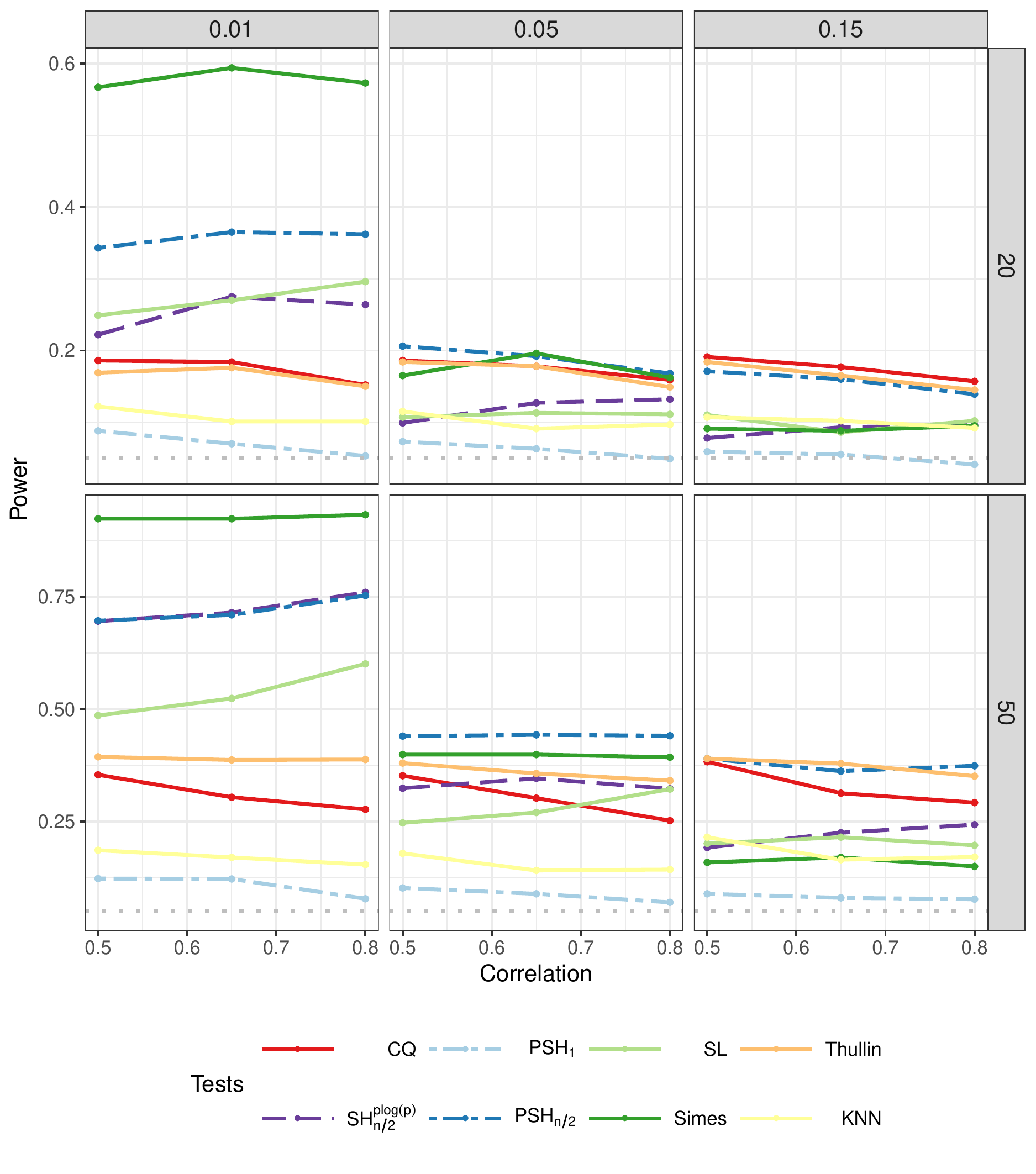}
			\caption{Non-parametric tests power comparison, \textit{block} covariance matrix, block size is 10, Exponential noise was added to each dimension independently.}
		\end{figure} \label{Block10_NonPara_Exp}

		\begin{figure} 
			\includegraphics[width=\linewidth,height=15.5cm,keepaspectratio]{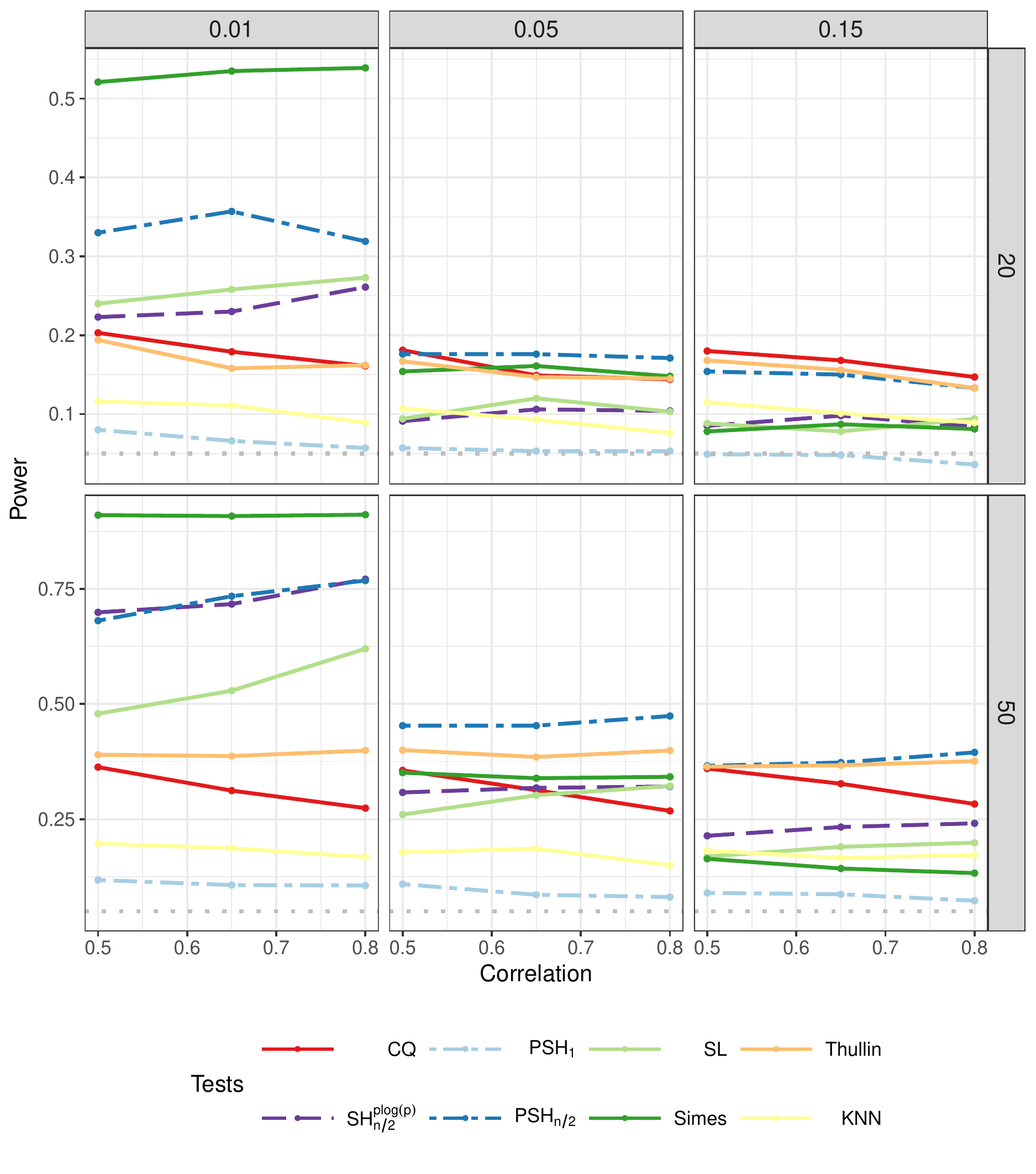}
			\caption{Non-parametric tests power comparison, \textit{block} covariance matrix, block size is 10, Laplace noise was added to each dimension independently.}
		\end{figure} \label{Block10_NonPara_Laplace}

		\begin{figure} 
			\includegraphics[width=\linewidth,height=15.5cm,keepaspectratio]{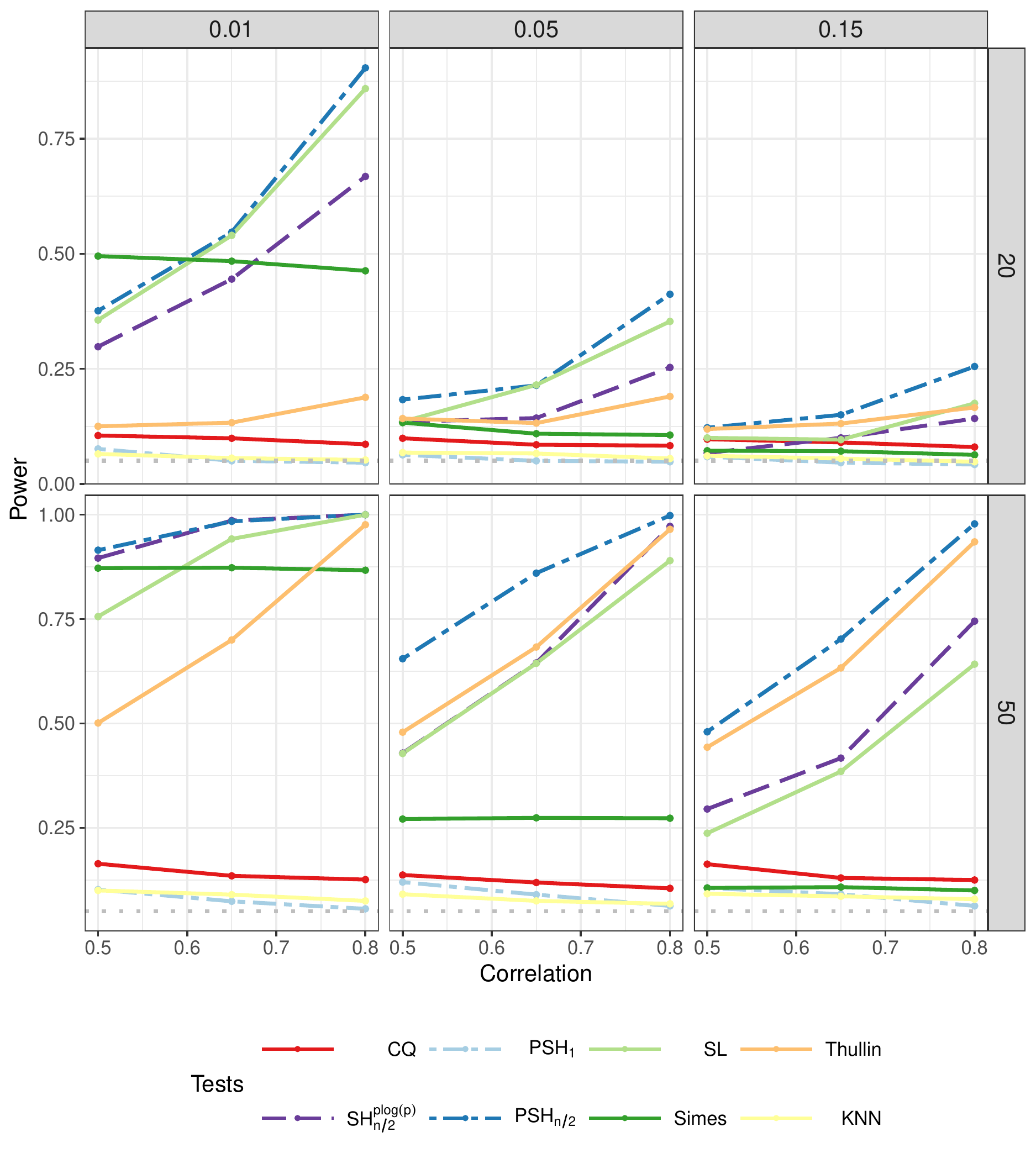}
			\caption{Non-parametric tests power comparison, \textit{block} covariance matrix, block size is 25, Gaussian noise was added to each dimension independently.}
		\end{figure} \label{Block25_NonPara_None}

		\begin{figure} 
			\includegraphics[width=\linewidth,height=16cm,keepaspectratio]{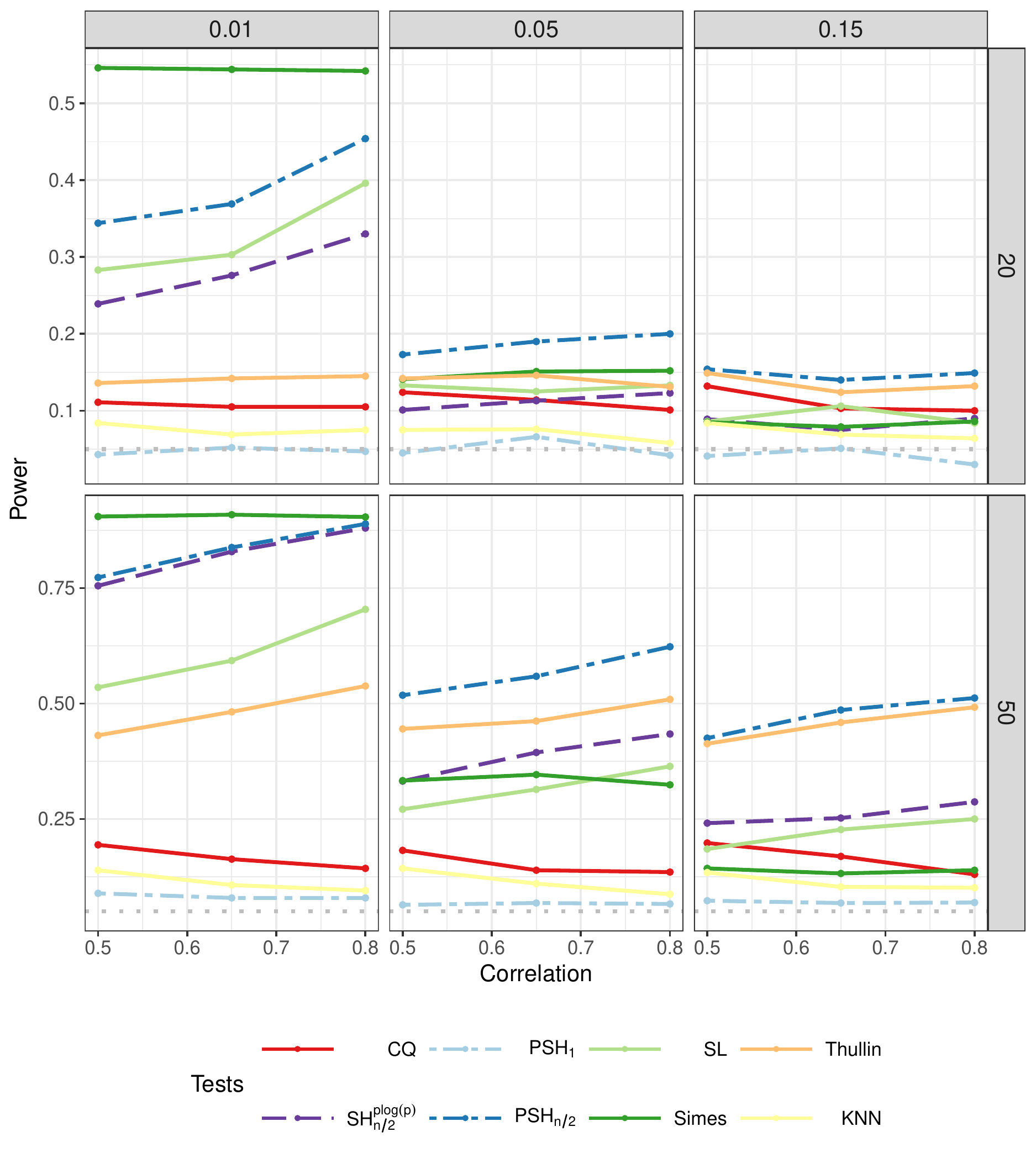}
			\caption{Non-parametric tests power comparison, \textit{block} covariance matrix, block size is 25, Exponential noise was added to each dimension independently.}
		\end{figure} \label{Block25_NonPara_Exp}

		\begin{figure} 
			\includegraphics[width=\linewidth,height=16cm,keepaspectratio]{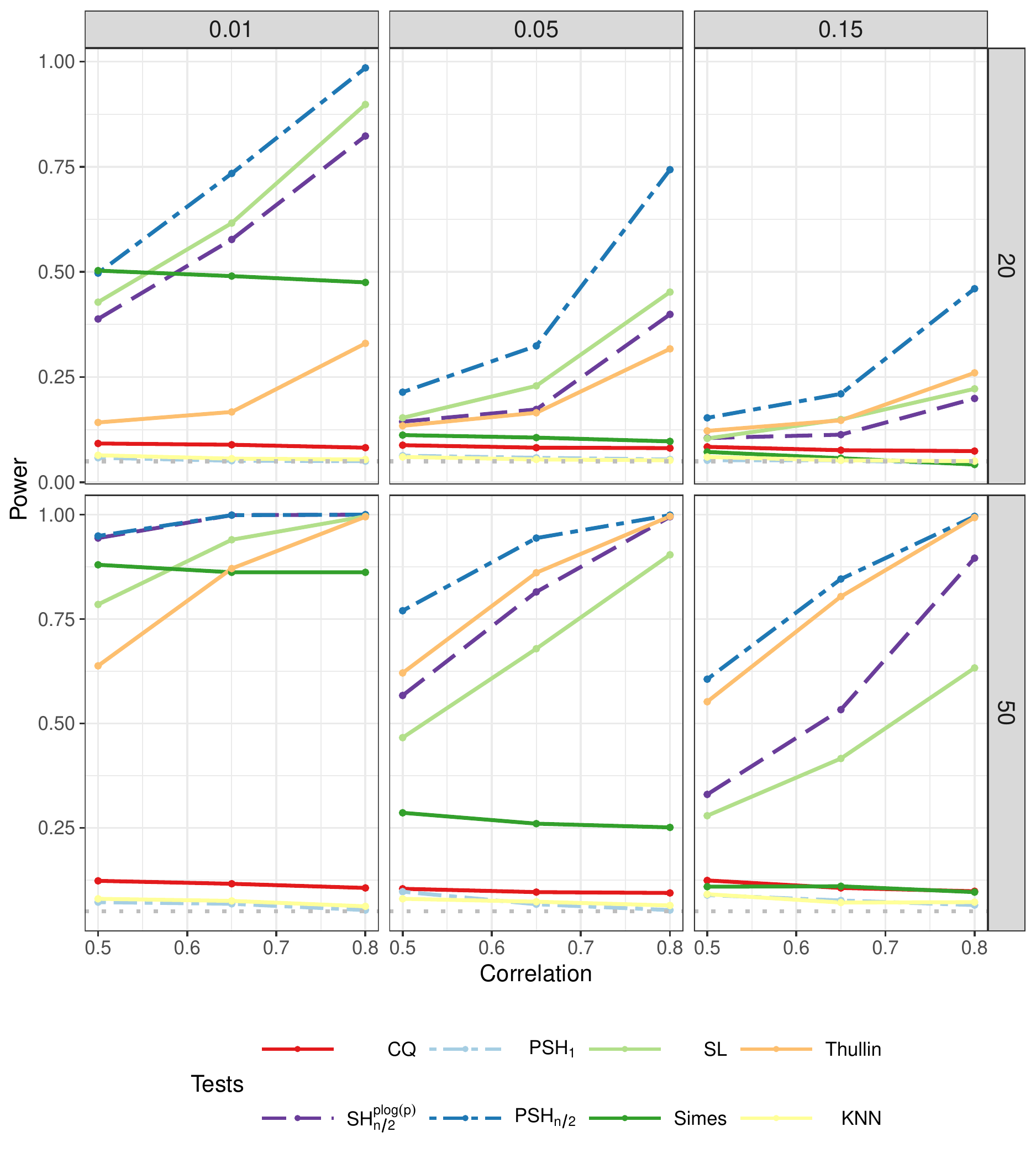}
			\caption{Non-parametric tests power comparison, \textit{block} covariance matrix, block size is 25, Laplace noise was added to each dimension independently.}
		\end{figure} \label{Block25_NonPara_Laplace}


		\begin{figure} 
			\includegraphics[width=\linewidth,height=16cm,keepaspectratio]{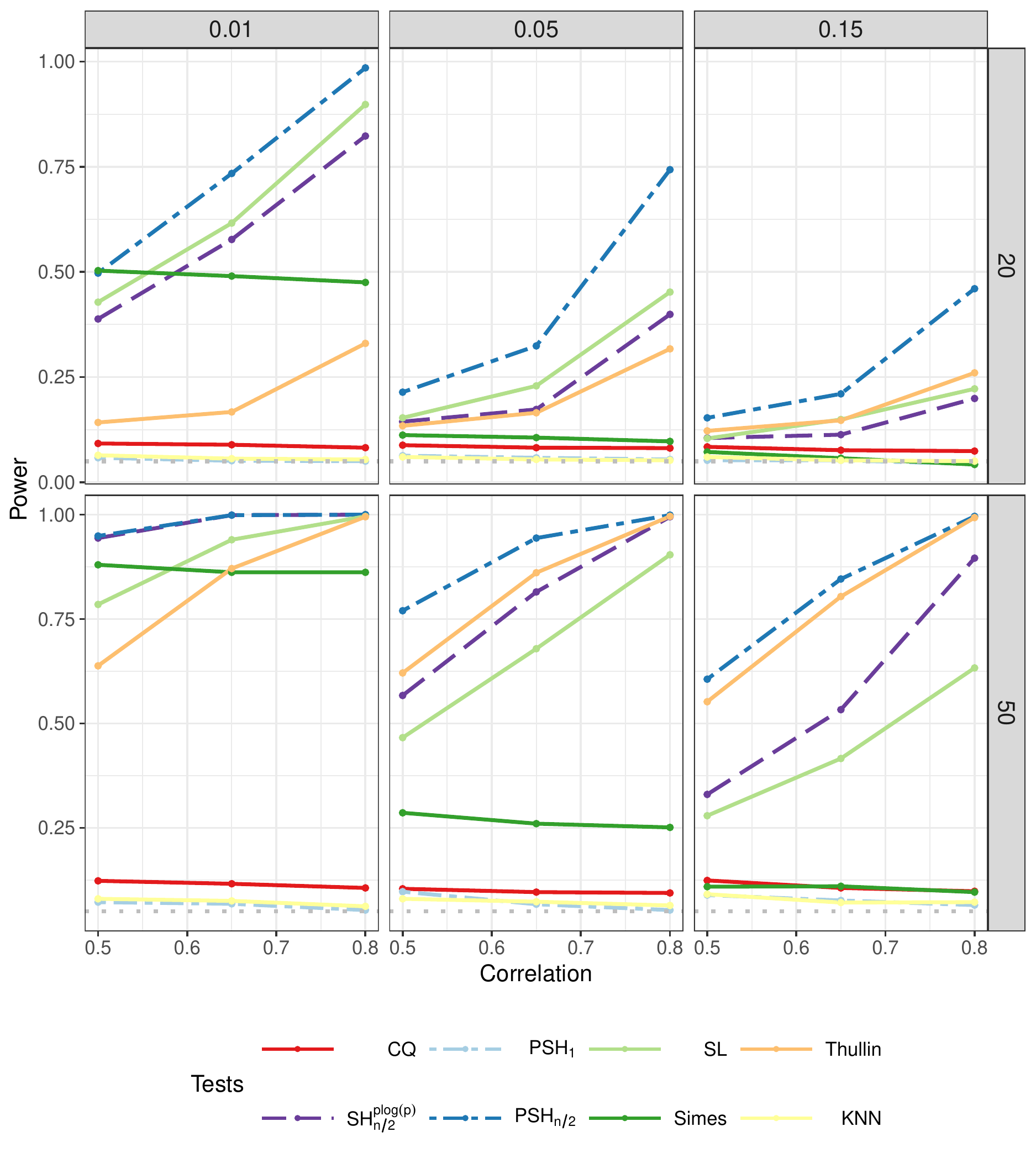}
			\caption{Non-parametric tests power comparison, \textit{block} covariance matrix, block size is 50, Gaussian noise was added to each dimension independently.}
		\end{figure} \label{Block50_NonPara_None}

		\begin{figure} 
			\includegraphics[width=\linewidth,height=16cm,keepaspectratio]{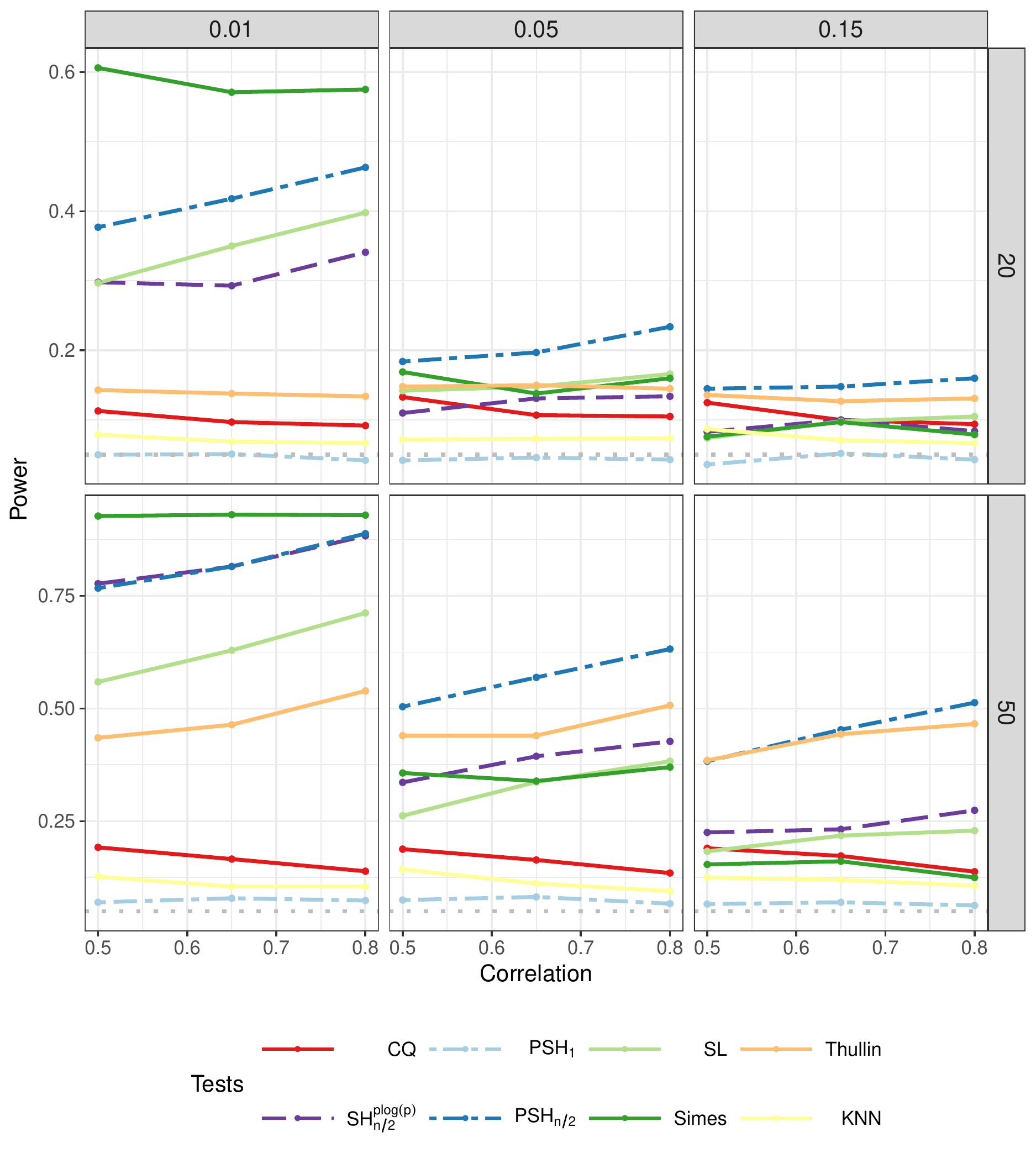}
			\caption{Non-parametric tests power comparison, \textit{block} covariance matrix, block size is 50, Exponential noise was added to each dimension independently.}
		\end{figure} \label{Block50_NonPara_Exp}

		\begin{figure} 
			\includegraphics[width=15cm,height=16cm,keepaspectratio]{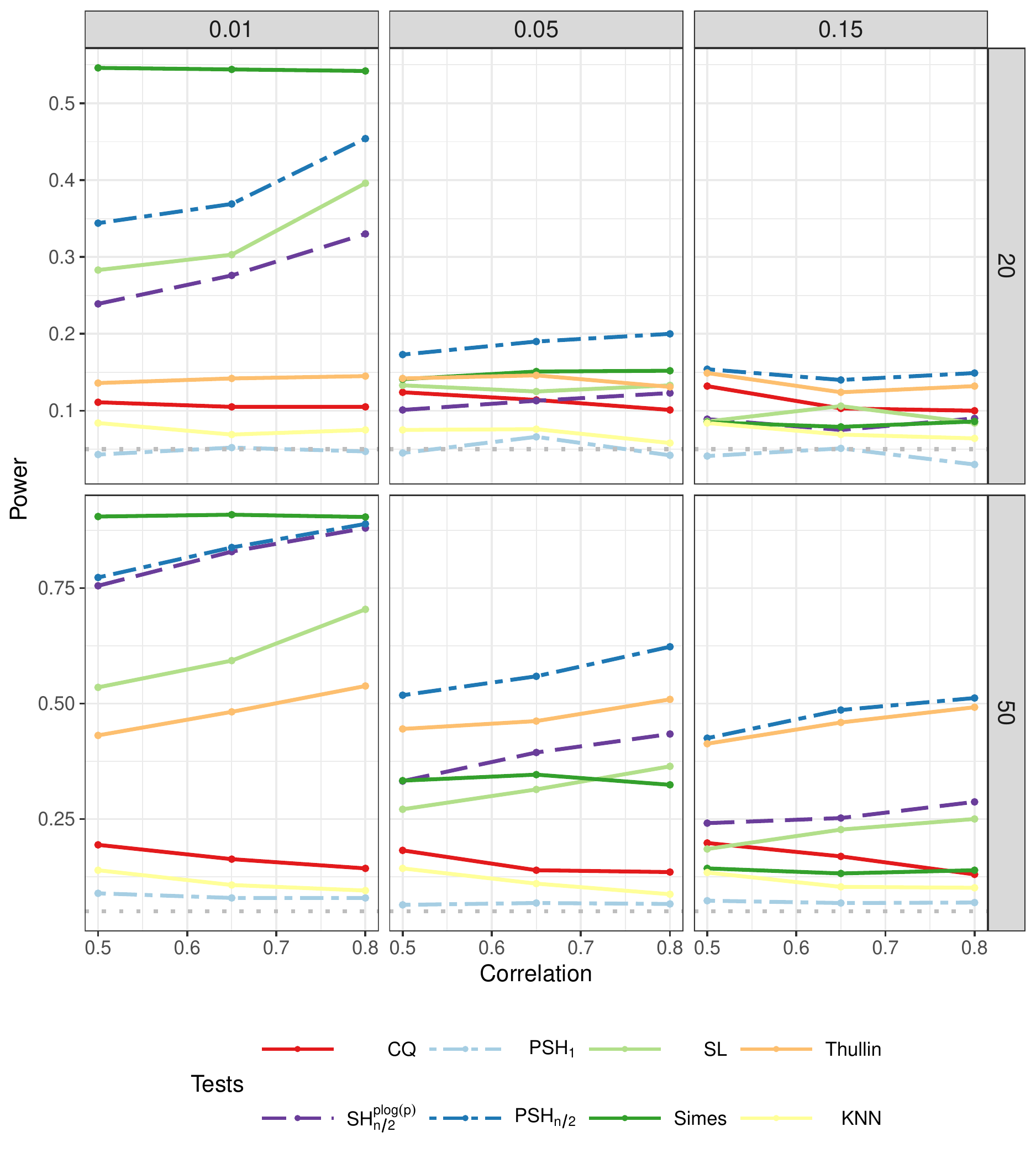}
			\caption{Non-parametric tests power comparison, \textit{block} covariance matrix, block size is 50, Laplace noise was added to each dimension independently.}
		\end{figure} \label{Block50_NonPara_Laplace}

	\end{document}